\documentclass[12pt,letterpaper]{article}
\usepackage{jheppub}

\usepackage{graphicx}
\usepackage{bbm}
\usepackage{amsmath}
\usepackage{amssymb}
\usepackage{mathtools}
\usepackage{amsthm}
\usepackage{marvosym}
\usepackage{dsfont}
\usepackage[labelformat=simple]{subcaption}
\usepackage{xcolor}

\usepackage{mdframed}

\allowdisplaybreaks

  \definecolor{dark-gray}{gray}{0.20}
  \definecolor{gray}{gray}{0.30}
  \definecolor{light-gray}{gray}{0.80}
  \definecolor{dark-red}{rgb}{0.7,0,0}
  \definecolor{dark-green}{rgb}{0.1,0.4,0}
  \definecolor{dark-blue}{rgb}{0.3,0.3,0.7}
  \definecolor{light-blue}{rgb}{0.8,0.8,1}
      \definecolor{swamp}{RGB}{240, 199, 197}

\newcommand{\be}{\begin{equation}}
\newcommand{\ee}{\end{equation}}

\def\be{\begin{equation}}
\def\ee{\end{equation}}

\newcommand{\beq}{\begin{equation}}  \newcommand{\eeq}{\end{equation}}

\newcommand{\cM}{\mathcal M}

\newcommand{\mKK}{m_{\textrm{KK}}}

\newcommand{\dd}{\mathrm{d}}

\newcommand{\Tr}{\mathrm{Tr}}

\numberwithin{equation}{section}

\def\simleq{\; \raise0.3ex\hbox{$<$\kern-0.75em
      \raise-1.1ex\hbox{$\sim$}}\; }
   \def\simgeq{\; \raise0.3ex\hbox{$>$\kern-0.75em
      \raise-1.1ex\hbox{$\sim$}}\; }

\numberwithin{equation}{section}

\hypersetup{
	colorlinks=true,
	linkcolor=dark-blue,
	citecolor=dark-red,
	urlcolor=dark-green,
	linktoc=page
}

\theoremstyle{remark}

\newtheoremstyle{named}{0.75\baselineskip}{0.75\baselineskip}{\itshape}{}{\bfseries}{.}{.5em}{#3}
\theoremstyle{named}
\newtheorem*{namedconjecture}{Conjecture}
\newtheorem*{prominentconjecture}{Conjecture}

\mdfsetup{skipabove=\topskip,skipbelow=\topskip}
\surroundwithmdframed[linewidth=1pt,innertopmargin=-5pt]{prominentconjecture}

\newcommand{\Mpld}{M_{\rm Pl;d}}
\newcommand{\Mten}{M_{\rm Pl;10}}
\newcommand{\Mnine}{M_{\rm Pl;9}}

\title{\centering Running Decompactification, Sliding Towers, and the Distance Conjecture}

\author{Muldrow Etheredge,$^{1}$}\author{Ben Heidenreich,$^{1}$}\author{Jacob McNamara,$^{2}$}\author{Tom Rudelius,$^{3}$} \author{Ignacio Ruiz,$^{4}$} 
\author{Irene Valenzuela$^{4,5}$} \affiliation{${}^{1}$Department of Physics, University of Massachusetts, Amherst, MA 01003, USA}
\affiliation{${}^{2}$Walter Burke Institute for Theoretical Physics,
California Institute of Technology, Pasadena, CA 91125, USA}
\affiliation{${}^{3}$Department of Physics, University of California, Berkeley, CA 94720, USA}
\affiliation{${}^{4}$Instituto de F\'isica T\'eorica UAM-CSIC and Departamento de F\'isica T\'eorica, Universidad Aut\'onoma de Madrid, Cantoblanco, 28049 Madrid, Spain}
\affiliation{${}^{5}$CERN, Theoretical Physics Department, 1211 Meyrin, Switzerland}

\emailAdd{metheredge@umass.edu}
\emailAdd{bheidenreich@umass.edu}
\emailAdd{jmcnamar@caltech.edu}
\emailAdd{t.rudelius@gmail.com}
\emailAdd{ignacio.ruiz@uam.es}
\emailAdd{irene.valenzuela@cern.ch}

\abstract{  
 We study towers of light particles that appear in infinite-distance limits of moduli spaces of 9-dimensional $\mathcal{N}=1$ string theories, some of which notably feature decompactification limits with running string coupling. The lightest tower in such decompactification limits consists of the non-BPS Kaluza-Klein modes of Type I$'$ string theory, whose masses depend nontrivially on the moduli of the theory.   We work out the moduli-dependence by explicit computation, finding that despite the running decompactification the Distance Conjecture remains satisfied with an exponential decay rate $\alpha \ge \frac{1}{\sqrt{d-2}}$ in accordance with the sharpened Distance Conjecture.  The related sharpened Convex Hull Scalar Weak Gravity Conjecture also passes stringent tests. Our results non-trivially test the Emergent String Conjecture, while highlighting the important subtlety that decompactification can lead to a running solution rather than to a higher-dimensional vacuum.
}

\hypersetup{pageanchor=false}

\begin{document}
\makeatletter
\let\old@fpheader\@fpheader
\renewcommand{\@fpheader}{\old@fpheader \vspace*{-0.1cm} \hfill
{ACFI-T23-02}\\ \vspace*{-0.1cm} \hfill CERN-TH-2023-121\\ \vspace*{-0.1cm} \hfill IFT-UAM/CSIC-23-78}
\makeatother

\maketitle

\section{Introduction}
\label{sec:intro}

While much of quantum gravity remains shrouded in mystery, some corners of the landscape are relatively well understood. Asymptotic regimes of known moduli spaces appear to display universal behavior, which has led to the development of a number of quantum gravity conjectures (known as Swampland conjectures) regarding such behavior.
The oldest of these is the Distance Conjecture of Ooguri and Vafa \cite{Ooguri:2006in}, which states:
\begin{prominentconjecture}[The Distance Conjecture]
Let $\cM$ be the moduli space of a quantum gravity theory in $d \geq 4$ dimensions, parametrized by vacuum expectation values of massless scalar fields. Fixing a point $p_0 \in \mathcal{M}$, the theory at a point $p \in \mathcal{M}$ sufficiently far away in the moduli space has an infinite tower of light particles, each with mass  in Planck units $(\kappa_d^2 = \Mpld^{2-d} = 1)$ scaling as 
\be
m(p) \sim \exp( -\alpha \,d(p, p_0) )\quad \text{as}\quad  d(p, p_0)\rightarrow \infty \,,
\label{DCdef}
\ee 
where $d(p, p_0)$ is the length of the shortest geodesic in $\mathcal{M}$ between $p$ and $p_0$, and $\alpha>0$ is some order-one number.
\end{prominentconjecture}
\noindent
This conjecture has been extensively tested in a plethora of string theory compactifications \cite{Grimm:2018ohb,Blumenhagen:2018nts,Grimm:2018cpv,Corvilain:2018lgw,Joshi:2019nzi,Marchesano:2019ifh,Font:2019cxq,Erkinger:2019umg, Buratti:2018xjt, Heidenreich:2018kpg, Gendler:2020dfp, Lanza:2020qmt, Klaewer:2020lfg} and plays a key role in the Swampland program \cite{vanBeest:2021lhn, Palti:2019pca, Vafa:2005ui,Blumenhagen:2017cxt}, which aims to determine the constraints that any EFT must satisfy to be UV completed in quantum gravity.

Significant effort has been invested in sharpening and refining the Distance Conjecture, both as a means to test it more stringently as well as to expand its consequences.
One notable refinement of the Distance Conjecture, proposed by~\cite{Lee:2019wij}, constrains the microscopic nature of the tower of states:\footnote{Substantial work has also been done on a different class of refinements constraining the distance travelled before light towers appears, see \cite{Baume:2016psm, Klaewer:2016kiy,Blumenhagen:2018nts,Erkinger:2019umg,  Rudelius:2023mjy}, but we will not discuss these conjectures further in the present work.}

\begin{namedconjecture}[The Emergent String Conjecture]
Every infinite-distance limit in the moduli space of a quantum gravity theory is either an emergent string limit (featuring a fundamental string with a weakly coupled tower of  string oscillator modes) or a decompactification limit (featuring a tower of Kaluza-Klein modes).
\end{namedconjecture}
\noindent
The Emergent String Conjecture is supported by all known string theory examples in flat space \cite{Lee:2018urn,Lee:2019xtm, Lanza:2021udy, Baume:2019sry, Xu:2020nlh} and holographic AdS compactifications \cite{Baume:2020dqd,Perlmutter:2020buo,Baume:2023msm}, though it needs to be slightly modified to account for the non-holographic AdS cases where infinite-distance limits associated to free points in the dual conformal manifold feature a tower of higher spin operators that  are not necessarily dual to the fundamental string (see \cite{Perlmutter:2020buo}).

Another notable refinement of the Distance Conjecture, proposed by \cite{Etheredge:2022opl}, places a sharp lower bound on the possible values of the exponential rate $\alpha$ in \eqref{DCdef}: 

\begin{namedconjecture}[The Sharpened Distance Conjecture]
The Distance Conjecture remains true with the added requirement that
\beq
\alpha \geq \frac{1}{\sqrt{d-2}},
\eeq
where $d$ is the spacetime dimension.
\end{namedconjecture}

\noindent
In fact, the sharpened Distance Conjecture and the Emergent String Conjecture are closely related, since $\alpha_{\rm osc}= \frac{1}{\sqrt{d-2}}$ is precisely the exponential rate of the tower of oscillator modes of a perturbative fundamental string, whereas Kaluza-Klein (KK) modes typically have a larger exponential rate.

 More concretely, the exponential rate for a KK tower in a toroidal compactification is given by (see, e.g., \cite{Etheredge:2022opl})
\beq
\alpha_{\rm KK}^{(n)}=\sqrt{\frac{d+n-2}{n(d-2)}} \,,
\label{KKalpha}
\eeq
where $d$ is the space-time dimension and $n$ the number of decompactifying dimensions, which is indeed larger than $\alpha_{\rm osc}= \frac{1}{\sqrt{d-2}}$ for $d>2$.

Since \eqref{KKalpha} applies equally to the overall volume modulus of an arbitrary Ricci-flat manifold, it is tempting to conclude that $\alpha_{\rm osc}$ and $\alpha_{\rm KK}^{(n)}$ are the only possible values for $\alpha$ compatible with the Emergent String Conjecture.\footnote{To be precise, these $\alpha$ values are associated to certain ``pure'' emergent string and/or decompactification limits; there are often ``mixed'' limits that continuously interpolate between these, which have intermediate values $\alpha$ as well. A sharper statement would be that $|\vec\zeta|$ as defined in~\S\ref{sec:review_of_swgc} remains fixed at one of these special values for each tower satisfying the Distance Conjecture. We will show, however, that even this is false.} However, we will see that this is incorrect, since the exponential rate of a KK tower can differ and possibly become smaller than \eqref{KKalpha} when the compactification metric is not a direct product but instead involves warping. Warped compactifications have been extensively studied in the context of string theory, but have not been discussed yet in the context of the Distance Conjecture to the best of our knowledge.

In this paper, we will explore the Emergent String Conjecture and the sharpened Distance Conjecture in the moduli space of $\mathcal{N}=1$ supersymmetric string theories in nine dimensions, which arise from heterotic string theory compactified on a circle.  Our results are consistent with both conjectures provided that we clarify the Emergent String Conjecture in an important way: some infinite-distance limits in moduli space do not lead to either an emergent tensionless string or a higher-dimensional vacuum, but rather to a higher-dimensional running solution. In other words, the decompactification limits specified by the Emergent String Conjecture may or may not lead to vacuum solutions of the higher-dimensional theory. The decompactification limit of Type I$'$ string theory in nine dimensions with a nontrivial dilaton profile is a prototypical example of the latter. 

Nonetheless, we will show in simple examples that these running solutions still feature Kaluza-Klein towers which satisfy the Distance Conjecture with $\alpha \geq \frac{1}{\sqrt{d-2}}$, consistent with the sharpened Distance Conjecture of \cite{Etheredge:2022opl}. This is made possible by the fact that the corresponding Kaluza-Klein modes are not BPS and consequently their masses are a rather complicated function of moduli space position, so the exponential rate changes depending on the asymptotic geodesic trajectory that bring us to infinite distance. By careful computation, we determine this function and show that the exponential rate for the KK tower can get as small as $\frac{5}{2\sqrt{7}}$ in these nine dimensional examples, which is still compatible with the bound  $\alpha \geq \frac{1}{\sqrt{d-2}}=\frac{1}{\sqrt{7}}$ in $d=9$, but does not correspond to one of the special values \eqref{KKalpha} for any integer $n$.

We will also show that these nine dimensional compactifications satisfy a ``convex hull'' version of the Scalar Weak Gravity Conjecture (SWGC)  \cite{Palti:2017elp, Calderon-Infante:2020dhm, Etheredge:2022opl, Etheredge:SWGCasDC} (reviewed below in \S\ref{sec:review_of_swgc}). This is especially remarkable in light of the aforementioned moduli-dependence of the masses of the non-BPS particles, which implies that the convex hull varies as a function of the moduli. We will further see that the Distance Conjecture itself resembles a convex hull condition in each asymptotic region of moduli space (as proposed previously in \cite{Calderon-Infante:2020dhm} under the name of the Convex Hull Distance Conjecture), but this requires different convex hulls in different region of moduli space that do not obviously combine into a single global picture.

The remainder of this paper is structured as follows: in \S\ref{sec:review_of_swgc}, we review the sharpened Distance Conjecture and the Convex Hull SWGC, introducing the machinery we will need for our subsequent analysis and applying it to Type II string theory on a circle as a warm-up example. In \S\ref{sec:puzzle_in_9d}, we take a first look at heterotic string theory on a circle, noting how the self-T-duality of the theory leads to a puzzle. In \S\ref{sec:decompactification}, we review how decompactification limits of Type I$'$ string theory introduce new complications, leading to ten-dimensional running solutions. We then explicitly compute the spectrum of Kaluza-Klein modes for Type I$'$ string theory on a circle and show how this resolves the puzzle mentioned above in a manner consistent with the sharpened Distance Conjecture and the SWGC (leaving the details of the calculation to Appendices~\ref{app:dual}--\ref{app:metric}). In \S\ref{sec: other mod spaces} we extend our analysis to other nine-dimensional theories of lower rank, thereby checking the Distance Conjecture and its various refinements in a wide range of 9d theories with sixteen supercharges. In \S\ref{CONC}, we conclude by summarizing our results and highlighting interesting directions for future research.

\section{The Distance Conjecture and Convex Hulls}
\label{sec:review_of_swgc}

Consider a theory in $d$ dimensions with a set of massless scalar fields (moduli) $\phi^i$ weakly coupled to gravity, with action given by
\beq\label{eq:gravity_with_massless_scalars}
S=M_{\rm Pl;d}^{d-2}\int \dd^dx \sqrt{-h}\left(\frac{R}2-\frac12 \mathsf{G}_{ij} \partial_\mu\phi^i\partial^\mu\phi^j+\dots\right),
\eeq
where $\mathsf{G}_{ij}$ is the field space metric and the geodesic field distance is given in Planck units by
\beq
d(\phi_0,\phi)=\int_{\phi_0}^\phi \sqrt{\mathsf{G}_{ij}\dd\phi^i \dd\phi^j}\ .
\eeq

According to the Distance Conjecture, there will be a tower of particles that becomes exponentially light at every infinite-distance limit in this moduli space. To understand precisely how this occurs as a function of the moduli, it is convenient to define the scalar charge-to-mass vector of a particle of mass $m$ as
\be
\zeta_i \equiv - \frac\partial{\partial \phi^i}\log m\,, \label{zetavec}
\ee
following, e.g., \cite{Calderon-Infante:2020dhm,Etheredge:2022opl}, where the derivative is evaluated with the $d$-dimensional Planck mass held fixed. Associated to the moduli-space one-form $\zeta_i$ there is a dot product $\zeta^2 = \vec\zeta \cdot \vec\zeta \equiv \mathsf{G}^{ij} \zeta_i \zeta_j$ defined by the inverse of the metric on moduli space $\mathsf{G}^{ij}$. In practice, we pick an $n$-bein $e_a^i e_b^j \delta^{ab}=\mathsf{G}^{ij}$ and write $\zeta_a = e_a^i \zeta_i$ in orthonormal components, which has the advantage that the dot product of $\vec\zeta$ vectors is the Cartesian dot product, but at the expense of having to choose a frame at each point in moduli space.

To understand why $\vec\zeta$ is the scalar charge-to-mass ratio, note that the moduli mediate long-range forces between particles whose masses depend their vacuum expectation values. The strength of these interactions is proportional to $\partial_{\phi^i} m$, as can be read off from the Lagrangian expanded about a given point in the moduli space
\beq
\mathcal{L}\supset m^2(\phi)\chi^2=(m_0^2+2m_0(\partial_{\phi^i} m)\phi^i)\chi^2+\dots \,.
\eeq
Thus, by direct analogy with gauge charges, $\mu_i = - \partial_{\phi^i} m$ are the scalar charges (the sign being purely conventional) and $\frac{\mu_i}{m} = - \partial_{\phi^i} \log m = \zeta_i$ is the vector of scalar charge-to-mass ratios.

Now consider the vicinity of some infinite-distance locus, commonly known as an \emph{asymptotic region}. Given a particle that is exponentially light in accordance with the Distance Conjecture \eqref{DCdef}, the exponential rate at which its mass decreases is given by the projection of the scalar charge-to-mass vector $\vec\zeta$ along the corresponding geodesic trajectory approaching the infinite-distance limit, i.e.
\beq
\label{SDCrate}
\alpha=\vec{\zeta}\cdot \hat{\tau},
\eeq
where $\hat{\tau}^a=e^a_i\frac{\partial_\lambda\phi^i(\lambda)}{\|\partial_\lambda\vec{\phi}\|}$ is the normalized tangent vector to the asymptotic geodesic trajectory $\vec{\phi}(\lambda)$.

On the other hand, working by analogy to the Weak Gravity Conjecture \cite{Arkanihamed:2006dz} and motivated by the connection to scalar forces, reference \cite{Palti:2017elp} proposed a Scalar Weak Gravity Conjecture (SWGC), as follows:
           \begin{namedconjecture}[Scalar Weak Gravity Conjecture] In a quantum gravity theory with massless scalar fields \eqref{eq:gravity_with_massless_scalars}, at every point in moduli space there exists a state with sufficiently large scalar charge-to-mass ratio $|\vec\zeta| \geq \alpha_{\text{min}}$ for some order-one constant $\alpha_{\text{min}}$.\footnote{More specifically, reference \cite{Palti:2017elp} required the existence of some (possibly higher dimensional) state upon which the scalar force would act more strongly than the gravitational force. Further refinements along these lines were proposed in \cite{Gonzalo:2019gjp,Freivogel:2019mtr,DallAgata:2020ino,Benakli:2020pkm,Gonzalo:2020kke}. In this paper, we are instead interested in the case where the state is a particle and $\alpha_{\text min}$ is fixed not by a force condition but by its relationship to the exponential rate in the (sharpened) Distance Conjecture.} 
            \end{namedconjecture}
\noindent
The SWGC is a local statement in moduli space, as it only involves the first derivatives of the masses of states with respect to the moduli fields. Comparing with~\eqref{SDCrate}, it is evident that there is some connection between this conjecture and the Distance Conjecture, see, e.g., \cite{Palti:2017elp, Calderon-Infante:2020dhm,Grimm:2018ohb}. For instance, the Distance Conjecture implies that a tower version of the SWGC holds at least asymptotically, with $\alpha_{\text{min}}$ equal to the minimum allowed exponential rate (believed to be $\frac{1}{\sqrt{d-2}}$ per the sharpened Distance Conjecture \cite{Etheredge:2022opl}). Conversely, given a tower of particles satisfying the SWGC with this value of $\alpha_{\text{min}}$, $|\vec\zeta| \ge \alpha_{\text{min}}$ is the exponential rate at which the tower becomes light along its own gradient flow trajectory (i.e. $\partial_\lambda\phi^i\propto-\partial^{i}m(\phi)$), and we recover the Distance Conjecture for this particular asymptotic limit.

However, even in its tower form, the original version of the SWGC is too weak to make a useful connection with the Distance Conjecture in theories with multiple moduli, since only one particle/tower is required by the conjecture, and this is not enough to satisfy the Distance Conjecture in all possible asymptotic limits. To address this weakness, the conjecture has to be strengthened with some kind of convex hull condition to account for the various directions in which different asymptotic limits lie.

As discussed below, there have been two notable attempts to do so~\cite{Calderon-Infante:2020dhm,Etheredge:2022opl}, with different strengths and weaknesses. The first of these---the Convex Hull Distance Conjecture---relies on certain global properties of the moduli space in asymptotic limits and straightforwardly implies the Distance Conjecture. By contrast, the second---the (sharpened) Convex Hull SWGC---is a purely local statement like the original SWGC, relying on few preconditions, but it requires us to consider of both light and heavy towers and the connection to the Distance Conjecture is non-trivial (see, e.g.,~\cite{Etheredge:SWGCasDC}).
    
 To introduce these conjectures, first note that in the presence of multiple moduli fields, there can be asymptotic geodesics (say, with normalized tangent vector $\hat \tau$) that are not parallel to the scalar charge-to-mass vector $\vec \zeta$ of any tower. When this happens, the exponential rate $\alpha_{\textrm{max}}(\hat\tau)$ of the leading (i.e., lightest) tower along such a geodesic will be given as in \eqref{SDCrate} by the maximum value of $\vec \zeta \cdot \hat \tau$ among the different towers that exist in this asymptotic regime. Thus, the Distance Conjecture holds with minimum rate $\alpha_{\textrm{min}}$ if and only if we have $\alpha_{\textrm{max}}(\hat \tau) \geq \alpha_{\textrm{min}}$ for all asymptotic directions $\hat \tau$.\footnote{The general procedure followed in this paper is to choose a slice of the tangent space of the moduli space which has dimension equal to the codimension of the infinite-distance loci. This way all radial vectors in the slice correspond to tangent vectors of geodesics approaching the infinite-distance loci. However, it is also interesting to analyze higher dimensional slices such that not all the vectors are associated to geodesics, and use the convex hull condition as a criterium to select what trajectories could become geodesics in the IR upon adding a scalar potential. This has been used to put constraints on the scalar potential from using only the Distance Conjecture (see \cite{Calderon-Infante:2020dhm}).}
    
In all examples checked so far in the literature \cite{Gendler:2020dfp, Calderon-Infante:2020dhm,Grimm:2022sbl,Etheredge:2022opl}, the convex hull of the $\vec\zeta$-vectors of the towers that become light remains unchanged as we move in a given asymptotic region of the moduli space, even if the individual $\vec\zeta$-vectors move. When this happens, the Distance Conjecture can be reformulated as in  \cite{Calderon-Infante:2020dhm} as the following convex hull condition:

\begin{namedconjecture}[Convex Hull Distance Conjecture] In any given asymptotic region of a quantum gravity theory, the outside boundary of the convex hull generated by the $\vec\zeta$-vectors \eqref{SDCrate} of all light towers must remain outside the ball of radius $\alpha_{\textrm{min}}$ in the range of directions defining the asymptotic region.
\end{namedconjecture}

\noindent This formulation of the Distance Conjecture is powerful because it encodes global information about the different infinite-distance limits in a given asymptotic region rather than each asymptotic geodesic independently.\footnote{It can also be used to either predict the existence of new light towers of states in an EFT or to constrain the possible trajectories along which the Distance Conjecture is satisfied, and therefore, the level of non-geodesicity that should be allowed in the asymptotic valleys of the scalar potential  \cite{Calderon-Infante:2020dhm}.} We will see that it also captures the information needed to derive the weakly coupled dual description that emerges at infinite field distance.  

However, it is not obvious at all whether this formulation of the Distance Conjecture makes sense when the convex hull of the $\vec \zeta$-vectors of the towers changes as we move in the moduli space. When this happens, it is useful to consider a closely related statement that makes sense locally at any point in moduli space rather than in an entire asymptotic region:

\begin{namedconjecture}[Convex Hull SWGC] In a quantum gravity theory with massless scalar fields \eqref{eq:gravity_with_massless_scalars}, at every point in moduli space, the convex hull generated by the $\vec\zeta$-vectors \eqref{SDCrate} of all massive states contains a ball of radius $\alpha_{\textrm{min}}$ centered at the origin of the scalar charge-to-mass vector space.
\end{namedconjecture}

\noindent       
This conjecture has been extensively discussed and tested in~\cite{Etheredge:2022opl} with the specific choice $\alpha_{\textrm{min}} = \frac{1}{\sqrt{d-2}}$ (as motivated by the sharpened Distance Conjecture). Note that the Convex Hull SWGC (applied to towers of states) differs from the Convex Hull Distance Conjecture because it involves both light and heavy states, and is required to hold everywhere in moduli space. In the remainder of this paper, we will refer to the Convex Hull SWGC as simply the SWGC; the reader should take care not to confuse this with the related but distinct version of the SWGC originally proposed in \cite{Palti:2017elp}.
                        
            The goal of this paper is precisely to consider examples in which the $\vec\zeta$-vectors change dramatically as we move in the moduli space and to determine the fate of the various conjectures described above. In particular, we will explore in detail the case of heterotic string theory compactified on a circle, where we will see that the $\vec\zeta$-vectors of certain non-BPS towers are highly moduli-dependent in regions of the moduli space corresponding to warped compactifications. Interestingly, we will see that all the above conjectures still hold in a non-trivial way with $\alpha_{\rm min}=\frac1{\sqrt{d-2}}$. Moreover, we will see that in each asymptotic region, the Distance Conjecture will still resemble a convex hull condition, but will require different convex hulls in different regions that do not obviously combine into any single global picture.
            
            \begin{figure}
\begin{center}
\begin{subfigure}{0.475\textwidth}
\includegraphics[width = 70mm]{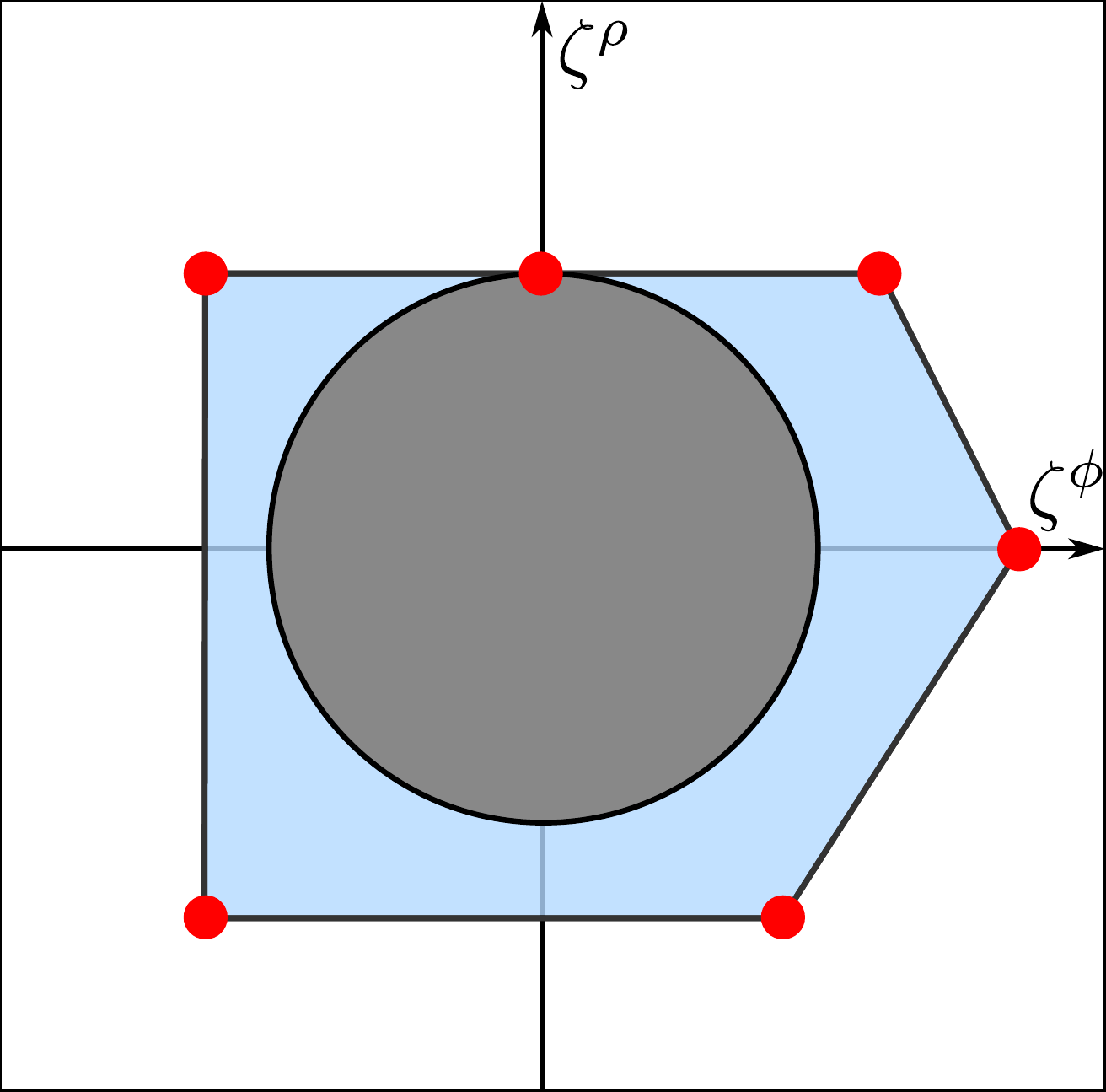} 
\caption{Convex Hull SWGC Plot}
\label{subfig:CHSWGC}
\end{subfigure}
\hspace{0.5cm}
\begin{subfigure}{0.475\textwidth}
\includegraphics[width = 70mm]{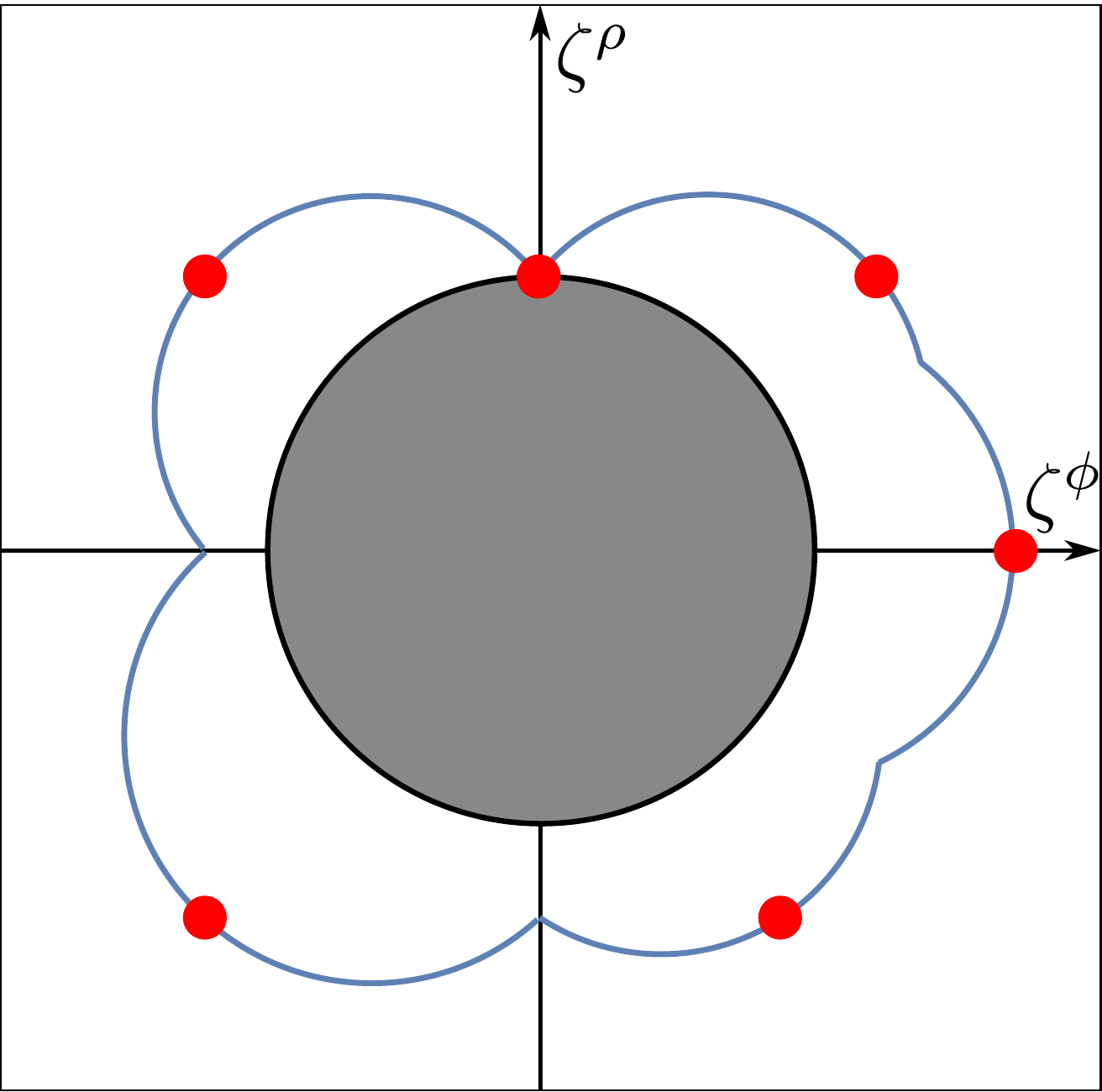}
\caption{Max-$\alpha$ Plot}
\label{subfig:MaxAlpha}
\end{subfigure}
\end{center}
\caption{{\bf (a)} Convex hull of the scalar charge-to-mass vectors. The Convex Hull SWGC holds that this convex hull must contain the ball of radius $\alpha_{\textrm{min}}$, which we take to be $\frac{1}{\sqrt{d-2}}$ in this paper. Points at which the boundary of the convex hull meets the boundary of the sphere correspond to points at which the SWGC is saturated, and in known examples these points are always populated by towers of string oscillator modes. {\bf (b)} Exponential rate $\alpha_{\rm max}(\hat \tau)$ (in blue arcs) of the leading tower in the direction $\hat \tau$. We can see that $\alpha_{\rm max}(\hat \tau) \geq \alpha_{\rm min}$ if and only if the convex hull of $\vec \zeta$-vectors of towers contains the ball of radius $\alpha_{\rm min}$. Each of the bubbles correspond to regions of asymptotic directions in the moduli space with different leading towers, with the boundary between them corresponding to directions along which two or more towers decay at the same rate.}
\label{fakefig}
\end{figure}
            
            To illustrate our examples, we will employ two different types of plots. The first of these is the SWGC plot, where we plot the various $\vec \zeta$-vectors of towers of states and draw their convex hull. An example of a SWGC plot is illustrated in Figure \ref{subfig:CHSWGC}, where the towers are indicated by red dots and the convex hull is indicated by the light blue region. The SWGC plot is defined at any fixed point in moduli space, and can change as we move from point to point. The second plot is the max-$\alpha$ plot, where we plot the exponential rate $\alpha_{\rm max}(\hat \tau)$ of the leading tower as a function of the asymptotic direction $\hat \tau$. We illustrate a max-$\alpha$ plot in Figure \ref{subfig:MaxAlpha}, with the function $\alpha_{\rm max}(\hat \tau)$ plotted in blue. Notice that the exponential rate of a given tower is a function $\alpha(\theta)$ of the angle $\theta$ between $\vec \zeta$ and $\tau$, and is given by a sphere of radius $|\vec \zeta|/2$ that goes through the point $\vec \zeta$ and the origin, so that $\alpha(\theta)$ varies between $0$ and $|\vec \zeta|$. The max-$\alpha$ plot is defined globally in the moduli space, and doesn't depend on any reference point.

\subsection{Example: Type IIB on a circle\label{sec: IIB circ}}

As an illustrative example, we consider the case of Type IIB string theory compactified on a circle. This theory was previously shown to satisfy the sharpened Distance Conjecture and the SWGC in \cite{Etheredge:2022opl}, and it will serve as a useful warmup for our primary case of interest, namely heterotic string theory on a circle.

For simplicity, we will set the Type IIB axion $C_0$ to vanish. Upon compactification to nine dimensions, this leaves a flat two-dimensional moduli space parametrized by the 10D dilaton $\Phi=\log g_s$ and the radius $R_{\rm IIB}$ of the circle. We define a canonically normalized dilaton by setting $\phi = - \sqrt{2} \Phi$ (we include a minus sign so that large $\phi$ corresponds to weak IIB string coupling) and radion $\rho=\sqrt{\frac{8}{7}}\log R_{\rm IIB}$. The 9d action can be obtained from dimensionally reducing the Einstein-dilaton part of the 10d Type IIB effective action as follows:
\begin{align}\label{eq: IIB action}
	S_{\rm IIB}&\supset\frac{1}{2\kappa_{10}^2}\int \dd^{10}x\sqrt{-G}e^{-\phi/\sqrt{2}}\left\{R_G+2\partial_M\phi\partial^M\phi\right\}\notag\\
	&=\frac{1}{2\kappa_{9}^2}\int\dd^9 x\sqrt{-g}\left\{R_g-(\partial\phi)^2-\frac{8}{7}(\partial\log R_{\rm IIB})^2\right\}\notag\\
	&=\frac{1}{2\kappa_{9}^2}\int\dd^9 x\sqrt{-g}\left\{R_g-(\partial\phi)^2-(\partial \rho)^2\right\},
\end{align}
where $G_{MN}$ and $g_{\mu\nu}$ are respectively the 10-dimensional string frame and the 9-dimensional Einstein frame metrics.

Type IIB string theory in ten dimensions features a fundamental string whose tension is given by
\begin{equation}
T =  \frac{ 2 \pi \Mten^2}{ (4 \pi)^{1/4} } \exp\left( -  \frac{\phi}{\sqrt{2}}\right)\,.
\end{equation}
There is also a D-string with tension given by
\begin{equation} 
\tilde T = \frac{ 2 \pi  \Mten^2 }{ (4 \pi)^{1/4} } \exp\left( + \frac{\phi}{\sqrt{2}}\right)\,.
\end{equation}
Upon dimensional reduction, each of these strings gives rise to a tower of string oscillator modes as well as a tower of string winding modes. The former towers have characteristic mass scales
\begin{equation} \label{eq: IIB 1}
m_\text{osc} = \frac{ 2 \pi  \Mnine }{ (4 \pi)^{1/7} } \exp\left( - \frac{ \phi  }{ \sqrt{7} }- \frac{\rho}{\sqrt{56}} \right)\,,~~~ \tilde m_\text{D-osc} = \frac{ 2 \pi  \Mnine }{ (4 \pi)^{1/7} }  \exp\left( + \frac{\phi  }{ \sqrt{7} }- \frac{\rho}{\sqrt{56}} \right)\,,
\end{equation}
while the latter towers have characteristic mass scales
\begin{equation} \label{eq: IIB 2}
m_{\text{w}} = \frac{ 2 \pi  \Mnine }{ (4 \pi)^{1/7} } \exp\left( - \frac{\phi  }{ \sqrt{2} }+ \frac{ 3 \rho}{\sqrt{14}} \right)\,,~~~ \tilde m_{\text{D-w}}  = \frac{ 2 \pi  \Mnine }{ (4 \pi)^{1/7} } \exp\left( +\frac{ \phi  }{ \sqrt{2} } + \frac{3 \rho}{\sqrt{14}} \right)\,,
\end{equation}
There is also a tower of Kaluza-Klein modes with associated mass scale 
\begin{equation} \label{eq: IIB 3}
m_{\text{KK}} = \frac{ 2 \pi  \Mnine }{ (4 \pi)^{1/7} } \exp\left( - \sqrt{\frac{ 8 }{ 7 }} \rho\right)\,.
\end{equation}
These five towers of particles yield scalar charge-to-mass vectors given by
\begin{align}\label{eq: IIB S1 vecs}
\vec{\zeta}_{\text{osc}}  = \left(\frac{1}{\sqrt{8}} , \frac{1}{\sqrt{56}}\right) ~~~&, ~~~
\vec{\zeta}_{\text{D-osc}}  = \left(-\frac{1}{\sqrt{8}} , \frac{1}{\sqrt{56}}\right) \notag \\ 
\vec{\zeta}_{\text{w}}  = \left(\frac{1}{\sqrt{2}} , -\frac{3}{\sqrt{14}}\right)~~~
&,~~~\vec{\zeta}_{\text{D-w}}  = \left(-\frac{1}{\sqrt{2}} , -\frac{3}{\sqrt{14}}\right) \\\vec{\zeta}_{\text{KK}}  &= \left(0, \sqrt{\frac{8}{7}} \right) \,. \notag
\end{align}
Notably, these vectors are independent of the vacuum expectation values of the dilaton and the radion, so they do not change as we move in the moduli space. Relatedly, all of the particles in these towers are BPS. The scalar charge-to-mass ratio of the KK modes and winding modes becomes
\beq\label{eq: bps KK lenght}
|\vec\zeta_{\text{KK}}|=|\vec\zeta_{\text{w}}|=|\vec\zeta_{\text{D-w}}|=\sqrt{\frac87}
\eeq
as expected from decompactifying one extra dimension \cite{Etheredge:2022opl}, while
\beq
|\vec\zeta_{\rm osc}|=|\vec\zeta_{\text{D-osc}}|=\frac1{\sqrt{7}}
\eeq
corresponds to the expected result for the oscillation modes of a critical perturbative string \cite{Etheredge:2022opl}.

\begin{figure}
\begin{center}
\includegraphics[width = 80mm]{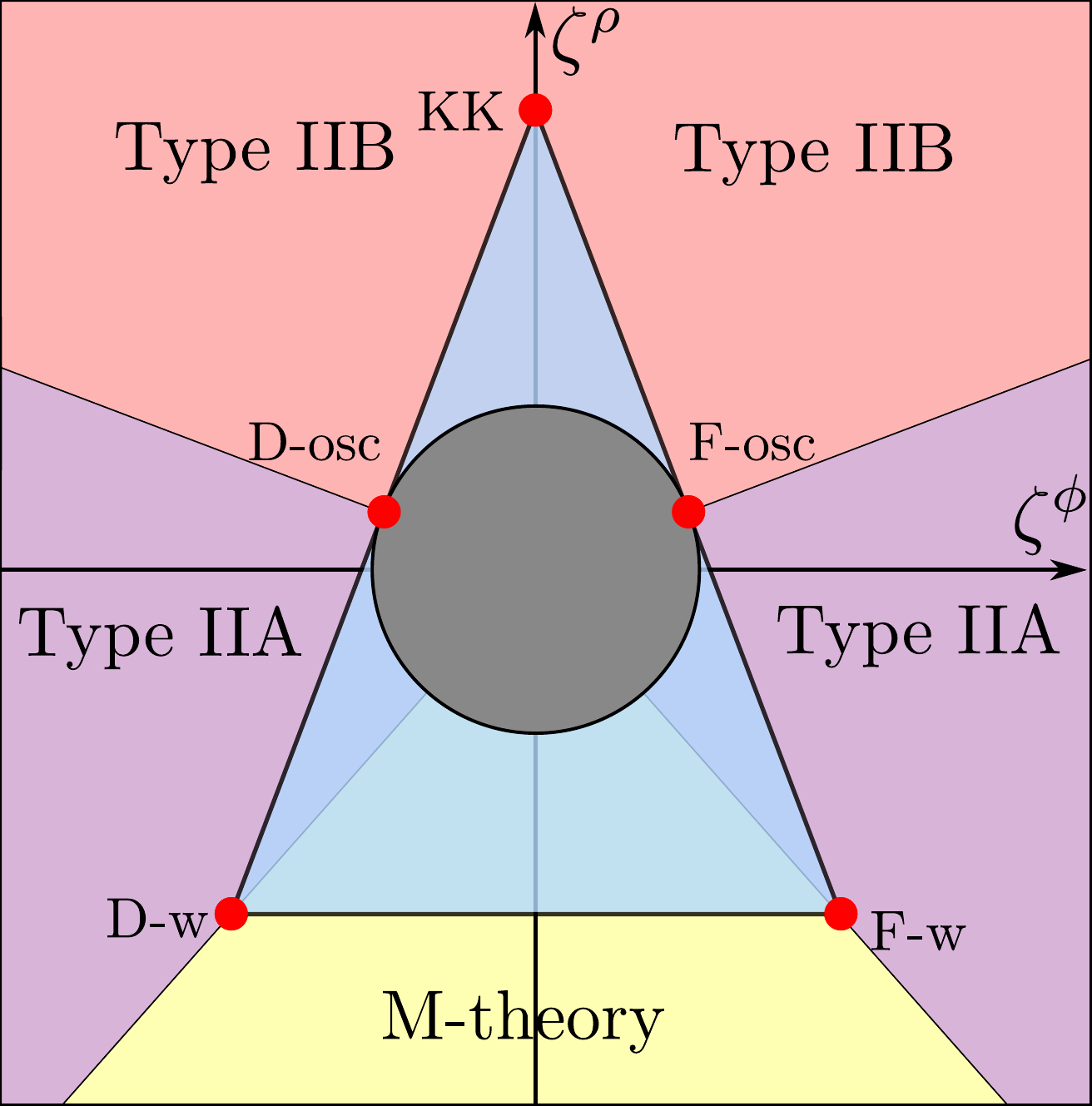}
\end{center}
\caption{Convex hull of the scalar charge-to-mass vectors for Type IIB string theory on a circle. The convex hull generated by the Kaluza-Klein modes, the fundamental string winding modes, and the D-string winding modes contains the ball of radius $\frac{1}{\sqrt{d-2}}$ (gray), ensuring that the SWGC is satisfied. The five different duality frames of the radion-coupling moduli space are coloured depicted in different shades, with the vertical axis corresponding with the self-dual line.}
\label{CHII}
\end{figure}

These five scalar charge-to-mass vectors (and their convex hull) are plotted in Figure \ref{CHII}. One can see  that the convex hull contains the ball of radius $\frac{1}{\sqrt{d-2}}$, ensuring that the SWGC is satisfied along these directions in moduli space. The points of tangency, where the convex hull condition is only marginally satisfied, correspond to emergent string limits.

\begin{figure}
\begin{center}
\includegraphics[width = 90mm]{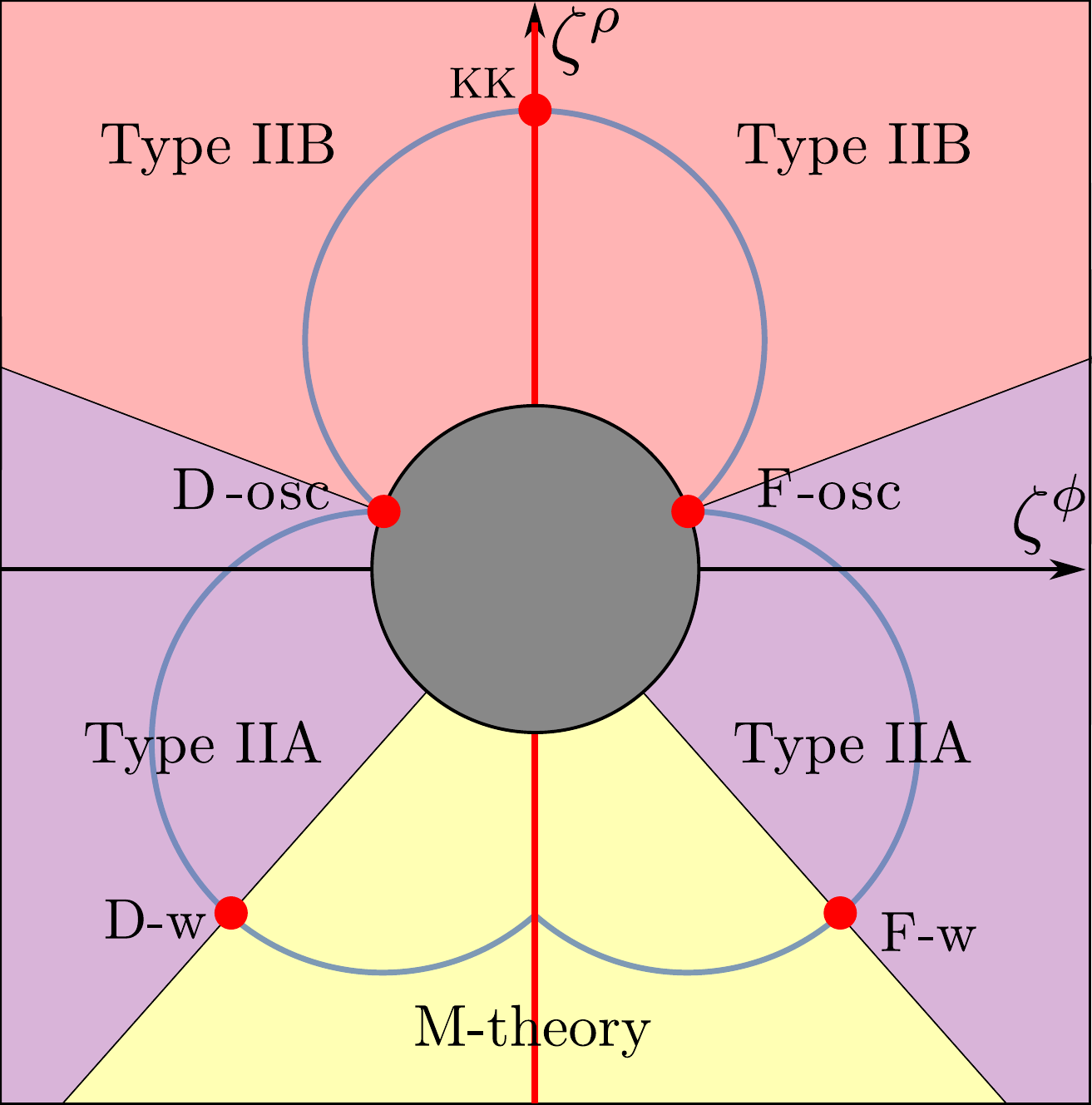}
\end{center}
\caption{Max-$\alpha$ hull for Type IIB string theory on a circle (blue). Because this hull contains the ball of radius $\frac{1}{\sqrt{d-2}}$ (shaded gray), every infinite-distance limit in the dilaton-radion plane features a tower of particles with $\alpha \geq\frac{1}{\sqrt{d-2}}$. Limits with $\alpha_{\text{max}} = \frac{1}{\sqrt{d-2}}$ represent emergent string limits, while limits with $\alpha > \frac{1}{\sqrt{d-2}}$ represent decompactification limits in some duality frame. There are five duality frames in the dilaton-radion plane, two of Type IIB string theory (shaded red), two of Type IIA string theory (shaded purple), and one of M-theory (shaded yellow). The red vertical line depcts the Type IIB self S-dual line. This extends to the M-theory frame, corresponding to decompactifying opposite cycles.}
\label{MLII}
\end{figure}

Figure \ref{MLII} depicts the function $ \alpha_{\text{max}}(\hat \tau)$ corresponding to the exponential rate of the leading light tower along each asymptotic geodesics moving in each direction $\hat \tau$. As discussed above, the sharpened Distance Conjecture requires $\alpha_{\text{max}}(\hat \tau) \geq \frac{1}{\sqrt{d-2}}$ for all $\hat \tau$, which is equivalent to the statement that $ \alpha_{\text{max}}(\hat \tau)$ must lie outside the ball of radius $\frac{1}{\sqrt{d-2}}$ for all $\hat \tau$. This is indeed satisfied in our example. It is no coincidence that the region bounded by $\alpha_{\text{max}}(\hat \tau)$ here strictly contains the convex hull of the generators, shown in Figure \ref{CHII}.

Figures \ref{CHII} and \ref{MLII} also depict the various duality frames in the theory as a function of location in moduli space. The region with $\phi >0$, $\rho > \frac{1}{\sqrt{7}}\phi$ corresponds to weakly-coupled Type IIB string theory compactified on a large circle,\footnote{Here, a large circle is one whose Kaluza-Klein scale $m_{\text{KK}}$ is lighter than the string scale, $m_\text{s}$.} as does the S-dual region with $\phi <0$, $\rho > -\frac{1}{\sqrt{7}}\phi$. The region with $-\frac{3}{\sqrt{7}}\phi < \rho < \frac{1}{\sqrt{7}}\phi$ admits a (T-dual) description as Type IIA string theory compactified on a large circle, as does the region with $\frac{3}{\sqrt{7}} \phi < \rho < -\frac{1}{\sqrt{7}}\phi$. Finally, the region with $ \rho < \frac{3}{\sqrt{7}}\phi$ and $ \rho < -\frac{3}{\sqrt{7}}\phi$ is described by 11-dimensional M-theory compactified on $T^2$. In summary, the dilaton-radion moduli space is divided into five duality frames: two of Type IIA, two of Type IIB, and one of M-theory.

Within each duality frame, an infinite-distance limit with $\alpha_{\text{max}}(\hat \tau) > \frac{1}{\sqrt{d-2}}$ corresponds to a decompactification limit of the corresponding string/M-theory. Meanwhile, a limit with $\alpha_{\text{max}}(\hat \tau) = \frac{1}{\sqrt{d-2}}$ corresponds to an emergent string limit. Every infinite-distance limit falls into one of these two categories, as predicted by the Emergent String Conjecture. Notably, the scalar charge-to-mass vectors are located precisely at the interfaces between the different duality frames, so that we have as many leading towers as boundaries between different duality frames. 

This concludes our brief review of Type II string theory in nine dimensions. In what follows, we will carry out a similar analysis for heterotic string theory in nine dimensions, and we will see that the story is far more subtle due to the importance of non-BPS particles.

\section{Heterotic String Theory in Nine Dimensions}

In this section, we test the sharpened Distance Conjecture in the moduli space of heterotic string theory compactified on a circle to nine dimensions, a theory with 16 supercharges and $r = 17$ vector multiplets. This theory has an 18-dimensional moduli space of the form
\begin{equation}\label{eq:het_moduli_space}
\mathcal{M} = \hat{\mathcal{M}} \times \mathbb{R}, \quad \hat{\mathcal{M}} =  SO(17, 1; \mathbb{Z}) \backslash SO(17, 1) / SO(17),
\end{equation}
where $\mathbb{R}$ parametrizes the dilaton and the Narain moduli space $\hat{\mathcal{M}}$ parametrizes the radius of the circle and the 16 Wilson lines for the heterotic gauge fields.

\begin{figure}
\begin{center}
\includegraphics[width =70mm]{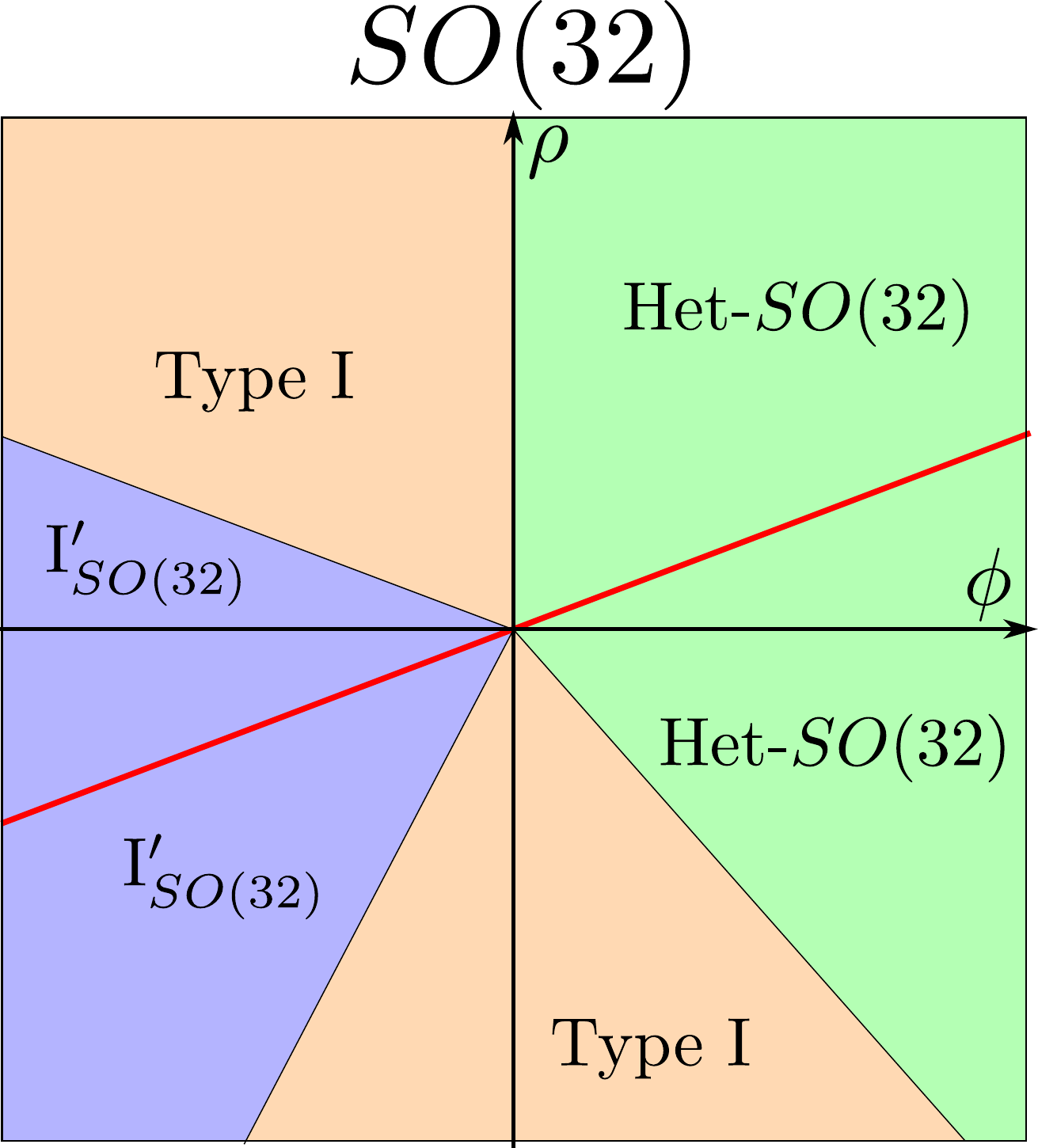}
\hspace{.5cm}
\includegraphics[width = 70mm]{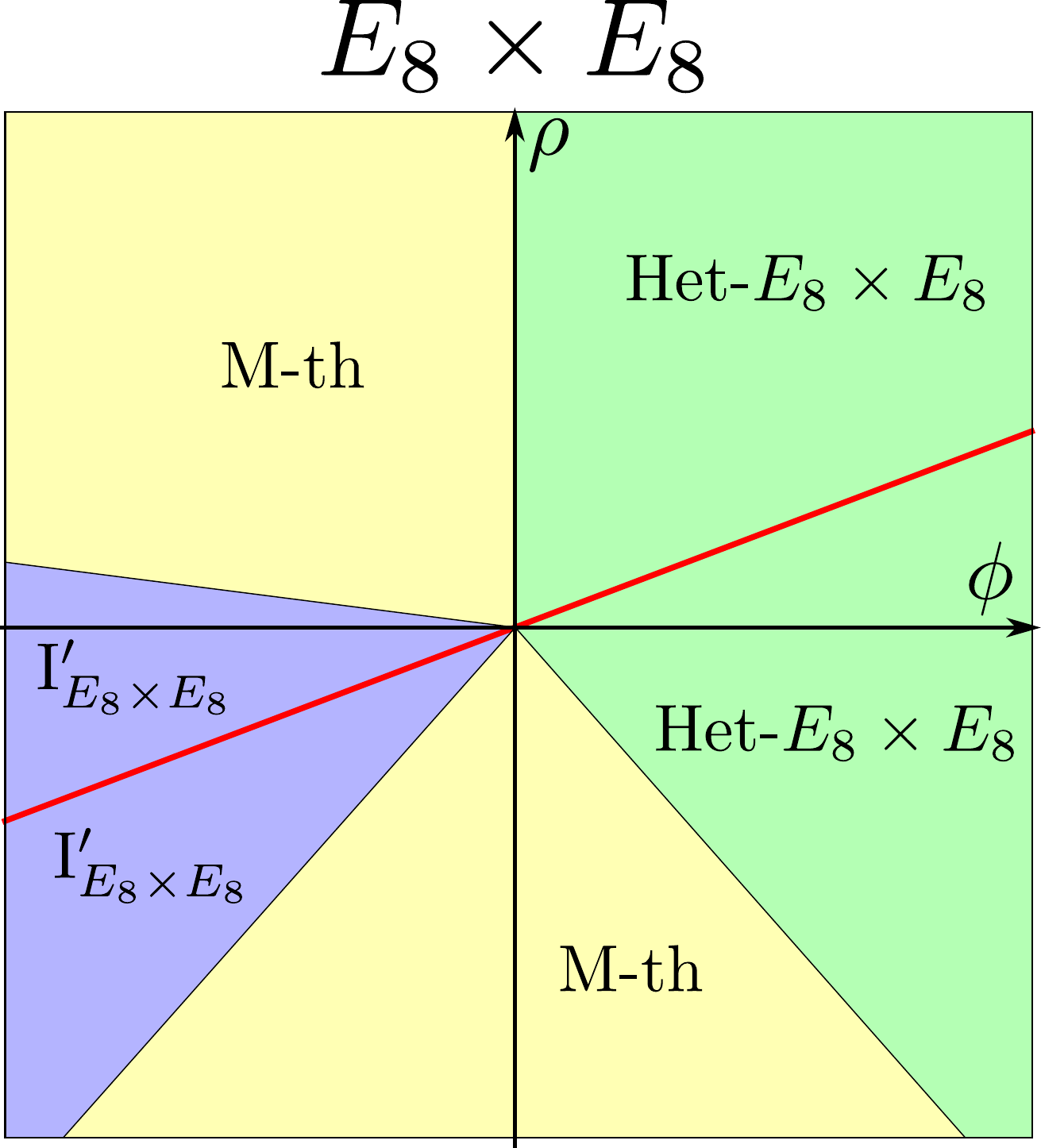}
\end{center}
\caption{Relevant $SO(32)$ and $E_8\times E_8$ slices of the moduli space of the $\mathcal{N}=1$ 9-dimensional theory with $r=17$, parametrized by the canonically normalized heterotic 10-dimensional dilaton and radion, $\phi$ and $\rho$. The regions best described by heterotic, type I$'$, and type I string theories, as well as M-theory, are respectively colored green, blue, orange and yellow. The self-dual line is depicted in red, with $\rho = \frac{1}{\sqrt{7}} \phi$. The boundaries between the different regions have the same direction as several of the towers depicted in Figures \ref{SO32max} and \ref{E8E8max}.\\\emph{\textbf{Note}: The above color code for the different duality frames will be used in later figures throughout this paper, though we will omit the labels.}} 
\label{modspace}
\end{figure}

We will be primarily interested in two particular slices of this moduli space, depicted in Figure \ref{modspace}, obtained by compactifying either the $SO(32)$ or $E_8 \times E_8$  ten-dimensional heterotic theory on a circle with all Wilson lines turned off.\footnote{\label{other_so32} Note that there is an additional slice with $SO(32)$ enhanced gauge symmetry, obtained by turning on a Wilson line in the $\mathbb{Z}_2$ center of the global gauge group ${\rm Spin(32)}/\mathbb{Z}_2$. We will comment briefly on this additional slice below in \S\ref{sec: other mod spaces}.} Each slice is two dimensional and flat, so it is parametrized by two moduli which we take, without loss of generality, to be the heterotic dilaton $\phi$ (as in Section \ref{sec: IIB circ}, $\phi=-\sqrt{2}\log g_s$, so that the weak coupling limit corresponds to $\phi\gg 1$) and the radion $\rho$ associated to the heterotic circle compactification (both being canonically normalized). 

Depending on the values taken by these fields, the theory is best described by different dual descriptions, so we can split the moduli space into different duality frames associated to the different weakly coupled (perturbative) descriptions that arise asymptotically (see Figure \ref{modspace}). Starting in the heterotic frame in the upper right-hand side corner of the plots in Figure \ref{modspace}, we can move to smaller values of the radion and dilaton and reconstruct the other duality frames by performing a series of  T- and S-dualities. A very detailed description of all these dualities can be found in \citep{Aharony:2007du}. As the heterotic dilaton $\phi$ decreases (i.e., as we go to larger values for the coupling $g_s$), the theory is better described by its S-dual theory, which is Type I on a circle for the case of $SO(32)$ or M-theory on a torus for the case of $E_8\times E_8$. If we then also decrease the radius, it is convenient to perform a T-duality and describe the theory in terms of Type I$'$ on a circle.  
Moreover, these slices of the moduli space are self-dual, which means that they exhibit a self-dual line below which the moduli space is a copy of the moduli space above. In the above coordinates, the self-dual line occurs at $\rho =\frac{1}{\sqrt{7}} \phi$ (red line in Figure \ref{modspace}), where the Kaluza-Klein photon enhances to an $SU(2)$ gauge symmetry. This self-duality corresponds to a T-duality in the heterotic frame. 

\subsection{A Puzzle in the $SO(32)$ Slice of the Moduli Space}
\label{sec:puzzle_in_9d}

We begin by analyzing the different tower of states that emerge in the subspace of the moduli space which has $SO(32)$ gauge symmetry. 

\begin{figure}
\begin{center}
\includegraphics[width = 70mm]{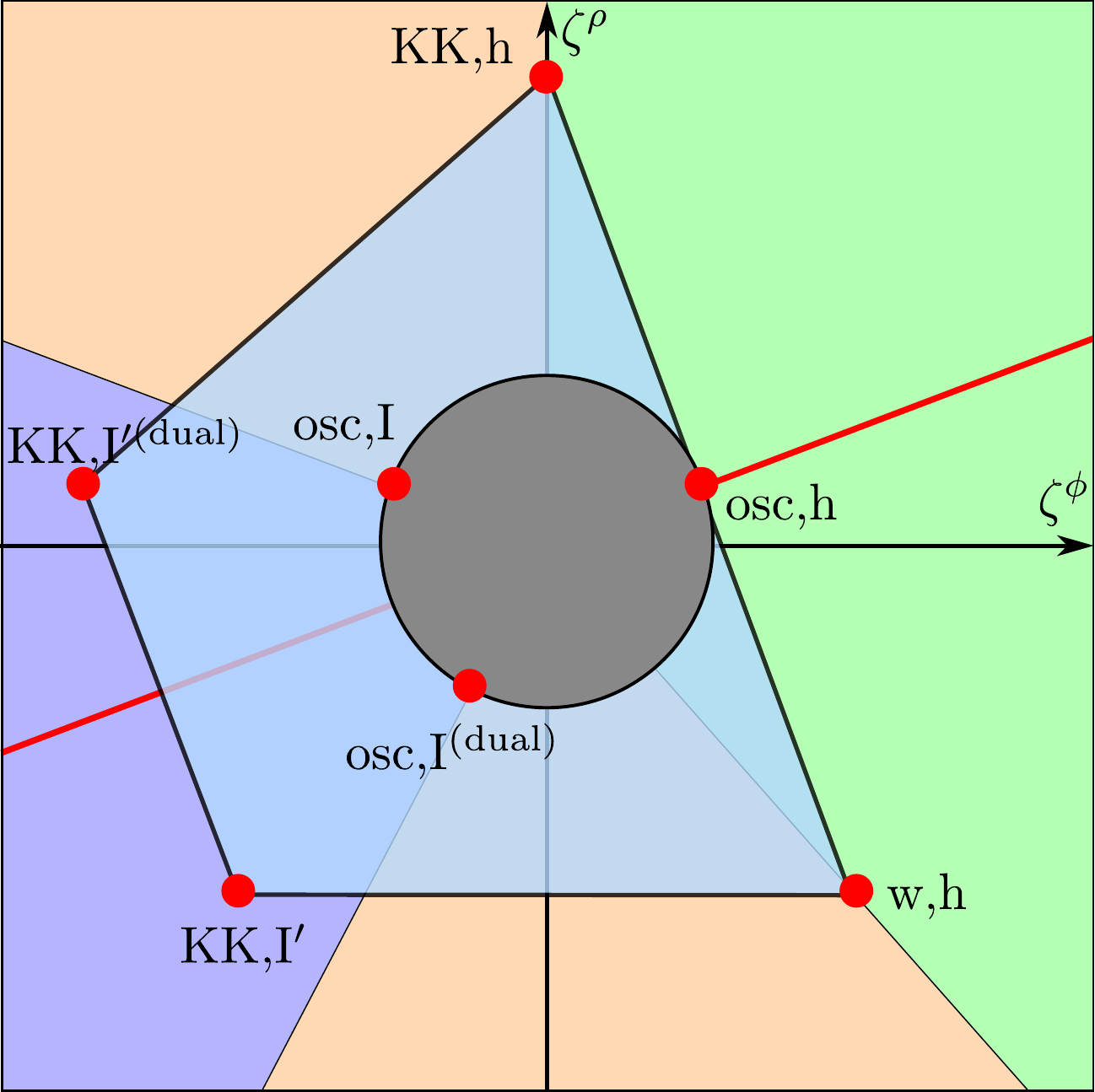}\hspace{0.5cm}
\includegraphics[width = 70mm]{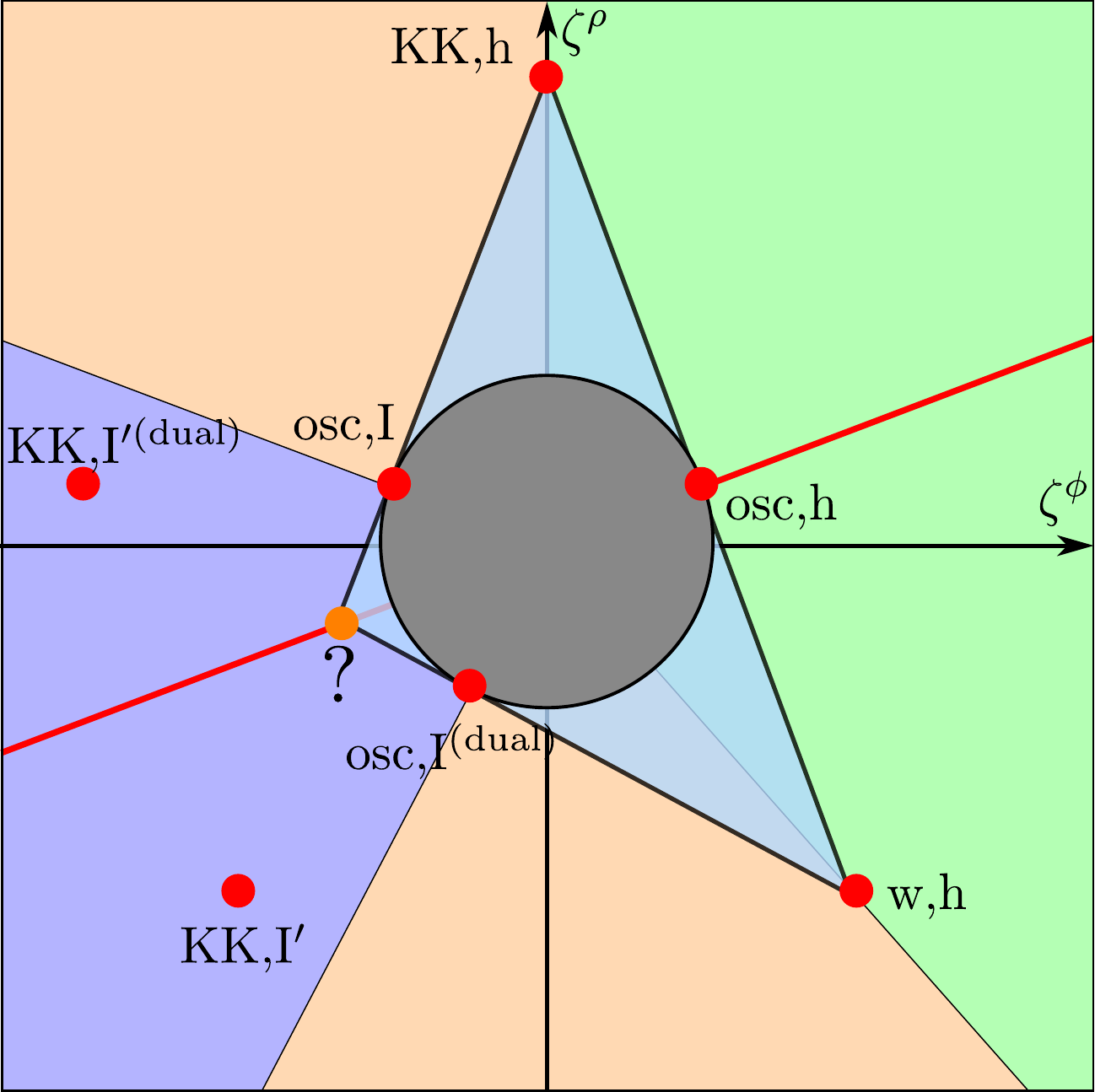}
\end{center}
\caption{Two naive (i.e., incorrect) convex hulls of the scalar charge-to-mass vectors in $SO(32)$ heterotic string theory. The left convex hull is incompatible with the fact that a tensionless Type I string emerges in the limit $\phi \rightarrow -\infty$, $\rho/\phi =  - \frac{1}{\sqrt{7}}$. The right convex hull is incompatible with the existence of the Type I winding modes and requires an unidentified mystery tower (shown in orange). The resolution to this puzzle lies in the fact that the scalar charge-to-mass ratio vectors for Type I$'$ KK modes  varies as a non-trivial function of moduli space, as we will see in \S \ref{sec:decompactification}.}
\label{naivefig}
\end{figure}

This theory features (among others) a tower of BPS Kaluza-Klein modes, a tower of heterotic string oscillator modes, and a tower of BPS heterotic string winding modes. These have the same dilaton and radion dependence as the Kaluza-Klein modes, fundamental string oscillator modes, and the fundamental string winding modes in Type II string theory discussed above, i.e.,
\begin{align}
\vec{\zeta}_{\text{osc,h}}  = \left(\frac{1}{\sqrt{8}} , \frac{1}{\sqrt{56}}\right) \,,~~~
\vec{\zeta}_{\text{w,h}}  = \left(\frac{1}{\sqrt{2}} , -\frac{3}{\sqrt{14}}\right) \,,~~~ \vec{\zeta}_{\text{KK,h}}  &= \left(0, \sqrt{\frac{8}{7}} \right) \,.
\end{align}
The $SO(32)$ heterotic string is S-dual to Type I string theory. Thus, the strongly coupled regime of the heterotic string features a tower of Type I string oscillator modes and Type I string winding modes, with moduli dependence matching that of the D-string in Type IIB string theory, i.e., 
\begin{align}
\vec{\zeta}_{\text{osc,I}}  = \left(-\frac{1}{\sqrt{8}} , \frac{1}{\sqrt{56}}\right) \,,~~~\vec{\zeta}_{\text{w,I}}  = \left(-\frac{1}{\sqrt{2}} , -\frac{3}{\sqrt{14}}\right) \,.
\end{align}
Further, as mentioned above, heterotic string theory has the property of self-T-duality; a circle compactification of $SO(32)$ heterotic string theory 
with Wilson lines turned off is T-dual to another $SO(32)$ heterotic string theory, under which Kaluza-Klein modes and winding modes of the heterotic string theory exchange. This implies the existence of a dual phase of Type I string theory, with particles whose scalar charge-to-mass vectors are related to those of the original Type I phase by reflection across the self-duality line, $\rho = \frac{1}{\sqrt{7}}\phi$:
\begin{align}
\vec{\tilde \zeta}_{{\text{osc,I}^{\rm (dual)}}}  = \left(-\frac{1}{\sqrt{32}} , -\frac{5}{\sqrt{224}}\right) \,,~~~\vec{\tilde\zeta}_{\text{w,I}^{\rm (dual)}}  = \left(-\frac{3}{\sqrt{8}} , \frac{1}{\sqrt{56}}\right)\,.
\end{align}
These scalar charge-to-mass vectors are plotted in Figure \ref{naivefig}.

Here, a puzzle presents itself: Figure \ref{naivefig} (left-hand plot) suggests that along the infinite-distance geodesic with $\rho \rightarrow \infty$, $\rho/ \phi = - \frac{1}{\sqrt{7}}$ (i.e. with tangent vector parallel to $\vec\zeta_{\text{osc,I}}$), the lightest tower of particles should be the tower of winding modes associated with the dual Type I string. In reality, however, we know that this limit is actually an emergent string limit well described by perturbative Type I string theory on a circle, which means that the lightest tower of particles is the tower of Type I string oscillator modes, with $\alpha = \frac{1}{\sqrt{7}}$. Our naive picture is wrong!

In fact, we can argue that the decompactification limit associated to the Type I winding modes is obstructed in certain regimes. To see this, it is more convenient to switch to the T-dual theory of Type I, which is Type I$'$. The regime of validity of the Type I$'$ region is $\frac{1}{\sqrt{7}}\phi\leq \rho\leq -\frac{1}{\sqrt{7}}\phi$ with $\phi<0$ from the perspective of the heterotic variables, which is equivalent to weak coupling and small radius for Type I.
The Type I$'$ theory is an orientifold of Type IIA on a circle, and has two orientifold planes $O8^-$ sitting at the endpoints of an interval, together with 16 D8-branes to cancel the D-brane charge. The background which is dual to Type I with no Wilson lines (i.e. such that the gauge group is $SO(32)$) has all the D8-branes sitting on top of one of the orientifolds. The Type I$'$ string coupling then grows as we go from this orientifold to the other one. Hence, for a given value of the Type I$'$ string coupling near the $O8^-+D8's$, there is maximum value for the length of the interval, as otherwise the string coupling would diverge at some regular point in between the orientifolds. Thus, the decompactification limit is obstructed unless we also send the Type I$'$ string coupling to zero fast enough. The limiting case occurs if we move along the self T-dual line, for which the string coupling diverges precisely at the location of the $O8^-$ without the branes.\footnote{The diverging coupling at the $O8^-$ leads to the enhanced $SU(2)$ gauge symmetry along the self-dual line, as described in \citep{Aharony:2007du}.} This implies, in particular, that the theory does not decompactify if we move along an asymptotic geodesic whose tangent vector is parallel to $\vec\zeta_{\text{I-osc}}$, as this would correspond to increasing the radius but also increasing the string coupling from the Type I$'$ perspective (recall that winding modes of Type I are dual to KK modes of Type I$'$). 
 In particular, this means that the tower of dual Type I winding states do not exist along the asymptotic trajectory in the direction of $\vec\zeta_{\text{I-s}}$.

Taking this reasoning into account, we could plot a new convex hull including only the BPS states and the string oscillator modes while ignoring the Type I winding modes (see Figure \ref{naivefig}, right-hand plot). However, this convex hull would not contain the ball of radius $\frac{1}{\sqrt{d-2}}$ unless there is a new mystery tower with $\vec\zeta=\left(-\frac{2}{3\sqrt{2}},-\frac{2}{3\sqrt{14}}\right)$ (orange point in Figure \ref{naivefig}, right-hand plot). If we were to take this mystery tower seriously, it would have $|\vec\zeta|=\frac{4}{3 \sqrt{7}}$, which by \eqref{KKalpha} is equal to the exponential rate of a Kaluza-Klein tower for a decompactification to an 18-dimensional vacuum. However, this convex hull is also incorrect, since it is well known that the resolution of taking the infinite-distance limit along the self-dual line is not a new 18 dimensional vacuum, but a running solution of Type I$'$ in 10 dimensions. As explained above, this corresponds to the limiting case in which the string coupling diverges in one of the orientifolds, and we simply recover 10 dimensional massive Type IIA with a running dilaton in the decompactification limit.

In what follows, we will explain how the apparent contradiction is resolved due to the fact that the Type I winding modes are not BPS and their scalar charge-to-mass vectors vary across moduli space. Furthermore, the fact that we are decompactifying to a warped, running solution will also change the result for the scalar charge to mass ratio of the KK towers, deviating from the unwarped result of $|\vec{\zeta}_{\rm KK}|=\sqrt{\frac{8}{7}}$, \eqref{eq: bps KK lenght}.

\subsection{The Resolution: Sliding and Decompactification to a Running Solution}
\label{sec:decompactification}

\begin{figure}
\begin{center}
\begin{subfigure}{0.475\textwidth}
\center
\includegraphics[width=70mm]{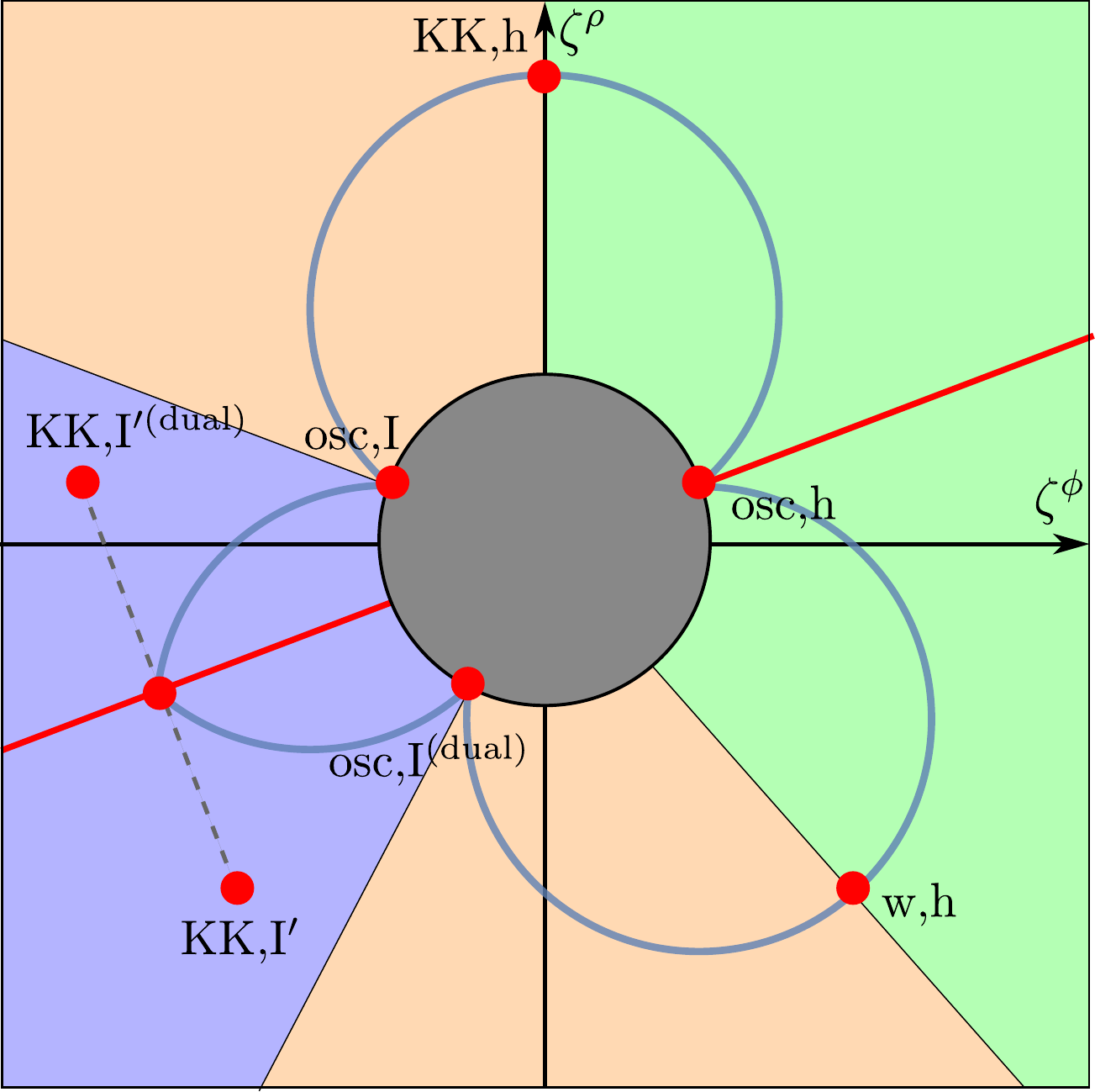}
\caption{$SO(32)$} \label{SO32max}
\end{subfigure}
\begin{subfigure}{0.475\textwidth}
\center
\includegraphics[width=70mm]{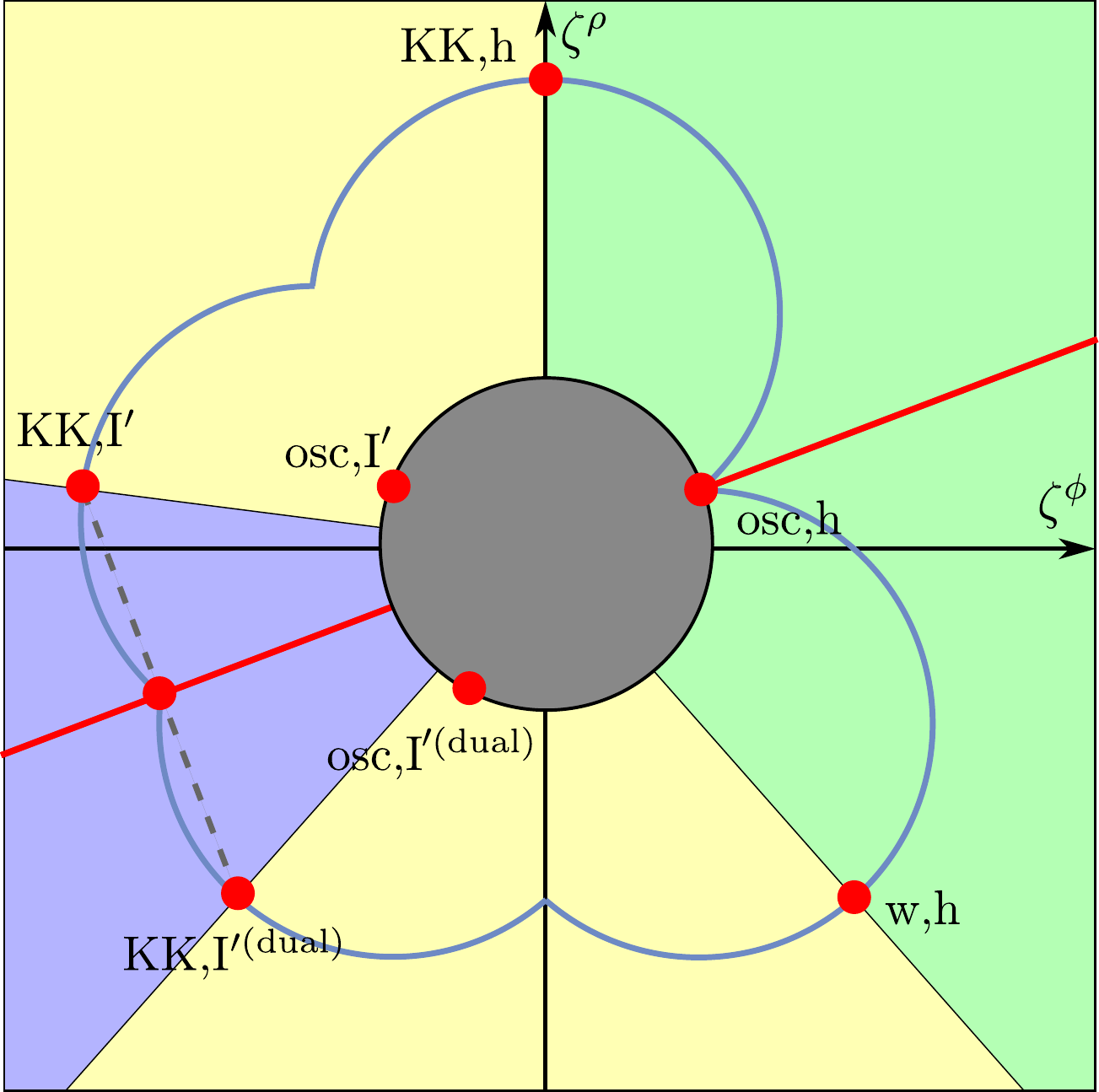}
\caption{$E_8\times E_8$} \label{E8E8max}
\end{subfigure}
\caption{Max-$\alpha$ hull for $SO(32)$ and $E_8\times E_8$ heterotic string theories on a circle. Everywhere except the Type I$'_{SO(32)}$ region, the arc of the hull is determined by the $\vec{\zeta}$-vector of the tower contained in said arc. In the Type I$'_{SO(32)}$ regions, however, the leading tower is more subtle: in the Type I$'_{SO(32)}$ region above the self-dual line, the leading tower is the Type I$'$-KK tower whose $\vec{\zeta}$-vector is located \emph {below} the self dual-line. In the dual I$'_{SO(32)}$ region below the self dual line, the leading tower is the Type I$'$-KK tower located \emph{above} the line.}
\end{center}
\end{figure}

To begin resolving the puzzle outlined in the previous section, we will focus on one of the Type I$'$ regions of moduli space (shaded blue in Figure \ref{SO32max}). Taking an infinite-distance limit $\phi \rightarrow -\infty$ with fixed $\frac{1}{\sqrt{7}}\phi\leq \rho  \leq -\frac{1}{\sqrt{7}}\phi$ corresponds to a decompactification limit of weakly coupled Type I$'$ string theory to ten dimensions. However, as we now review, such a decompactification limit leads not to a ten-dimensional vacuum, but rather to a running solution.

\begin{figure}[h]
\begin{center}
\includegraphics[width = 0.5\textwidth]{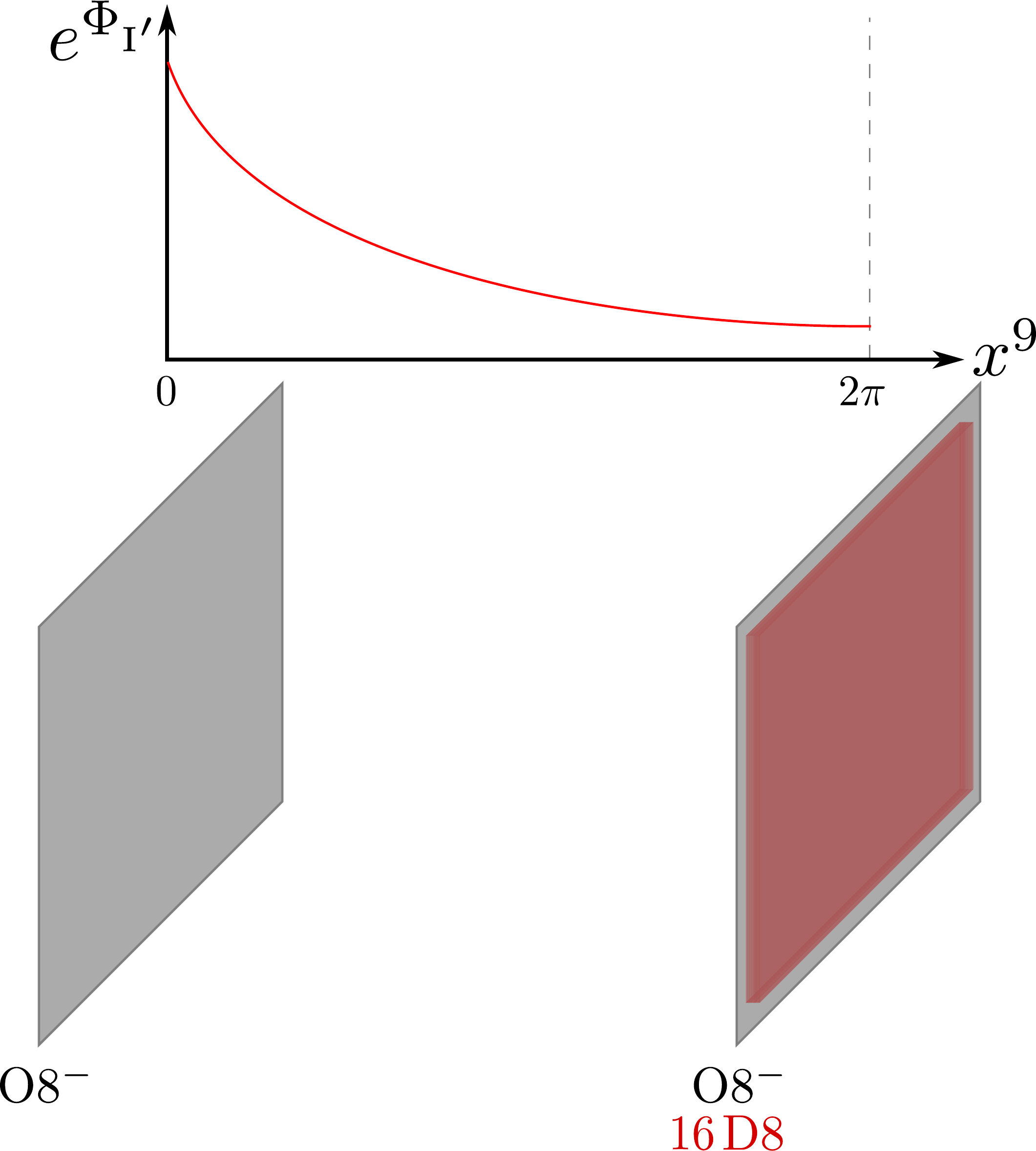}
\end{center}
\caption{Sketch of the $O8$-($O8$+16 D8) brane-orientifold configuration needed to obtain the $SO(32)$ gauge group in Type I$'$, as well as the dilaton profile for $B>0$.}
\label{fig: so32 branes}
\end{figure}

As explained in \cite{Polchinski:1995df} and rederived in Appendix \ref{app:eom}, type I$'$ theory compactified on a interval $x^9\in[0,2\pi]$, with two orientifolds located at its extremes, $x^9_{O8}=0,\, 2\pi$, and 16 D8-branes located at $\{x^9_i\}_{i=1}^{16}\subset[0,2\pi]$ (see Figure \ref{fig: so32 branes}) results in a warped metric $G_{MN}=\Omega^2(x^9)\eta_{MN}$ and running dilaton $g_{s,\rm I'}=e^{\Phi_{\rm I'}(x^9)}$ given by
\begin{equation}
\label{warping}
	e^{\Phi_{\rm I'}(x^9)}=z(x^9)^{-5/6}\qquad \Omega(x^9)=Cz(x^9)^{-1/6},
\end{equation}
with
\begin{equation}\label{eq: z gen}
	z(x^9)=\sqrt{\frac{180}{41}}(\alpha_{\rm I'}')^2\mu_8 C\left\{B+\int_{x_0^9}^{x^9}\sum_{i=1}^{16}\tau\delta(\tau-x_i^9)\dd\tau-\left(\int_0^{x^9}\sum_{i=1}^{16}\delta(\tau-x ^9_i)\dd \tau-8\right)x^9\right\}.
\end{equation}
Here,  $\mu_8$ is the coupling to the 9-form potential, and $\nu_0$, given by $F_{10}\wedge\star F_{10}=\star\nu_0^2$ is proportional to the Romans mass of the massive Type IIA theory that arises between the D8-branes. As explained in Appendix \ref{app:eom}, the expression of $z(x^9)$ greatly simplifies for the brane configuration leading to the $SO(32)$ and $E_8\times E_8$ gauge sectors. In this first case,
$$
z_{SO(32)}(x^9)=\sqrt{\frac{180}{41}}(\alpha_{\rm I'}')^2\mu_8 C(B+8x^9),\quad \text{with }B>0\,.
$$

For later convenience, it is useful to define the dimensionless quantities $\hat \Omega =\Omega (\alpha'_{\rm I'})^{-1/2}$ and $\hat C= C(\alpha'_{\rm I'})^{-1/2}$, so that $\hat C,B$ are both dimensionless fields, and $z(x^9)\sim \hat C (B + 8 x^9)$ up to global numerical factors.

This non-trivial warping and dilaton background is important because it modifies the masses (and therefore the scalar charge-to-mass ratio) of the Type I$'$ Kaluza-Klein modes. We have seen previously that circle reduction of a 10-dimensional vacuum solution leads to Kaluza-Klein modes whose masses scale with the radion $\rho$ as
\begin{equation}
\mKK \sim \exp\left(- \sqrt{\frac{8}{7}}\rho \right)\,,
\end{equation}
which implies $\partial_\rho \log \mKK = - \sqrt{\frac{8}{7}}$ \emph{everywhere} in moduli space.

In the case of Type I$'$ string theory at hand, however, this simple calculation no longer applies. Instead, the moduli dependence of the Kaluza-Klein mass must be computed via a careful dimensional reduction of the 10-dimensional theory, which requires an explicit computation of the Laplacian spectrum taking into account the non-trivial warping. We present the computation in detail in Appendix \ref{app:KK}, while here we only show the final result for the Type I$'$ KK mass:
\begin{equation}\label{eq: gen KK mass}
	m_{\rm KK,\rm I'}=\left(\int_0^{2\pi}\dd x^9\hat\Omega^8e^{-2\hat\Phi_{\rm I'}}\right)^{-1/7}M_{\rm Pl;9}.
\end{equation}
This is valid for both the $SO(32)$ and $E_8\times E_8$ slices of moduli space.

As derived in Appendix \ref{app:dual}, the oscillator modes of the Type I$'_{SO(32)}$ string have a mass of order
\beq
m_{\rm osc, I'}=\left(\int_0^{2\pi}\dd x^9\hat\Omega^2\right)^{1/4}
	\left(\int_0^{2\pi}\dd x^9\hat\Omega^8 e^{-2\Phi_{\rm I'}}\right)^{-11/28}
	\left(\sum_{i=1}^{16}\left.\hat\Omega^5 e^{-\Phi_{\rm I'}}\right|_{x^9=x^9_i}\right)^{1/2}M_{\rm Pl; 9}\\
\eeq
As a side remark, let us mention that this decompactification limit was also recently discussed in \cite{Bedroya:2023tch} to argue from the bottom-up that this asymptotic limit in the 9d supergravity moduli space should correspond to decompactifying to Type I$'$ string theory, although their estimation of the KK mass and string scale disagree with our results. Eqs.(25) and (26) of \cite{Bedroya:2023tch} imply $m_{\rm KK,\rm I'}\sim C^{-25/24}$ and $m_{\rm osc, I'}\sim C^{-5/24}$, while the correct result is $m_{\rm KK,\rm I'}\sim C^{-25/21}$ and $m_{\rm osc, I'}\sim C^{-5/14}$, which are obtained after plugging \eqref{warping} into our results for the masses above (see e.g. \ref{KKmassappendix}). In any case, this does not change the fact that the KK mass is lighter than the string mass along the self-dual line, so the qualitative results of \cite{Bedroya:2023tch} remain unchanged.

In order to plot the convex hull of all states, we also need to express the BPS masses in the (B,C) variables. This can be done by identifying the 9-dimensional actions and the microscopic interpretation of the BPS states from both the heterotic and Type I$'$ perspectives. This is done in Appendix \ref{app:dual}, and the result for the heterotic radius and the heterotic dilaton in terms of the warping factor is given by
	\begin{align}
R_{\rm h}
&\sim \left(\int_0^{2\pi}\dd x^9\hat\Omega^2 \right)^{-1} \left(\int_0^{2\pi}\dd x^9\hat\Omega^8e^{-2\hat\Phi_{\rm I'}}\right)^{1/7}M_{\rm Pl; 9}^{-1}\\
g_{\rm h}&\sim\frac{\sqrt{2}}{\pi}\left(\int_0^{2\pi}\dd x^9\hat\Omega^2\right)^{-1/2}
	\left(\int_0^{2\pi}\dd x^9\hat\Omega^8 e^{-2\Phi_{\rm I'}}\right)^{3/2}
	\left(\sum_{i=1}^{16}\left.\hat\Omega^5 e^{-\Phi_{\rm I'}}\right|_{x^9=x^9_i}\right)^{-2}
	\end{align}
This, together with the relation between the string scales $m_{\rm osc,h}=g_h^{1/2}m_{\rm osc,I'}$, is enough to obtain the BPS heterotic KK and winding modes, $m_{\rm KK,h}=R_h^{-1}$ and $m_{\rm w,h}=m_{\rm osc,h}^2R_h$, in terms of the $(B,C)$ fields. The explicit expressions are presented in the Appendices in \ref{app: dual rel} and \ref{sec: BPS mass}.

The final piece of information that we need to compute the scalar charge-to-mass ratios is the moduli space metric $\mathsf{G}_{ab}$.  This can be either computed from dimensionally reducing the 10-dimensional Type I$'$ action, or more easily, by imposing that the scalar charge-to-mass ratio of the BPS heterotic states which are purely KK or winding should remain fixed at any point of the moduli space. The field space metric is computed using both methods in  Appendix \ref{app:metric}, which combined with the final expression for the different tower masses in terms of $B$ and $C$ (see Appendix \ref{app.sliding}) and using  \eqref{zetavec}, leads to the following result for the scalar charge to mass ratio vectors in the flat coordinates\footnote{As obtained in Appendix \ref{app:flat}, $\{\phi_B,\phi_C\}$ are flat coordinates such that $\dd s^2_{\mathcal{M}_{SO(32)}}=\dd \phi_B^2+\dd\phi_C^2$ given by \eqref{eq: flat SO32} \begin{subequations}
	\begin{align*}
		\phi_C&=\frac{10}{3\sqrt{7}}\log C+\frac{5}{2\sqrt{7}}\log \left[(B+16\pi)^{4/3}-B^{4/3}\right]\\
		\phi_B&=\frac{1}{2}\log\frac{(B+16\pi)^{2/3}+B^{2/3}}{(B+16\pi)^{2/3}-B^{2/3}}\,,
	\end{align*}
\end{subequations}
with the peculiarity that the sliding will occur in the $\phi_B$ direction, and with the $\phi_C$ axis corresponding with the self-dual line. Any other flat frame will be related by a $O(2)$ transfromation.} $(\phi_B,\phi_C)\in\mathbb{R}_{>0}\times\mathbb{R}$: 
\begin{align}\label{eq: vecs fixed}
		\vec{\zeta}_{\rm osc, I'}=\left(\frac{1}{4},\frac{3}{4\sqrt{7}}\right),\, ~
		&\vec{\zeta}_{\rm KK,h}=\left(1,-\frac{1}{\sqrt{7}}\right),\notag\\
		\vec{\zeta}_{\rm osc,h}=\left(0,-\frac{1}{\sqrt{7}}\right),\, ~
		&\vec{\zeta}_{\rm w,h}=\left(-1,-\frac{1}{\sqrt{7}}\right),
\end{align}
while for the Type I$'$ KK mode we have a slightly more complicated expression
	\begin{align}\label{eq: sliding BC}
		\vec{\zeta}_{\rm KK, I'}&=\left(-\frac{3}{2}\left[\frac{2}{\sqrt{1-e^{-4\phi_B}}}+1\right]^{-1},\frac{5}{2\sqrt{7}}\right)\notag\\		
		&=\left\{
		\begin{array}{ll}
		\left(-\frac{3}{2}\sqrt{\phi_B}+\mathcal{O}(\phi_B),\frac{5}{2\sqrt{7}}\right)&\text{for }\phi_B\ll 1,\\
		\left(-\frac{1}{2}+\frac{1}{6}e^{-4\phi_B}+\mathcal{O}\left(e^{-8\phi_B}\right),\frac{5}{2\sqrt{7}}\right)&\text{for }\phi_B\gg 1.
		\end{array}\right.
	\end{align}
This formula (see \eqref{eq: sliding BC-het} for its expression in the $(\phi,\rho)$ flat frame) is one of the most important results of this paper. Unlike the previous towers, the scalar charge-to-mass ratio of the Type I$'$ KK modes change as move in the moduli space, in such a way that $\vec{\zeta}_{\rm osc, I'}$ slides continuously along a segment of length $\frac{1}{2}$ in the $\phi_B$ (equivalently $B$) direction of the tangent space of the moduli space. In doing so, it interpolates between the unwarped result $\left(-\frac{1}{2},\frac{5}{2\sqrt{7}}\right)$ when $\phi_B,\,B\rightarrow \infty$ and the highly warped result $\left(0,\frac{5}{2\sqrt{7}}\right)$ at the self-dual line when $\phi_B,\,B\rightarrow 0$.

Since the above scalar charge to mass ratio is given in the flat coordinates $(\phi_B,\phi_C)$, we still need to make a change of coordinates to write them in terms of the flat frame associated to the heterotic dilaton and radius, in order to compare the results with Figure \ref{naivefig}. The flat frames for the tangent spaces in different coordinates are simply related by an $O(2)$ transformation. Knowing that in this case the Jacobian matrix of the coordinate change is positive definite, said transformation will be part of $SO(2)$, that is, a rotation. We can determine the transformation matrix $M_\vartheta$ by imposing that $\vec{\zeta}_{\rm osc,h}=\left(0,\sqrt{\frac{8}{7}}\right)$ in the $(\phi,\rho)$ flat frame, which leads to  the rotation angle $\vartheta=\arccos\left(-\frac{1}{2\sqrt{2}}\right)$, and therefore
\be
M_\vartheta=\begin{pmatrix}
	 -\frac{1}{2 \sqrt{2}} & -\sqrt{\frac{7}{8}} \\
 \sqrt{\frac{7}{8}} & -\frac{1}{2 \sqrt{2}} \\
\end{pmatrix}\,,
\ee

which finally allows us to write $\vec{\zeta}_{\rm KK, I'}$ from \eqref{eq: sliding BC} in the $(\phi,\rho)$ flat frame:
	\begin{align}\label{eq: sliding BC-het}
		\vec{\zeta}_{\rm KK, I'}&=\left(-\frac{5}{4\sqrt{2}}+\frac{3}{4\sqrt{2}}\left[1+\frac{2}{\sqrt{1-e^{-4\phi_B}}}\right]^{-1},-\frac{5}{4\sqrt{14}}-\frac{3}{4}\sqrt{\frac{7}{2}}\left[1+\frac{2}{\sqrt{1-e^{-4\phi_B}}}\right]^{-1}\right)\notag\\
		&=\left\{
		\begin{array}{ll}
		\left(\frac{-5}{4\sqrt{2}}+\frac{3}{4\sqrt{2}}\sqrt{\phi_B},\frac{-5}{4\sqrt{14}}-\frac{3}{4}\sqrt{\frac{7}{2}}\sqrt{\phi_B}\right)+\mathcal{O}\left(\phi_B\right)&\text{for }\phi_B\ll 1,\\
		\left(-\frac{1}{\sqrt{2}}-\frac{1}{12\sqrt{2}}e^{-4\phi_B},-\frac{3}{\sqrt{14}}+\frac{1}{12}\sqrt{\frac{7}{2}}e^{-4\phi_B}\right)+\mathcal{O}\left(e^{-8\phi_B}\right)&\text{for }\phi_B\gg 1\,.
		\end{array}\right.
	\end{align}

\begin{figure}[h]
\begin{center}
\includegraphics[width = 90mm]{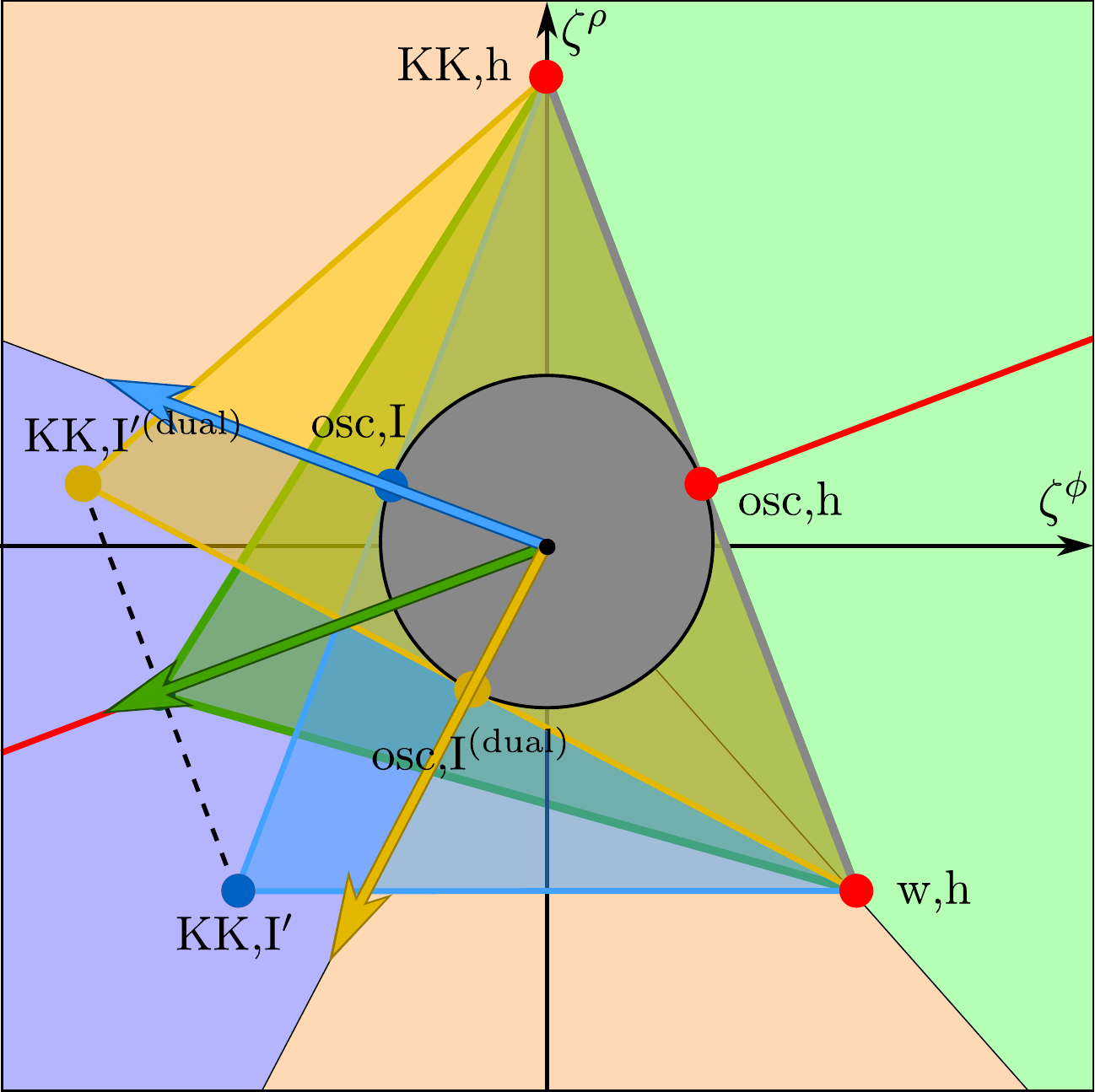}
\end{center}
\caption{Convex hulls in the two asymptotic limits for $B$ in the I$'_{SO(32)}$ regions. The convex hull for the limit $B\to \infty$ is shown in blue; this corresponds to the limit of the I$'_{SO(32)}$ string with no warping. The convex hull for the analogous zero-warping limit of the dual I$'_{SO(32)}$ string is shown in yellow. The convex hull for the limit $B\to 0$ limit is shown in green; this corresponds to decompactification to the 10-dimensional running solution. The sliding segment is depicted by a black dashed line. Note that the heterotic towers remain fixed in any limit.
}
\label{SO32slide}
\end{figure}

The final result for the scalar charge-to-mass ratios of the towers in the heterotic variables is plotted in Figure \ref{SO32slide}. The upshot of this result is that the scalar charge-to-mass vector representing the Type I$'$ Kaluza-Klein modes varies continuously as a function of the moduli, sliding along the black dashed line  in Figure \ref{SO32slide} on one side of the self-dual line. Similarly, the Kaluza-Klein modes of the dual Type I$'$ string slide along the black dashed line on the other side.\footnote{The fate of the states when one crosses the self dual line is not entirely clear from our analysis. It may be that the two towers of states are one and the same, or that one tower becomes unstable at the self-dual line and decays into the other.} This implies that the convex hull of the towers indeed changes as we me move in the moduli space.

\begin{figure}[h]
\begin{center}
\includegraphics[width = 1\textwidth]{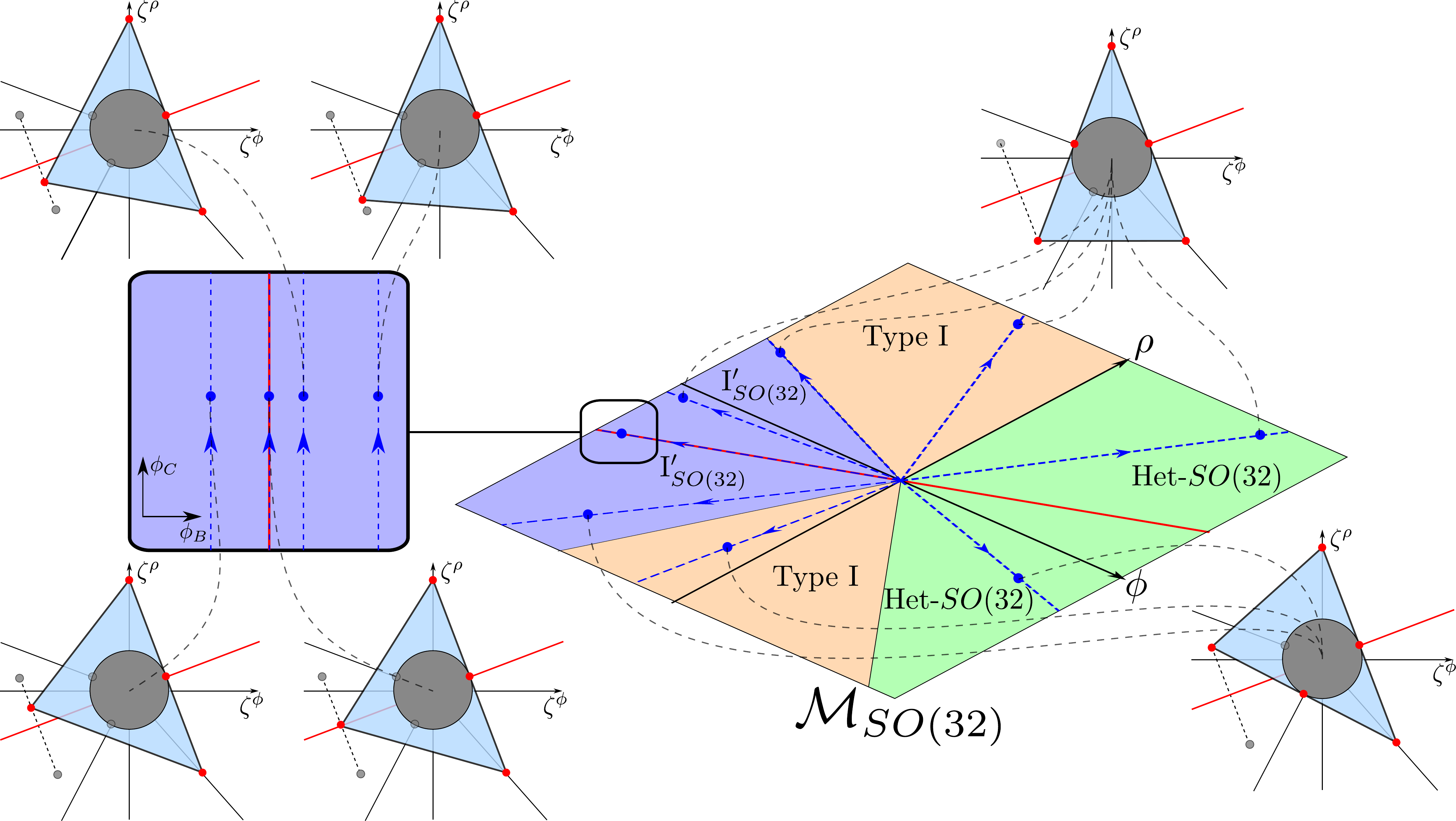}
\end{center}
\caption{Sketch of the behavior of the convex hull (which is defined in the tangent space, $T_p\mathcal{M}_{SO(32)}$) for points along trajectories moving to different limits of the $SO(32)$ slice of the moduli space, $\mathcal{M}_{SO(32)}$, parameterized by the canonically normalized heterotic $(\phi,\rho)$. Note that in the Type I and Heterotic regions the convex hull is the same as in the unwarped limit of the Type I$'$ region (i.e. $B,\phi_B\to \infty$), which is also the same shape as in the Type IIB string compactification on a circle, Figure \ref{CHII}. For fixed $\phi_B$, the tangent vector  is parallel to the self-dual line, but the shape of the convex-hull depends on the distance to it, given by $\phi_B$. As we move closer, the figure is deformed by the sliding of the Type I$'$ KK scalar charge-to-mass vector sliding, while still containing the $\frac{1}{\sqrt{d-2}}$ radius ball.}
\label{SO32tangent mod}
\end{figure}

A crucial consequence of the formula \eqref{eq: sliding BC} is that the sliding of $\vec{\zeta}_{\rm KK, I'}$ occurs entirely as a function of the flat coordinate $\phi_B$, which has the interpretation as the perpendicular distance to the self-dual line. Thus, if we move along any asymptotic geodesic that is not perpendicular to the self-dual line, $\phi_B$ will grow arbitrarily large as we move towards the asymptotic region, and $\vec{\zeta}_{\rm KK, I'}$ will approach the unwarped result $\left(-\frac{1}{2},\frac{5}{2\sqrt{7}}\right)$ exponentially quickly. If we are only interested in tracking the dependence of the exponential rate as a function of direction, then, the sliding will happen instantaneously right as our asymptotic geodesic becomes parallel to the self-dual line, as depicted in Figure \ref{SO32tangent mod}.

However, there is a two-parameter family of asymptotic geodesics, parametrized by both the direction as well as the ``impact parameter,'' or the initial displacement in the perpendicular direction. While for most geodesics the impact parameter will not affect the value of $\vec{\zeta}_{\rm KK, I'}$ in the asymptotic regime, for geodesics parallel to the self-dual line we see that the value of $\vec{\zeta}_{\rm KK, I'}$ depends very strongly on the impact parameter (in this case, $\phi_B$), even asymptotically, as depicted in the left part of Figure \ref{SO32tangent mod}. We can see that there is an order-of-limits issue regarding the asymptotic value of $\vec{\zeta}_{\rm KK, I'}$: the limits of taking our geodesic parallel to the self-dual line and moving infinitely far along our geodesic do not commute.

With the dependence of $\vec{\zeta}_{\rm KK, I'}$ on our asymptotic trajectory in hand, let us now derive the $\alpha_{\text{max}}$-plot which provides the value of the exponential rate for the lightest tower for every asymptotic geodesic of the moduli space. For the purposes of computing the $\alpha_{\text{max}}$ as a function of direction in the Type I$'$ phases of moduli space, we can imagine that the Kaluza-Klein scalar charge-to-mass vector jumps discontinuously from $\vec{\zeta}_{\text{KK,I$'$}}  = (-\frac{1}{\sqrt{2}} , -\frac{3}{\sqrt{14}})$ to $\vec{\zeta}_{\text{KK,I$'^{\rm (dual)}$}}  = (-\frac{3}{\sqrt{8}} , \frac{1}{\sqrt{56}})$ as one crosses from one Type I$'$ phase into the other. Note that for geodesics parallel to the self-duality line, i.e., with $\rho_1 = \frac{1}{\sqrt{7}}\phi_1 < 0$, we have $\alpha_{\text{max}} =  \frac{5}{\sqrt{28}}$ independently of the values of $\phi_0$ and $\rho_0$. This independence is a bit surprising given the nontrivial sliding of the Type I$'$ Kaluza-Klein modes' scalar charge-to-mass vector that occurs as $\phi_0$ and $\rho_0$ are shifted, but this shifting turns out to have no effect on $\alpha_{\text{max}}$ for the simple reason that the self-duality line is orthogonal to the line segment on which the sliding occurs, as can be seen from Figure \ref{SO32slide}.

The result of all this is the max-$\alpha$ hull shown in Figure \ref{SO32max}. Clearly, the sharpened Distance Conjecture is satisfied, as $\alpha_{\text{max}} \geq \frac{1}{\sqrt{d-2}}$ in every direction in the dilaton-radion plane. The limits in which this bound is saturated correspond to the three emergent string limits, one of which is a heterotic string limit, the other two of which are Type I string limits. Our initial puzzle is resolved: the Type I$'$ KK modes exist as long as we stay in one Type I$'$ region, but do not obstruct the other Type I emergent string limit since their scalar charge-to-mass vector varies as a function of moduli space.

An interesting consequence of our results is that the maximum value of the exponential rate for the Type I$'$ KK tower occurs precisely along the self-dual and it is smaller than the naive unwarped result, $\alpha_{\text{max}} =   \frac{5}{\sqrt{28}}<\sqrt{\frac87}$. Hence, one has to be careful when assuming \eqref{KKalpha} for a KK tower. This raises an obvious question: how small the can the exponential rate of a KK tower become due if decompactifying to a running solution? Could it get even smaller than the one corresponding to the fundamental string oscillator modes? If so, this would violate the sharpened Distance Conjecture but not the Emergent String Conjecture, which in particular shows that the latter conjecture does not necessarily imply the former. Clearly, in the case under consideration, this does not happen, and the sharpened Distance Conjecture is still satisfied in a non-trivial way, but this possibility opens interesting avenues to explore in the future.

\subsection{The $E_8 \times E_8$ Slice of Moduli Space }\label{sec:E8E8 slice}

A similar analysis to that presented above can be carried out for $E_8 \times E_8$ heterotic string theory on a circle. Recall that the different duality frames arising at different regions of the moduli space were shown in Figure \ref{modspace}.
Once again, this moduli space features a pair of Type I$'$ phases, which are related to each other by the self-T-duality of the $E_8 \times E_8$ heterotic string. To get the $E_8$ gauge group in this theory, we need to put 7 D8-branes on an orientifold plane, and one D8-brane away from it, precisely at a distance that will maintain the infinite string coupling at the $O8^-$ plane, as explained in \cite{Aharony:2007du}. If we do this at both ends, we get the vacuum of type I$'$ string theory that is dual to the heterotic $E_8 \times E_8$ string with no Wilson line turned on. We will see that the Kaluza-Klein modes for the two Type I$'$ phases will again slide along a line segment between $(-\frac{3}{\sqrt{8}} , \frac{1}{\sqrt{56}})$ and $(-\frac{1}{\sqrt{2}} , -\frac{3}{\sqrt{14}})$ as a function of position in moduli space. However, they will slide in the opposite direction from the case of the $SO(32)$ slice previously considered!

The dimensional reduction of the Type I$'$ theory is analogous to the case of $SO(32)$, with the exception that the warping and dilaton running along the compact direction are now given in terms of the $z_{E_8\times E_8}(x^9)$ functions by
 \begin{equation}
 	z_{E_8\times E_8}(x^9)\sim\left\{
 	\begin{array}{ll}
 	\hat Cx^9&\text{if } 0\leq x^9\leq B\\
 	\hat CB&\text{if } B\leq x^9\leq 2\pi-B\\
 	\hat C(2\pi-x^9)&\text{if } 2\pi-B\leq x^9\leq 2\pi
 	\end{array}
 	\right.,
 \end{equation}
 where, in the Type I$'$ frame, $B$ denotes the location, at $x^9=B\,,2\pi-B$, of the two D8-branes which are not located in the $O8$-planes.  Note that here $B\in[0,\pi]$, so that the two corresponding limits are: $B\to 0$, which corresponds with a low warping limit and a $E_8\times E_8\to SO(16)\times SO(16)$ enhancement, and $B\to \pi$ where the two bulk branes coincide in the middle of the interval, leading to the enhancement $E_8\times E_8\to E_8\times E_8\times SU(2)$ along the self-dual line \cite{Aharony:2007du}.

In the same way as in the $SO(32)$ slice, the Type I$'$ KK tower mass is given by \eqref{eq: gen KK mass}, with $\hat{\Omega}$ and $e^{\hat{\Phi}_{\rm I'}}$ given in terms of $z_{E_8\times E_8}(x^9)$ this time. Furthermore, as computed in Appendix \ref{app:dual}, the oscillator modes of the Type I$'_{E_8\times E_8}$ string read
\begin{equation}
	m_{\rm osc,I'}^{E_8\times E_8}\sim \left(\int_0^{2\pi-B}\dd x^9\hat\Omega^2\right)^{-1/4}
	\left(\int_0^{2\pi}\dd x^9\hat\Omega^8 e^{-2\Phi_{\rm I'}}\right)^{-1/7}
	\left(\sum_{i=1}^{16}\left.\hat\Omega^5 e^{-\Phi_{\rm I'}}\right|_{x^9=x^9_i}\right)^{1/4} M_{\rm Pl; 9}
\end{equation}

\begin{figure}[h]
\begin{center}
\includegraphics[width = 0.5\textwidth]{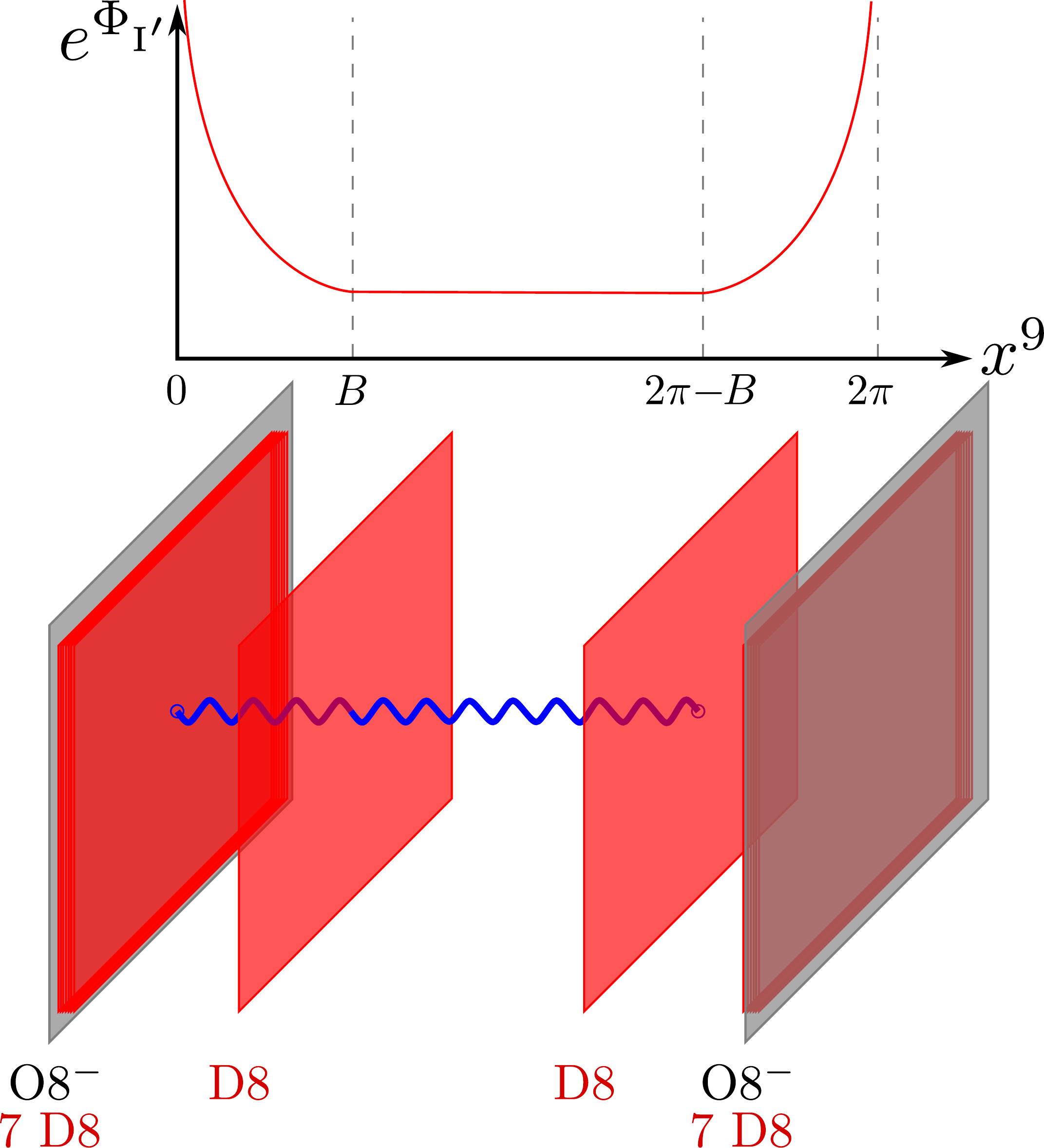}
\end{center}
\caption{Sketch of the ($O8$+7 D8)-D8-D8-($O8$+7 D8) brane-orientifold configuration needed to obtain the $E_8\times E_8$ gauge group in Type I$'$. The two stacks of seven D8-branes are located at the orientifolds at $x^9=0,\,2\pi$, with the other two branes at $x^9=B\, 2\pi-B$, such that the string coupling $e^{\Phi_{\rm I'}}$ diverges at each $O8^-$ plane. The string identified with the heterotic winding states in the Type I$'$ description is also depicted, stretching from one orientifold to the D8-brane located further from it along the interval.}
\label{fig: e8e8 branes}
\end{figure}
In order to obtain BPS masses in terms of the $(B,C)$ variables, we can identify the heterotic winding modes (which are wrapping M2-branes from the M-theory perspective) with the Type I$'$ string wrapping the interval, with endpoints at $x_I=0,\,2\pi-B$ (i.e. the Type I$'$  string is stretched between one $O8$-plane and the D8-brane  located further away in the interval\footnote{This identification of the heterotic winding states in the Type I$'$ description can be derived by matching charges of the states under the $E_8\times E_8\times U(1)\times U(1)$ gauge symmetry preserved at a generic point of this slice of moduli space.}, as in Figure \ref{fig: e8e8 branes}). This way, $
m_{\rm w, h}=\frac{R_{\rm h}}{2\pi\alpha'_{\rm h}}=m_{\rm w,I'}$. Using this, we obtain (see Appendix \ref{app: dual rel}) the following expression of the heterotic radius and dilaton,
\begin{align}
R_{\rm h}&\sim\left(\int_0^{2\pi-B}\dd x^9\hat\Omega^2\right)
	\left(\int_0^{2\pi}\dd x^9\hat\Omega^8 e^{-2\Phi_{\rm I'}}\right)^{-6/7}
	\left(\sum_{i=1}^{16}\left.\hat\Omega^5 e^{-\Phi_{\rm I'}}\right|_{x^9=x^9_i}\right)M_{\rm Pl; 9}^{-1}\\
		g_{\rm h}&\sim\left(\int_0^{2\pi-B}\dd x^9\hat\Omega^2\right)^{1/2}
	\left(\int_0^{2\pi}\dd x^9\hat\Omega^8 e^{-2\Phi_{\rm I'}}\right)
	\left(\sum_{i=1}^{16}\left.\hat\Omega^5 e^{-\Phi_{\rm I'}}\right|_{x^9=x^9_i}\right)^{-3/2},
\end{align}
The final result for the BPS states, $m_{\rm w, h}$ and $m_{\rm KK, h}\sim R_{\rm h}^{-1}$, is given in \eqref{eq: mKK E8E8} and \eqref{eq: mw E8E8}. Finally, using the field space metric of the $E_8\times E_8$ slice of the moduli space (see Appendix \ref{app:metric}), we can compute the scalar charge to mass ratio vectors of the above states in the flat $(\phi_B,\phi_C)\in\mathbb{R}_{>0}\times\mathbb{R}$ frame\footnote{The same way as in the $SO(32)$ case, we have $\dd s^2_{\mathcal{M}_{E_8\times E_8}}=\dd\phi_B^2+\dd\phi_C^2$, given by \eqref{eq: flat E8E8}
 \begin{subequations}
	\begin{align*}
		\phi_C&=\frac{10}{3\sqrt{7}}\log C+\frac{5}{6\sqrt{7}}\log\left[B(4\pi-B)^3\right]\\
		\phi_B&=-\frac{1}{2}\log\frac{3B}{4\pi-B},
	\end{align*}
\end{subequations}
again with the sliding only happening in the $\phi_B$ direction and $\phi_C$ axis being the self-dual line.}, as done in Appendix \ref{app.sliding}. All the towers have the same expression as the $SO(32)$ slice, i.e. \eqref{eq: vecs fixed}, except for the Type I$'$ KK tower, which reads
\begin{align}\label{eq:e8_kk_mass}
\vec{\zeta}_{\rm KK, I'}&=\left(\frac{1}{2}-\frac{2}{1+3e^{2\phi_B}},\frac{5}{2\sqrt{7}}\right)\notag\\&=\left\{\begin{array}{ll}
		\left(\frac{1}{2}-\frac{2}{3}e^{-2\phi_B}+\mathcal{O}\left(e^{-4\phi_B}\right),\frac{5}{2\sqrt{7}}\right)&\text{for }\phi_B\to\infty,\\
		\left(\frac{3}{4}\phi_B+\mathcal{O}\left(\phi_B^2\right),\frac{5}{2\sqrt{7}}\right)&\text{for }\phi_B\sim 0,
		\end{array}
		\right.
\end{align}
In this case, it interpolates between $\left(0,\frac{5}{2\sqrt{7}}\right)$ (highly warped along self-dual line) and $\left(\frac{1}{2},\frac{5}{2\sqrt{7}}\right)$ (unwarped), again solely as a function of the perpendicular distance to the self-dual line. The change of coordinates between $(\phi_B,\phi_C)$ and $(\rho,\phi)$ has positive-definite Jacobian, and the $SO(2)$ transformation is again given by $\vartheta=\arccos\left(-\frac{1}{2\sqrt{2}}\right)$. The same way as in the $SO(32)$ case, in the $(\phi,\rho)$ flat frame, we obtain
\begin{align}\label{eq:e8_kk_mass-het}
\vec{\zeta}_{\rm KK, I'}&=\left(-\frac{5}{4\sqrt{2}}-\frac{3}{4\sqrt{2}}\frac{e^{2 \phi_B}-1}{3 e^{2 \phi_B}+1},-\frac{5}{4\sqrt{14}}+\frac{3}{4}\sqrt{\frac{7}{2}}\frac{e^{2\phi_B}-1}{3e^{2\phi_B}+1}\right)\notag\\
&=\left\{\begin{array}{ll}
		\left(-\frac{3}{2\sqrt{2}}+\frac{1}{3\sqrt{2}}e^{-2\phi_B},\frac{1}{2\sqrt{14}}-\frac{1}{3}\sqrt{\frac{7}{2}}e^{-2\phi_B}\right)+\mathcal{O}\left(e^{-4\phi_B}\right)&\text{for }\phi_B\to \infty,\\
		\left(-\frac{5}{4\sqrt{2}}-\frac{3}{8\sqrt{2}}\phi_B,-\frac{5}{4\sqrt{14}}+\frac{3 }{8}\sqrt{\frac{7}{2}}\phi_B\right)+\mathcal{O}\left(\phi_B^2\right)&\text{for }\phi_B\sim 0.
		\end{array}
		\right.
\end{align}

\begin{figure}
\begin{center}
\includegraphics[width = 90mm]{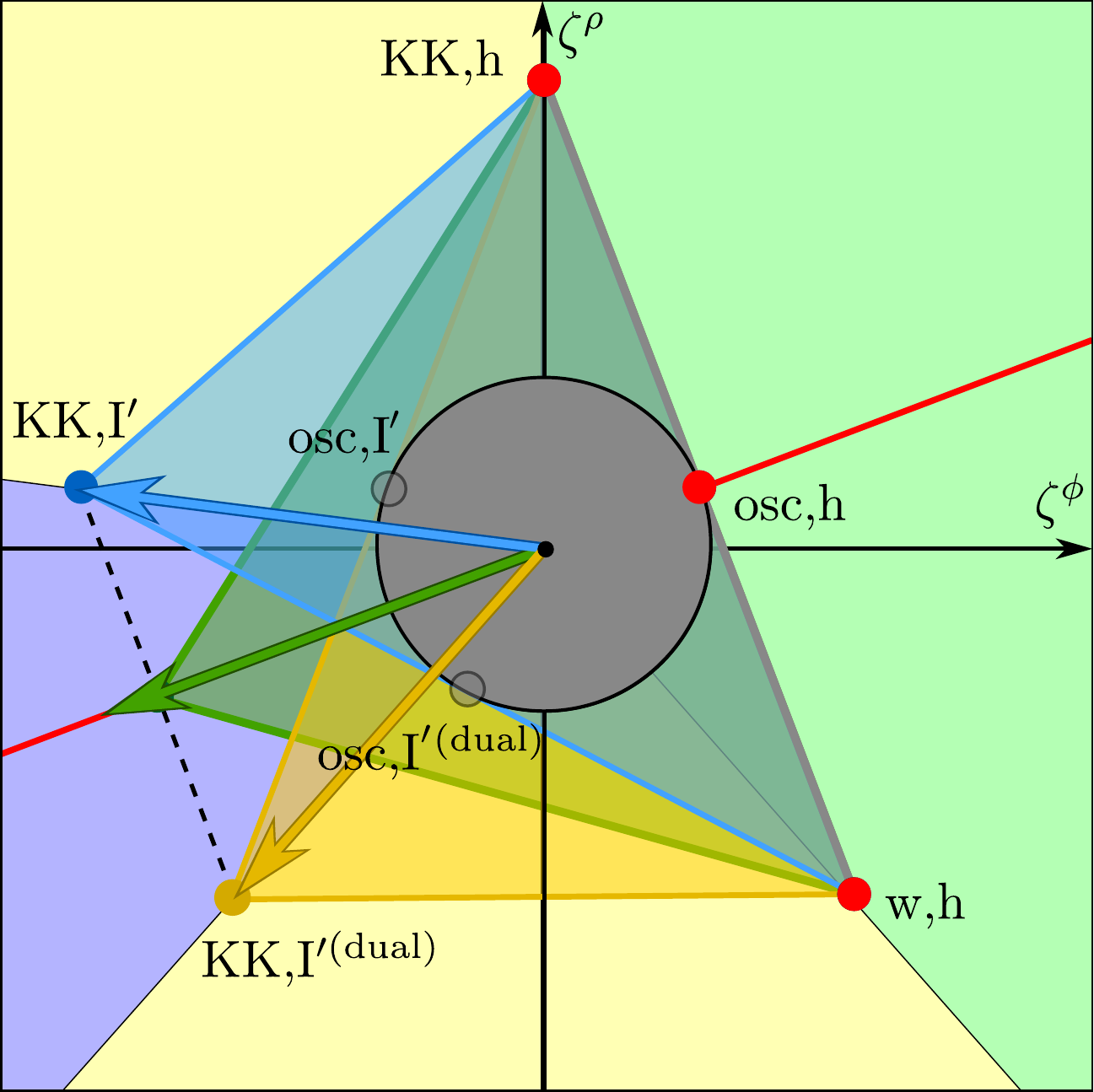}
\end{center}
\caption{Convex hulls in the two asymptotic limits for $B$ in the I$'_{E_8\times E_8}$ regions. The convex hull for the limit $B\to 0$ is shown in blue; this corresponds to the limit of the I$'_{E_8\times E_8}$ string with no warping and large string coupling between the $O8$'s. The convex hull for the analogous zero-warping, large coupling limit of the dual I$'_{E_8\times E_8}$ string is shown in yellow. The convex hull for the limit $B\to \pi$ limit is shown in green; this corresponds to decompactification to the 10-dimensional running solution. The sliding segment is depicted by a black dashed line. As in the $SO(32)$ case, the heterotic towers remain fixed in any limit. Note that the Type I$'$ emergent string limits are always obstructed.}
\label{E8E8sliding}
\end{figure}

The result for the SWGC convex hull is shown in Figure \ref{E8E8sliding}. The Type I$'$ KK modes again slide following the dashed black line. A key difference between $E_8 \times E_8$ heterotic string theory and its $SO(32)$ counterpart, however, is that the position of Type I$'$ KK scalar charge-to-mass vector approaches the value $(-\frac{1}{\sqrt{2}}, -\frac{3}{\sqrt{14}})$ in the \emph{bottom} Type I$'$ phase in Figure \ref{E8E8sliding}, whereas it approaches the value $(-\frac{3}{\sqrt{8}}, \frac{1}{\sqrt{56}})$ in the \emph{top} Type I$'$ phase (the $SO(32)$ case has top $\leftrightarrow$ bottom). Hence, as we move along the blue arrow in Figure \ref{E8E8sliding}, the convex hull is given by the blue triangle, while the yellow triangle arises when moving along the yellow arrow in the bottom Type I$'$ phase. As a result, these Kaluza-Klein modes obstruct the Type I$'$ emergent string limits,\footnote{This nicely reproduces the string theory expectations. From the string theory perspective, weak coupling implies moving the isolated D8's closer to each other, so there is a lower bound for the string coupling that gets saturated when the two D8's coincide at the middle of the interval. Hence, we cannot take the weak coupling limit while keeping the radius of the interval fixed, so the Type I$'$ emergent string limit is obstructed.} so there is only one emergent string limit in this moduli space, namely, the emergent heterotic string limit. On the contrary, the unwarped decompactification limit is not obstructed if we move along the direction of $\vec{\zeta}_{\rm KK, I'}=\left(\frac{1}{2},\frac{5}{2\sqrt{7}}\right)$.  

The result of this analysis for the Distance Conjecture is the max-$\alpha$ hull shown in Figure \ref{E8E8max}. Even if the towers are located at the same places than for $SO(32)$, the nature of the leading tower dominating along some of the asymptotic limits is different. Since the emergent string limit is obstructed, the yellow region now corresponds to decompactifying two dimensions (to M-theory). In the Type I$'$ blue region, we still decompactify to a 10-dimensional running solution (thereby the sliding of the KK modes), but only a finite region of the interval exhibits a non-vanishing Romans mass. The sharpened Distance Conjecture is satisfied, as the exponential rate of the leading tower satisfies $\alpha_{\text{max}} \geq \frac{1}{\sqrt{d-2}}$ in every direction, and saturation occurs only in the emergent string limit.

\subsection{The $SO(16)\times SO(16)$ Slice of Moduli Space}
For the sake of completeness, we consider one final slice of the moduli space of heterotic string theory compactified on $S^1$ to 9d: the slice with enhanced $SO(16)\times SO(16)$ gauge symmetry. From the Type I$'$ point of view, this corresponds \cite{Aharony:2007du} to having 8 D8-branes on each $O8^-$ plane located at the endpoints of the interval $S^1/\mathbb{Z}_2$. By \eqref{eq: z gen}, we have $z(x^9)\propto BC$, so there is no warping and the dilaton is constant along the interval. In this case $g_{\rm I'}\sim B^{-5/6}C^{-5/6}$ and $\Omega\sim B^{-5/6}C^{1/6}$, which implies that every point of moduli space has the same moduli space metric, and the scalar charge-to-mass vectors of the various towers do not slide.

\begin{figure}
\begin{center}
\includegraphics[width = 80mm]{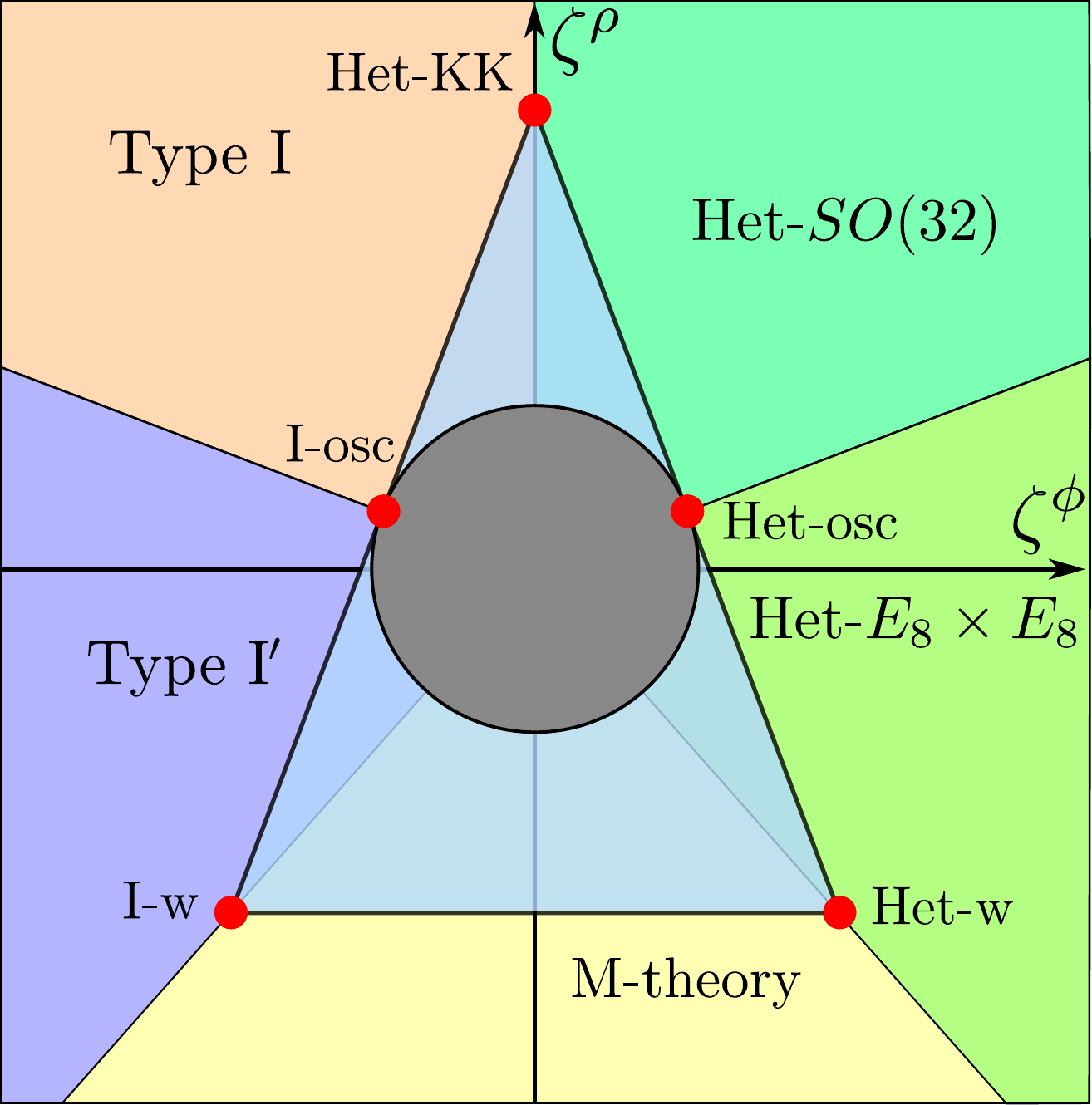}
\end{center}
\caption{The convex hull of the scalar charge-to-mass vectors for the $SO(16)\times SO(16)$ slice of 9d heterotic moduli space, given in terms of the canonically normalized dilaton and radion of the $SO(32)$ heterotic string. It is evident that the convex hull contains the ball of radius $\frac{1}{\sqrt{d-2}}=\frac{1}{\sqrt{7}}$. Note that this figure is essentially identical to Figure \ref{CHII}, but in this case the towers of Type I oscillation modes and winding modes are not BPS.}
\label{chso16so16}
\end{figure}

Starting from this Type I$'_{SO(16)\times SO(16)}$ frame, we can cover the entire two-dimensional slice of moduli space by a sequence of dualities \cite{Horava:1996ma,Polchinski:1995df, Horava:1995qa,Witten:1995ex,Ibanez:2012zz}. First, this Type I$'$ string theory is T-dual to Type I string theory with Wilson lines that break the gauge symmetry to $SO(16)\times SO(16)$. As in \S\ref{sec:puzzle_in_9d}, this is S-dual to $SO(32)$ heterotic string theory on $S^1$, again with Wilson lines preserving the $SO(16)\times SO(16)$ subgroup. This is in turn T-dual to $E_8\times E_8$ heterotic string theory on $S^1$ with Wilson lines preserving $SO(16)\times SO(16)$ (recall that that is the maximal common subgroup of $SO(32)$ and $E_8\times E_8$). Finally, as in \S \ref{sec:E8E8 slice}, the $E_8\times E_8$ heterotic string can be related via the Ho\v{r}ava-Witten construction to M-theory on $S^1/\mathbb{Z}_2\times S^1$, again with the Wilson lines on $S^1$ breaking $E_8$ to $SO(16)$. From the initial 9-dimensional Type I$'$ on $S^1/\mathbb{Z}_2$ point of view, the 10d bulk theory is Type IIA, which in the strong coupling limit can be lifted to M-theory by growing an additional $S^1$, thus completing the duality chain. The different regions, parametrized in terms of the canonically normalized $SO(32)$ heterotic dilaton and radion, are depicted in Figure \ref{chso16so16}.

Figure \ref{chso16so16} also depicts the scalar charge-to-mass vectors of the relevant towers of the theory. Beginning in $SO(32)$ heterotic string frame, we first have the KK modes (located on the $\rho$ axis) and the heterotic string oscillation modes (located on in the self-T-dual line), as well as the winding modes of the $SO(32)$ theory. Crossing to the Type I frame, we have the Type I KK and winding modes at each side of the T-duality line, along which the Type I string tower is located. The mass expressions in terms of $\{\phi,\rho\}$ and coordinates coordinates of the $\vec{\zeta}_I$ vectors in this $SO(32)$ heterotic frame are the same as in \eqref{eq: IIB 1}--\eqref{eq: IIB S1 vecs}, so the resulting convex hull, depicted in Figure \ref{chso16so16}, is the same as the Type IIB case shown in Figure \ref{CHII}. Notably, the convex hull remains unchanged as we move in the moduli space, which is possible because the heterotic theory is not self-T-dual in this slice of moduli space, so the convex hull is not symmetric under the T-dual heterotic line.

Similarly, the discussion about the leading tower decay rate $\alpha_{\max}(\hat \tau)$ along different directions $\hat \tau$ is the same as the Type IIB case, which was discussed above in \S\ref{sec: IIB circ}. Note, however, that in this case the towers of Type I string oscillation modes and winding modes are not BPS.

\section{Other Nine-Dimensional Moduli Spaces with 16 Supercharges\label{sec: other mod spaces}}

So far, our main focus has been testing the sharpened Distance Conjecture in nine-dimensional heterotic string theory. In this section, we will see that our results immediately generalize and allow us to describe multiple additional slices of the landscape of nine-dimensional quantum gravities with 16 supercharges. These additional slices can be viewed as ``frozen'' phases of the two slices we have considered with $SO(32)$ and $E_8 \times E_8$ gauge symmetry. We again refer the reader to \citep{Aharony:2007du} for a detailed account of these theories.

The first additional slice we can easily describe is the two-dimensional locus in the moduli space of the CHL string with enhanced $E_8$ gauge symmetry \cite{Aharony:2007du,Chaudhuri:1995fk,Chaudhuri:1995bf}. This theory has $r = 9$ vector multiplets, and is obtained by compactifying the $E_8 \times E_8$ heterotic string on a circle with a discrete Wilson line turned on for the gauged $\mathbb{Z}_2$ outer automorphism exchanging the two copies of $E_8$. The moduli space of this theory is identical to that of the $E_8 \times E_8$ slice we have considered previously (see Figure \ref{modspace}), with a self-T-duality line and duality frames given by $E_8 \times E_8$ heterotic string theory on a circle, M-theory on a M\"{o}bius band, and an unusual variant of Type I$'$ string theory.

To describe this variant, recall that the $E_8 \times E_8$ slice was given by a configuration of Type I$'$ string theory with 7 D8-branes on each $O8^-$ plane and two additional D8-branes placed precisely at a distance to maintain infinite string coupling on each $O8^-$. To obtain the $E_8$ locus of the CHL string, one must replace one of the $O8^-$ planes and the 7 D8-branes on top of it with an $O8^{(-1)}$ plane that is frozen to sit at infinite coupling. While this replacement changes the local dynamics on the orientifold plane, it does not change the bulk geometry whatsoever, and thus the scalar charge to mass ratio vector of the Type I$'$ KK modes is again given by \eqref{eq:e8_kk_mass}. As a result, the SWGC convex hulls for the $E_8$ slice of the CHL string moduli space will be identical to those for the corresponding point in the $E_8 \times E_8$ slice depicted in Figure \ref{E8E8sliding}, and the max-$\alpha$ hull will be the same as that depicted in Figure \ref{E8E8max}.

The next slice we can easily describe is the two-dimensional moduli space of the asymmetric orbifold of Type IIA string theory on a circle (AOA) by the action of $(-1)^{F_L}$ combined with a half-shift along the circle \cite{Aharony:2007du,Gutperle:2000bf,Hellerman:2005ja}, a theory with $r = 1$ vector multiplets. The moduli space of this theory is again identical to that of the $E_8 \times E_8$ slice, and has a self-T-duality line as well as duality frames given by the AOA theory, M-theory on a Klein bottle, and a configuration of Type I$'$ string theory. This configuration is obtained from the $E_8 \times E_8$ configuration by replacing both $O8^-$ planes and the 7 D8-branes on each with $O8^{(-1)}$ planes frozen to infinite coupling, leaving behind two D8-branes in the middle of the Type I$'$ interval. Again, the bulk geometry is identical to that of the $E_8 \times E_8$ configuration, and so the Type I$'$ KK scalar charge to mass ratios, SWGC convex hulls, and max-$\alpha$ plots will be identical to those for the $E_8 \times E_8$ slice considered previously.

The final slice we can easily describe is the two-dimensional moduli space of a similar asymmetric orbifold of Type IIB string theory on a circle (AOB) by $(-1)^{F_L}$ combined with a half-shift \cite{Aharony:2007du,Hellerman:2005ja,Gutperle:2000bf}, another theory with $r = 1$ vector multiplets. While the moduli space of the AOA theory is identical to that of the $E_8 \times E_8$ slice, the moduli space of the AOB theory is instead identical to that of the $SO(32)$ slice, with a self-T-duality line and duality frames given by the AOB theory, the Dabholkar-Park background \cite{Dabholkar:1996pc}, and a configuration of Type I$'$ string theory. This configuration is obtained by the $SO(32)$ configuration by replacing the $O8^-$ plane with 16 D8-branes on top of it with an $O8^+$ plane. Just as in the previous two cases, the bulk geometry remains identical (this time to that of the $SO(32)$ configuration), and so the Type I$'$ KK scalar charge to mass ratios, SWGC convex hulls, and max-$\alpha$ plots will be identical to those for the $SO(32)$ slice (given in \eqref{eq: sliding BC}, Figure \ref{SO32slide}, and Figure \ref{SO32max} respectively).

There are two additional slices of the landscape of nine-dimensional quantum gravity with 16 supercharges, to which our results do not quite apply verbatim, but for which we expect very similar (if not identical) results to hold. These are the additional slice with $SO(32)$ gauge symmetry mentioned in Footnote \ref{other_so32}, as well as its frozen phase, the new string theory with $r = 1$ vector multiplets described in reference \cite{Montero:2022vva} and obtained by turning on a discrete $\theta$-angle in the AOB theory. These theories are very similar to the first $SO(32)$ slice and the AOB theory respectively, but have additional $\mathbb{Z}_2$ restrictions on their charge lattices. Our expectation is that these differences will only change the prefactor of the masses of towers of states, and not the exponential rates, so it is our expectation that the SWGC convex hulls and max-$\alpha$ plots will be identical to those plotted in Figure \ref{SO32slide} and Figure \ref{SO32max} respectively.

\section{Discussion}\label{CONC}

In this paper, we have studied several noteworthy slices of the moduli space of quantum gravity theories in nine dimensions with 16 supercharges. Our findings have led to a striking confirmation of the sharpened Distance Conjecture and an important clarification for the Emergent String Conjecture. As demanded by the sharpened Distance Conjecture, every infinite-distance limit in moduli space considered above features at least one tower of light particles which decays with geodesic distance $\phi$ as $m \sim e^{ - \alpha \phi  } $, with $\alpha \geq \frac{1}{\sqrt{d-2}} = \frac{1}{\sqrt{7}}$. This bound is saturated only in emergent string limits, and is satisfied strictly in all other limits.

As demanded by the Emergent String Conjecture, all of these infinite-distance limits represent either emergent string limits or decompactification limits. However, in the case of the Type I$'$ decompactification limits, we found that the decompactification does not result in a 10-dimensional vacuum, but rather a running solution. The running of the dilaton in a Type I$'$ decompactification limit implies that the masses of the Type I$'$ Kaluza-Klein modes develop a non-trivial dependence on the moduli, which we computed explicitly by a careful dimensional reduction including the effects of a warped compactification. The possibility of a decompactification to a non-vacuum state is an important caveat to be considered when attempting to derive consequences from the Emergent String Conjecture (as in \cite{Bedroya:2023xue}), since it implies a possible suppression of the exponential rate of a KK tower due to the warping and a non-trivial variation of its value as we move in the moduli space.
Given this, it is perhaps a bit surprising that the sharpened Distance Conjecture continues to hold even in Type I$'$ decompactification limits, and more generally it is not obvious that the sharpened Distance Conjecture will remain valid once decompactifications to non-vacuum solutions are taken into account.

We also checked a version of the Scalar Weak Gravity Conjecture (SWGC) \cite{Palti:2017elp,Calderon-Infante:2020dhm} in these nine-dimensional theories, which implies a lower bound for the ratio $|\vec{\zeta}|=\frac{|\vec\nabla m|}{m}$ of the gradient of the mass to the mass of the tower of states, which is commonly known as the scalar charge-to-mass ratio. Unlike the Distance conjecture, the SWGC is a local condition of the moduli space, and we find that it is always satisfied in the asymptotic regimes if we take the bound to be $|\vec{\zeta}|\geq \frac{1}{\sqrt{d-2}}$. This holds thanks to the particular sliding behaviour of the non-BPS states. Notice that this version of the SWGC no longer has the interpretation of a balance of gravitational and scalar forces (as in the original SWGC proposal  \cite{Palti:2017elp}) since the numerical factor in the bound is different, and is instead fixed to coincide with the lower bound of the sharpened Distance conjecture.\footnote{This is why such type of bound was originally refered to in \cite{Calderon-Infante:2020dhm} as the Convex Hull Distance Conjecture and referred to in \cite{Etheredge:2022opl} as the Tower SWGC.}

\begin{figure}
\begin{center}
\begin{subfigure}{0.475\textwidth}
\center
\includegraphics[width=70mm]{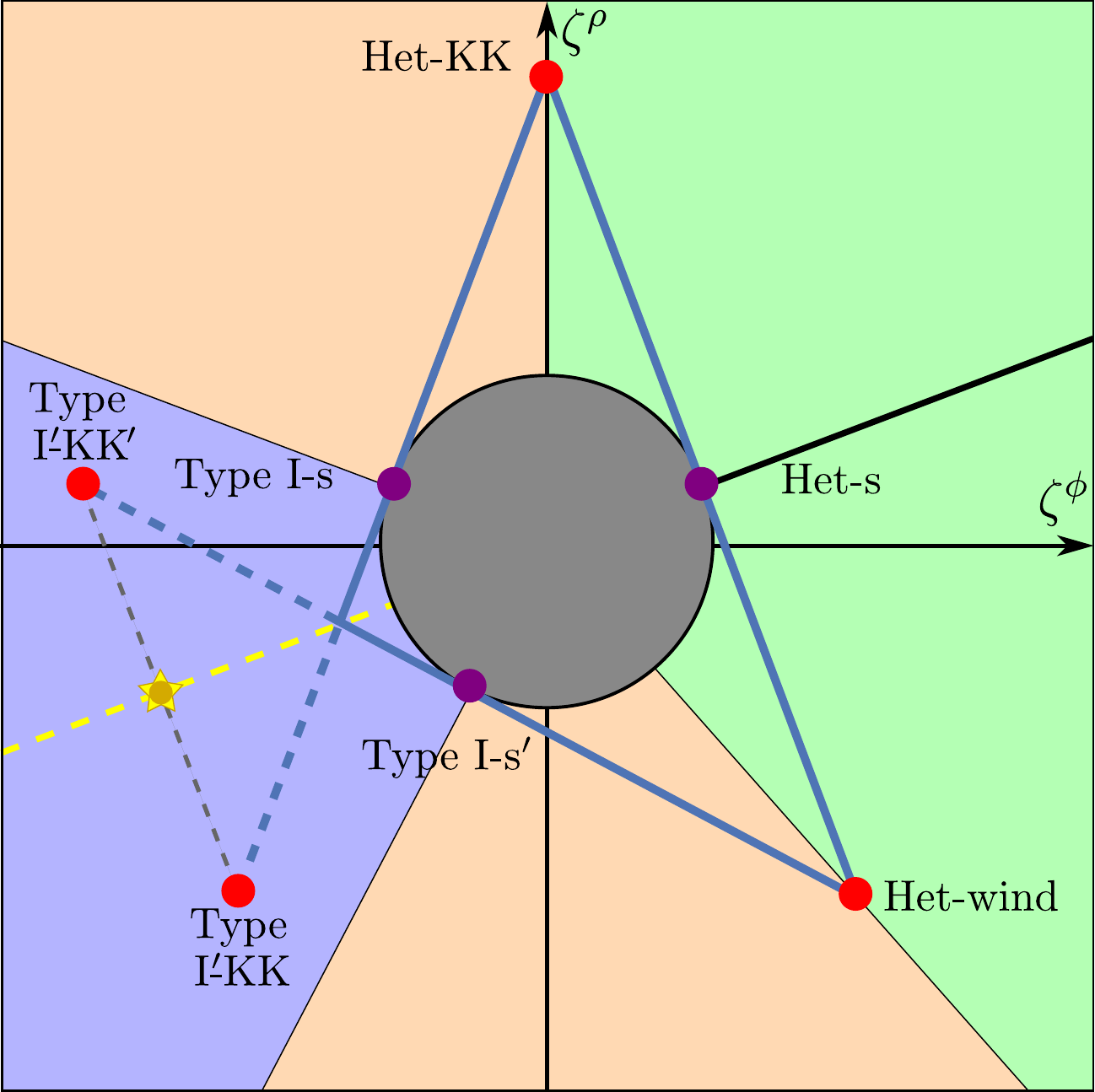}
\caption{$SO(32)$} \label{SO32ch}
\end{subfigure}
\begin{subfigure}{0.475\textwidth}
\center
\includegraphics[width=70mm]{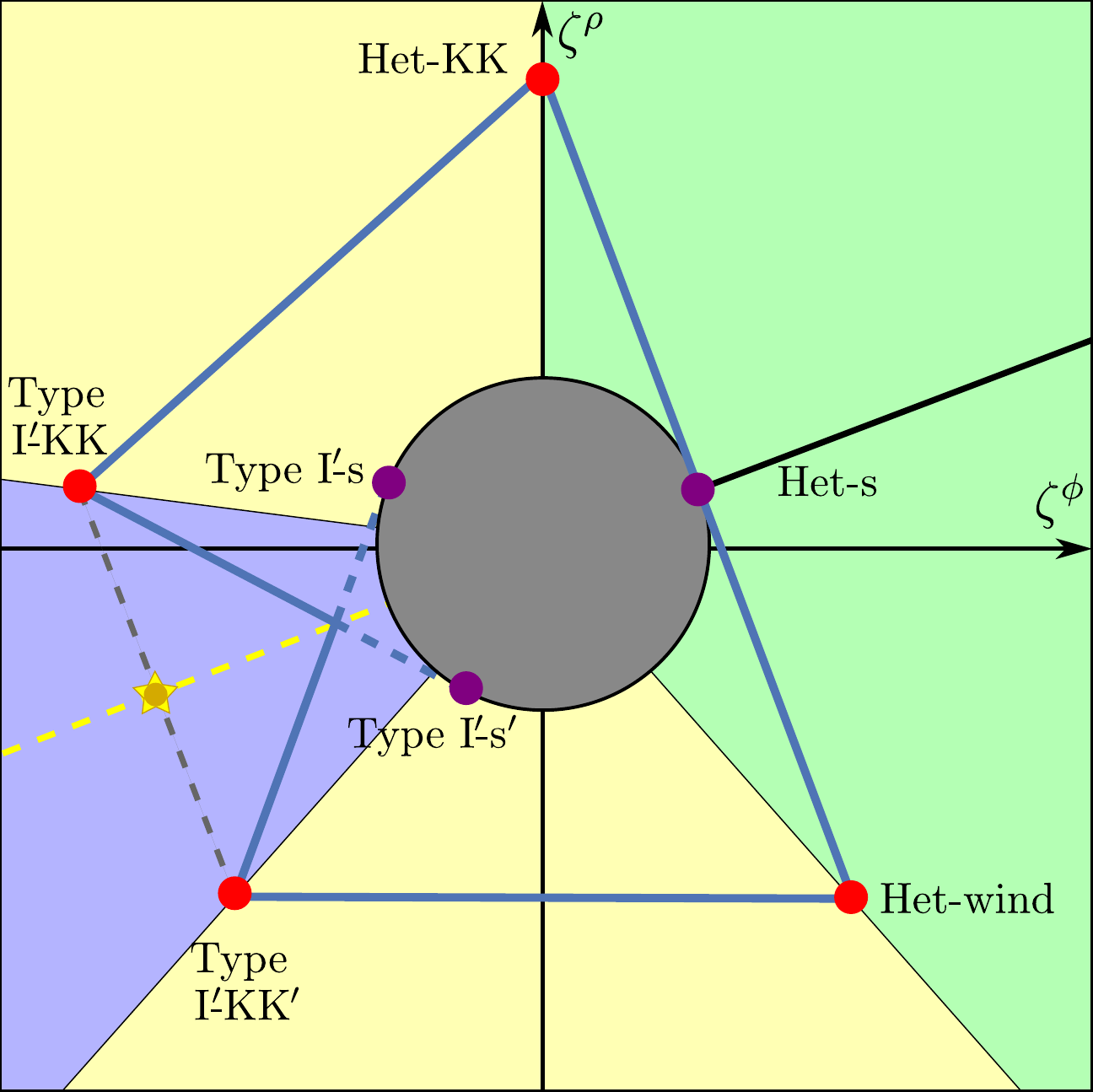}
\caption{$E_8\times E_8$} \label{E8E8ch}
\end{subfigure}
\caption{Sketch of the leading towers that fulfill the Convex Hull Distance Conjecture in the different asymptotic regions. Each facet is generated by two competing towers, each of which becomes light in an asymptotic region. The emergent string limits are depicted in purple. The self-dual line in the Type I$'$ region should be interpreted as a sort of \emph{branch cut}, in the sense that the scalar charge-to-mass ratios of the Type I$'$-KK towers jump depending on which side of the self-dual line the infinite-distance limit resides. This means that the competing towers in the Type I$'$ region are always the Type I$'$ KK modes and the string oscillator modes, but the location of these towers changes as we cross the self-dual line.\label{CHSDC}}
\end{center}
\end{figure}

Assuming that the scalar charge-to-mass ratio of the towers does not change in a given asymptotic regime, reference \cite{Calderon-Infante:2020dhm} showed that the Distance Conjecture is satisfied with minumum rate $\alpha_{\rm min}$ if and only if the convex hull of the towers of states includes the ball of radius $\alpha_{\rm min}$. In this paper, we find that this connection between the Distance Conjecture and this version of the SWGC still holds in the interior of any fixed asymptotic regime (i.e. for each of the dual regimes in Figure \ref{modspace}), even taking into account the sliding of the non-BPS states. This is possible thanks to the fact that the sliding of the Type I$'$ KK states occurs instantaneously as a function of the asymptotic direction, and so there is effectively no sliding as long as one considers asymptotic geodesics that have a different asymptotic tangent vector.\footnote{If we consider geodesics that are parallel to the self-dual line, the sliding occurs as a function of the distance to the self-dual line.} This implies that, in practice, one can draw the max-$\alpha$ plot by stitching together the max-$\alpha$ plots of the unwarped light towers in each region. However, the relevant light towers jump discontinuously as a function of direction when crossing the self-dual line, as represented in Figure \ref{CHSDC}. The jumping occurs in opposite directions for the case of $SO(32)$ or $E_8\times E_8$. As a result, the exponential rate of the Kaluza-Klein modes matches that of an unwarped circle compactification along the geodesics in the interiors of the Type I$'$ regions, even if it never reaches the unwarped rate of $\sqrt{\frac{8}{7}}$ in the $SO(32)$ case (recall Figure \ref{SO32max} for the max-$\alpha$ plot providing the exponential rate of the leading tower along each direction). It would be interesting to investigate whether this relationship between the Convex Hull SWGC and the Distance Conjecture holds more generally in any fixed asymptotic region (satisfying, therefore, the Convex Hull Distance Conjecture of \cite{Calderon-Infante:2020dhm}), or whether there are examples where the SWGC convex hull changes continuously as a function of asymptotic direction.

It is important to note that, while the max-$\alpha$ plot for each fixed asymptotic region arises from a convex hull, the resulting figure obtained from joining piece-wise the different convex hulls in each asymptotic regime is not a convex hull anymore, as depicted in see Figure \ref{CHSDC}. In each asymptotic regime, we draw the convex hull of the leading towers of states, which happen to be always a straight line between the two competing towers characterizing each asymptotic regime. In the Type I$'$ regime, the competing towers are always the Type I string oscillator modes and the Type I$'$ KK modes, but the KK modes jump as we cross the self-dual line. Figure \ref{CHSDC} also nicely captures the difference between the $SO(32)$ and $E_8\times E_8$ slices: even if the towers of states seem to be located at the same places, the Type I$'$ KK tower which is valid in a given Type I$'$ regime is located at opposite sides of the self-dual line, and the jumping occurs in opposite directions. This implies that the pure decompactification limit (or the emergent string limit) is obstructed for $SO(32)$ (or $E_8\times E_8$) respectively, since the relevant tower of states is not present when moving in the appropriate direction. This result nicely reproduces the string theory expectations.

An interesting exercise is to check how much we could have predicted about the weakly coupled descriptions that emerge in the infinite distance limits knowing only the towers of states (along the lines of \cite{Bedroya:2023xue}). First of all, we want to remark that knowing the leading tower along a particular asymptotic direction is not enough to find out the weakly coupled description that emerges asymptotically. For instance, consider an asymptotic trajectory on the upper left region of the moduli space in Figure \ref{CHSDC}.  The leading tower is the heterotic KK modes for $\rho>-\frac{3}{\sqrt{7}}\phi$ both in the case of  $SO(32)$ and $E_8\times E_8$. However, the emerging weakly coupled description is very different in the two cases, as one obtains 10d Type I string theory in the former and 11d M-theory in the latter. Hence, in general, it is necessary to have information about the multiple competing light  towers in a given asymptotic region, information that can be easily read from Figure \ref{CHSDC}. For the $SO(32)$ case, the upper left (orange) region is controlled by a KK tower of one extra dimension and a tower of string oscillator modes, so the emerging description is a string theory in 10 dimensions. Contrarily, for  $E_8\times E_8$,  the upper left (yellow) region is controlled by two KK towers of one extra dimension, so the resolution is an 11 dimensional theory (i.e., M-theory). Hence, one has to be careful when extracting conclusions from checking individual trajectories or neglecting the sliding of the towers in the moduli space.  Let us also mention that each dual frame (or equivalently, each face of Figure \ref{CHSDC}) is characterized by a unique result of the species scale hinting a particular weakly coupled description, as will be explored in more detail in \cite{madridSpeciesHullTBA, taxonomyTBA, patternTBA}.

One shortcoming of this work is that we have ignored periodic scalar fields, i.e., axions. This omission can be justified on the grounds that axions do not play a role in our discussion of the Distance Conjecture, since they may be taken to be constant along asymptotic geodesics in these slices.
However, axions do play an important role in the closely related Convex Hull SWGC \cite{Etheredge:2022opl, Calderon-Infante:2020dhm}, the refined Distance Conjecture \cite{Klaewer:2016kiy, Baume:2016psm}, and other attempts to extend the Distance Conjecture into the interior of moduli space \cite{Rudelius:2023mjy}. Furthermore, axions are more relevant for phenomenology than infinite-distance limits in moduli space. To this end, it would be worthwhile to study axion couplings to matter, both in the theories considered here and more generally. 

It has been over 17 years since Ooguri and Vafa first appreciated the appearance of universal behavior in infinite-distance limits of quantum gravity moduli spaces and proposed the celebrated Distance Conjecture. Yet even the last several years have seen remarkable progress in our understanding of these limits. The structures underlying the Distance Conjecture have come into focus, and the Distance Conjecture itself has attained a greater degree of precision and rigor. After the explosive activity of the past years, it is fair to say that we are now entering into a precision era in the Swampland program. In this paper, we have extended this program even further, and in the process we have demonstrated that even old and well-studied theories may hold new, important insights into old and well-studied Swampland conjectures. We hope that this work will inspire further exploration of uncharged territory in the Landscape and the Swampland, even more precisely.

\section*{Acknowledgements}

We are very thankful to Jos\'e Calder\'on-Infante, Alberto Castellano, Alek Bedroya and Miguel Montero for illuminating discussions.  IR also wishes to acknowledge the hospitality of the Department of Theoretical Physics at CERN during the development of this work. The work of BH and ME was supported by NSF grants PHY-1914934 and PHY-2112800. The work of JM is supported by the U.S. Department of Energy, Office of Science, Office of High Energy Physics, under Award Number DE-SC0011632. I.V. and I.R. acknowledge the support of the Spanish Agencia Estatal de Investigacion through the grant “IFT Centro de Excelencia Severo Ochoa CEX2020-001007-S and the grant PID2021-123017NB-I00, funded by MCIN/AEI/10.13039/ 501100011033 and by ERDF A way of making Europe. The work of IR is supported by the Spanish FPI grant No. PRE2020-094163. The work of I.V. is also partly supported by the grant RYC2019-028512-I from the MCI (Spain) and the ERC Starting Grant QGuide -101042568 - StG 2021.

\appendix

\section{Heterotic - Type I$'$ duality in nine dimensions\label{app:dual}}
In this appendix, we rederive the background of \cite{Polchinski:1995df}. Additionally, we compute the masses of the 1/2 BPS heterotic winding and KK modes. We defer to Appendix \ref{app:KK} to compute the masses of KK modes of Type I$'$ string theory. With these masses, we will compute in Appendix \ref{app.sliding} the sliding of the $\vec{\zeta}$-vectors for Type I$'$ KK modes.

\subsection{Equations of motion for the Type I$'$ dilaton and warp factor\label{app:eom}}
We begin by obtaining and solving the equations of motion for the massive Type I$'$ dilaton $\Phi_{\rm I'}$ and warp factor for the 10-dimensional string-frame metric, $g_{\rm I'}=\Omega^2\eta$. We will consider them to be dependent only on the internal dimension $x^9\in[0,2\pi]$. Along this interval we consider two orientifolds located at its extremes, $x^9_{O8}=0,\, 2\pi$, and 16 D8-branes localted at $\{x^9_i\}_{i=1}^{16}\subset[0,2\pi]$, with coupling $\mu_8$ to the 9-form potential. The bulk action (the brane terms will only account for the ``jumps'' of these functions and can be studied separately) is given by \cite{Ibanez:2012zz}
\begin{align}\label{e.10daction}
	S_{\rm I'}^{\rm (bulk)}&=\frac{1}{2\kappa_{10,I'}^2}\int\dd^{10}x\sqrt{-g_{\rm I'}}\left\{e^{-2\Phi_{\rm I'}}\left[R_{g_{\rm I'}}+4\partial_M\Phi_{\rm I'}\partial^M\Phi_{\rm I'}\right]+(\alpha_{\rm I'}')^4\nu_0^2\right\}\notag\\
	&=\frac{1}{2\kappa_{10,\rm I'}^2}\int\dd^{10}x e^{-4\Phi_{\rm I'}(x^9)}\Omega(x^9)^6\left\{(\alpha_{\rm I'}')^4\nu_0^2e^{4\Phi_{\rm I'}(x^9)}\Omega(x^9)^4+4\Omega(x^9)^2\partial_9e^{\Phi_{\rm I'}(x^9)}\right.\notag\\
	&\quad\left.-18e^{2\Phi_{\rm I'}(x^9)}\left[3\Omega'(x^9)^2+\Omega(x^9)\Omega''(x^9)\right]\right\}
\end{align}
where we have included an $(\alpha_{\rm I'}')^2$ term accompanying the Romans mass so that the $F_{10}\wedge \star_{g_{\rm I'}}F_{10}=\star_{g_{\rm I'}}\nu_0^2$ term has the correct units of $L^8$. Now, the solutions to the associated equations of motion are given by
\begin{equation}
	e^{\Phi_{\rm I'}(x^9)}=z(x^9)^{-5/6}\qquad \Omega(x^9)=Cz(x^9)^{-1/6},
\end{equation}
with\footnote{Note that the numerical coefficient $\sqrt{\frac{180}{41}}$ we obtain is \emph{slightly} different than the $\frac{3}{\sqrt{2}}=\sqrt{\frac{180}{40}}$ appearing in \cite{Polchinski:1995df}. While this difference will not have any further implication in our results, it is nonetheless an interesting observation.}
\begin{equation}
	z(x^9)=\sqrt{\frac{180}{41}}(\alpha'_{\rm I'})^2 C(B\mu_8\pm\nu_0 x^9),
\end{equation}
with $\nu_0$ constant between branes, such that there its value has a $\Delta(x_i^9)=n_i\mu_8$ jump at each stack of $n_i$ 8-branes located at $x_i^9$, resulting in $\nu_0(2\pi)=\nu_0(0)+16\mu_8$. As boundary conditions require $\nu_0(2\pi)=-\nu_0(0)=8\mu_8$, we end up having
\begin{equation}
	\nu_0(x^9)=\mu_8\left[\int_0^{x^9}\sum_{i=1}^{16}\delta(\tau-x ^9_i)\dd \tau-8\right].
\end{equation} 

 On the other hand, $B$ and $C$ are two functions with dimensions of $1$ and $L$ constant between branes. Following the discussion from \cite{Polchinski:1995df}, by requiring $\Omega$ and $\Phi_{\rm I'}$ to be continous, we have that $C$ must be constant and $\Delta B(x_i^9)=\mp n_i x_i^9$, so that
 \begin{equation}
 B(x^9)=B(x_0^9)\mp\int_{x_0^9}^{x^9}\sum_{i=1}^{16}\tau\delta(\tau-x_i^9)\dd\tau,
 \end{equation}
 where $x_0^9$ is some arbitrary position of the interval, $x_0^9\in[0,2\pi]$ for which $B(x_0^9)$ is finite. We will take $C$ and $B(x_0^9)$ (in our computations simply $B$) as moduli. Furthermore, positivity of the membrane tension will require taking the lower signs of the above expressions, finally reaching the following expression:
 \begin{equation}\label{eq:final z}
 	z(x^9)=\sqrt{\frac{180}{41}}(\alpha_{\rm I'}')^2\mu_8 C\left\{B+\int_{x_0^9}^{x^9}\sum_{i=1}^{16}\tau\delta(\tau-x_i^9)\dd\tau-\left(\int_0^{x^9}\sum_{i=1}^{16}\delta(\tau-x ^9_i)\dd \tau-8\right)x^9\right\}
 \end{equation}
 This greatly simplifies for the $SO(32)$, in which all the branes are located at $x_i^9=2\pi$, and we take $B=B(0)$, so that
\begin{equation}\label{eq zso32}
z_{SO(32)}(x^9)=\sqrt{\frac{180}{41}}(\alpha_{\rm I'}')^2\mu_8 C(B+8x^9),
\end{equation}
with the $B\to 0$ limit resulting in the string coupling diverging at $x^9=0$, while for $B\to \infty$ both $\Phi_{\rm I'}$ and $\Omega$ are approximately constant, corresponding with the low warping limit.

For $E_8\times E_8$ case, where we have 7 D8-branes at each $O8$-plane and two additional at two points $\pi\mp x_{\rm I'}^9$, with $B=B(\pi)$. If we further require the orientifold planes to have infinite coupling, we will need to impose $z(0)=z(2\pi)=0$, which using \eqref{eq:final z} amounts to $x_{\rm I'}^9=\pi-B$ (so that it is only valid for $B<\pi$), and thus
 \begin{equation}\label{eq zE8E8}
 	z_{E_8\times E_8}(x^9)=\left\{
 	\begin{array}{ll}
 	\sqrt{\frac{180}{41}}(\alpha_{\rm I'}')^2\mu_8 Cx^9&\text{if } 0\leq x^9\leq B\\
 	\sqrt{\frac{180}{41}}(\alpha_{\rm I'}')^2\mu_8 CB&\text{if } B\leq x^9\leq 2\pi-B\\
 	\sqrt{\frac{180}{41}}(\alpha_{\rm I'}')^2\mu_8 C(2\pi-x^9)&\text{if } 2\pi-B\leq x^9\leq 2\pi
 	\end{array}
 	\right.
 \end{equation}

As we will need to use it in the next subsection, we can obtain the 9-dimensional Einstein metric for our theory. For this we will write said 9-dimensional metric as $\mathsf{g}_{\mu\nu}=D^{-2}\eta_{\mu\nu}$, with $D$ some mass scale independent of $x^9$ we will soon determine, so that $g_{\rm I'}{}_{\mu\nu}=\left(\Omega D\right)^2\mathsf{g}_{\mu\nu}$. Doing this, we obtain
\begin{equation}\label{eq hash D}
S_{\rm I'}\supset\frac{1}{2\kappa_{10,\rm I'}^2}\int \dd^{10}x\sqrt{-g_{I'}}e^{-2\Phi_{\rm I'}}R_{g_{\rm I'}}\supset \frac{1}{2\kappa_{10,\rm I'}^2}\int\dd^9 x\sqrt{-\mathsf{g}}R_{\mathsf{g}}D^7\int_0^{2\pi}\dd x^9\Omega^8e^{-2\Phi_{\rm I'}},
\end{equation}
with additional terms contributing to the moduli space metric through the kinetic term $\frac 12 \mathsf{G}_{ij}\partial_\mu\varphi^i\partial^\mu\varphi^j$. Now, defining
\begin{equation}
r=\int_0^{2\pi}\dd x^9\Omega^8e^{-2\Phi_{\rm I'}},
\end{equation}
to go to Einstein frame we must use for $D$ the following value,
\begin{equation}
	D=\left(\frac{r}{r_0}\right)^{-1/7}r_0^{-1/8},
\end{equation}
where $r_0$ is some auxiliary scale, which will not have any implication in the final result and we just include to have dimensionally sensible expressions, we introduce to have a metric metric with the correct dimensions. This way, we get
\begin{equation}
	S_{\rm I'} \supset\underbrace{\frac{r_0^{1/8}}{2\kappa_{10,\rm I'}^2}}_{\frac{1}{2\kappa_{9,\rm I'}^2}}\int\dd^9\sqrt{-\mathsf{g}}R_{\mathsf{g}}\quad \Longrightarrow\quad M_{\rm Pl;9}^7=\frac{r_0^{1/8}}{\kappa_{10,\rm I'}^2}
\end{equation}
so that
\begin{align}\label{eq: kappa 9 I}
D^{-1}&=r^{1/7}r_0^{-1/56}=r^{1/7}\kappa_{10,\rm I'}^{-2/7}M_{\rm Pl;9}^{-1}=\frac{2^7}{2\pi(\alpha'_{\rm I'})^{4/7}}\left(\int_0^{2\pi}\dd x^9\Omega^8e^{-2\Phi_{\rm I'}}\right)^{1/7}M_{\rm Pl;9}^{-1}
\end{align}
and finally
\begin{equation}\label{eq: eins met 9}
	\mathsf{g}_{\mu\nu}=\left(\frac{r}{r_0}\right)^{2/7}r_0^{1/4}\eta_{\mu\nu}
\end{equation}

\subsection{Heterotic-Type I$'$ duality relations\label{app: dual rel}}

Once we have the equations of motion associated to the Type I$'$ dilaton and warping, we can obtain the heterotic and Type I$'$ radii and couplings in terms of the $B$ and $C$ moduli, from which we can obtain the KK and winding modes, which are 1/2 BPS, and emergent string towers of the heterotic theory. Anchoring the scalar-charge-to-mass ratios of these 1/2 BPS masses allow us to determine how the I$'$ KK modes slide. The strategy we employ is to compute the masses in heterotic string theory, and then express the masses in terms of Type I$'$ string theory's $B$ and $C$ fields and 9d Planck constant. Because this derivation involves translating between I$'$ string theory and heterotic string theory, we keep all dimensionful terms (such as kappas and $\alpha'$'s) explicit.

We start with the following terms appearing in the heterotic  10D action \cite{Ibanez:2012zz}:
\begin{equation}
	S_{\rm h}\supset\frac{1}{2\kappa_{10,\rm h}^2}\int\dd^{10} x\sqrt{-g_{\rm h}}e^{-2\Phi_{\rm h}}\left \{R_{g_{\rm h}}-\frac{\alpha'_{\rm h}}{4}\Tr_V F_2^2\right\},
\end{equation}
with $\kappa_{10,\rm h}^2=\frac{(2\pi)^{7}}{2}(\alpha'_{\rm h})^4$.
On the other hand, we find that for the Type I$'$ theory in the presence of D8-branes perpendicular to the $x^9$ direction and located at $\{x_i^9\}_{i=1}^{16}$ has
\begin{equation}
S_{\rm I'}\supset\frac{1}{2\kappa_{10,\rm I'}}\int\dd^{10}x\sqrt{-g_{\rm I'}}e^{-2\Phi_{\rm I'}}R_{g_{\rm I'}}-\frac{(\alpha'_{\rm I'})^{-5/2}}{8(2\pi)^6}\sum_{i=1}^{16}\int_{x^9=x^9_i}\dd^9 x\sqrt{-g_{\rm I'}^{(d)}}e^{-\Phi_{\rm I'}}\Tr_V F_2^2,
\end{equation}
where we have expanded the DBI action for D$p$-branes, with $B$-field set to zero
\begin{equation}
	S_{{\rm DBI},p}=-\frac{(\alpha')^{-\frac{p+1}{2}}}{(2\pi)^p}\int_{\Sigma^{p+1}}\dd^{p+1} x e^{-\Phi}\sqrt{-\det(g-2\pi\alpha'F_2)}
\end{equation}
up to $\mathcal{O}(\alpha')^2$ order and used that $\det(\mathds{1}+M)=\exp\left[\Tr_f\log(\mathds{1}+M)\right]$, with $\Tr_f=\frac{1}{2}\Tr_V$. Again, $\kappa_{10,\rm I'}^2=\frac{(2\pi)^{7}}{2}(\alpha'_{\rm I'})^4$.

We will consider that our metrics are conformally flat, with $g_{\rm I'}{}_{MN}=\Omega(x^9)^2\eta_{MN}$ in the 10-dimensional string frame, and $g_{\rm h}{}_{\mu\nu}=\mathsf{g}_{\rm I'}{}_{\mu\nu}=D^{-2}\eta_{\mu\nu}$ (so that the 9-dimensional Einstein frame metric is the same in both theories), with the compact dimension being along a circle of radius $R_h$. For the time being, we will not assume any specific form for the $\Omega$ and $e^{\Phi_{\rm I'}}$, only that they depend on the internal coordinate $x^9$.
In order to relate the parameters from the two theories, we can compare their actions. We start doing so with the gravitational terms:
\begin{align}
S_{\rm h}^{\rm (grav)}&=\frac{1}{2\kappa_{10,\rm h}^2}\int\dd^{10} x\sqrt{-g_{\rm h}}e^{-2\Phi_{\rm h}}R_{g_{\rm h}}\notag\\
&=\frac{1}{2\kappa_{10,\rm h}^2}\int\dd^9 x\int_0^{2\pi} \dd x^9 \sqrt{-g_{\rm h}^{(d)}}\underbrace{\sqrt{g_{\rm h}^{(9)}} }_{R_h}e^{-2\phi_{\rm h}}(R_{g_{\rm h}^{(d)}}+\underbrace{R_{g_{\rm h}^{(9)}}}_{0})\notag\\
&=\frac{2\pi  R_{\rm h}e^{-2\Phi_{\rm h}}}{2\kappa_{10,\rm h}^2}\int \dd^9 x \sqrt{-g_{\rm h}^{(d)}}R_{g_{\rm h}^{(d)}}
\end{align}
	Note that from the above we recover the usual expression for $\kappa_{9,\rm h}$,
	\begin{equation}
		\frac{1}{2\kappa_{9,\rm h}^2}=\frac{2\pi  R_{\rm h}e^{-2\Phi_{\rm h}}}{2\kappa_{10,\rm h}^2}\Longrightarrow \kappa_{9,\rm h}^2=\kappa_{10,\rm h}^2(2\pi R_{\rm h})^{-1}e^{2\Phi_{\rm h}}
	\end{equation}
	Now, by using that $\kappa_{9}^2=2 M_{\rm Pl;9}^{7}$, with the Planck mass theory-independent, we have that (using \eqref{eq: kappa 9 I}),
	\begin{equation}
		\left(\frac{\kappa_{10,\rm h}}{\kappa_{10,\rm I'}}\right)^2=\left(\frac{\alpha'_{\rm h}}{\alpha'_{\rm I'}}\right)^4=2\pi R_{\rm h}e^{-2\Phi_{\rm h}}\frac{\left(\frac{1}{2\pi}\int_0^{2\pi}\dd x^9\Omega\right)^7}{\int_0^{2\pi}\dd x^9\Omega^8e^{-2\Phi_{\rm I'}}},
	\end{equation}
	which we will later use. On the other hand, from the Type I$'$ action,
\begin{align}
S_{\rm I'}^{\rm (grav)}&=\frac{1}{2\kappa_{10,\rm I'}}\int\dd^{10}x\sqrt{-g_{\rm I'}}e^{-2\Phi_{\rm I'}}R_{g_{\rm I'}}\notag\\
&=\frac{1}{2\kappa_{10,\rm I'}^2}\int\dd^{9}x\int_0^{2\pi}\dd x^9\underbrace{\sqrt{-g_{\rm I'}^{(d)}}}_{\Omega^9D^9\sqrt{-g_{\rm I'}^{(d)}}}\underbrace{\sqrt{-g_{\rm I'}^{(9)}}}_{\Omega}e^{-2\Phi_{\rm I'}}(R_{g_{\rm I'}^{(d)}}+\underbrace{R_{g_{\rm I'}^{(9)}}}_{0})\notag\\
&=\frac{D^7}{2\kappa_{10,\rm I'}^2}\int \dd^9 x \sqrt{-g_{\rm h}^{(d)}}R_{g_{\rm h}^{(d)}}\int_0^{2\pi}\dd x^9\Omega^8e^{-2\Phi_{\rm I'}},
\end{align}
where we have used that in the 10-dimensional String frame, $g_{\rm I'}{}_{\mu\nu}=\Omega^2D^2g_{\rm h}{}_{\mu\nu}$, as as such, $R_{g_{\rm I'}^{(d)}}=\Omega^{-2}D^{-2}R_{g_{\rm h}^{(d)}}$ (with $\Omega$ is independent of the macroscopic coordinates). Now, by comparing the two actions, one gets
\begin{equation}\label{eq:grav}
	2\pi R_{\rm h}e^{-2\Phi_{\rm h}}=\left(\frac{\alpha'_{\rm h}}{\alpha'_{\rm I'}}\right)^4 D^7 \int_0^{2\pi}\dd x^9\Omega^8e^{-2\Phi_{\rm I'}}
\end{equation}
On the other hand, we can relate the gauge terms of both actions, as working in an analogous way as above
\begin{align}
S_{\rm h}^{\rm (gauge)}&=\frac{\alpha'_{\rm h}}{8\kappa^2_{10,\rm h}}\int \dd^{10} x\sqrt{-g_{\rm h}}e^{-2\Phi_{\rm h}}\Tr_VG_2^2=\frac{\alpha'_{\rm h}}{8\kappa^2_{10,\rm h}}2\pi R_{\rm h}e^{-2\Phi_{\rm h}}\int\dd^9 x\sqrt{-g_{\rm h}^{(d)}}\Tr_VF_2^2\\
S_{\rm I'}^{\rm (gauge)}&=\frac{(\alpha_{\rm I'})^{-5/2}}{8(2\pi)^6}\sum_{i=1}^{16}\int_{x^9=x^9_i}\dd^9 x\sqrt{-g_{\rm I'}^{(d)}}e^{-\Phi_{\rm I'}}\Tr_V F_2^2\notag\\
&=\frac{(\alpha_{\rm I'})^{-5/2}}{8(2\pi)^6}\sum_{i=1}^{16}\left.\Omega^5e^{-\Phi_{\rm I'}}\right|_{x^9=x^9_i}\int\dd^9 x\sqrt{-g_{\rm h}^{(d)}}\Tr_VF_2^2,
\end{align}
where in the last step one must take into account the use of the inverse metric to raise indices in $F_2$. This way, one obtains the following relation
\begin{equation}\label{eq:gauge}
2\pi R_{\rm h}e^{-2\Phi_{\rm h}}=\pi(\alpha_{\rm h}')^3(\alpha'_{\rm I'})^{-5/2}D^5\sum_{i=1}^{16}\left.\Omega^5e^{-\Phi_{\rm I'}}\right|_{x^9=x^9_i}
\end{equation}
This way, one can use eqs. \eqref{eq:grav} and \eqref{eq:gauge} to obtain the following value for $D$:
\begin{equation}
D=\sqrt{\pi}(\alpha'_{\rm h})^{-1/2}(\alpha'_{\rm I'})^{3/4}\sqrt{\frac{\sum_{i=1}^{16}\left.\Omega^5e^{-\Phi_{\rm I'}}\right|_{x^9=x^9_i}}{\int_0^{2\pi}\dd x^9\Omega^8 e^{-2\Phi_{\rm I'}}}}
\end{equation}
One can see that, in order for eqs. \eqref{eq:grav} and \eqref{eq:gauge} to make sense, we have that $[D]=L^{-1}$ and $[\Omega]=L$. 
The above relations are valid for both the $SO(32)$ and $E_8\times E_8$ heterotic string theories. In order to find a third relation that allows us to obtain the expression of $D$, $R_h$ and $g_h$, we need to identify different string states between the heterotic and Type I$'$ theories.

First of all, for the $SO(32)$ theories, we have that we can identify the masses of the heterotic KK and Type I$'$ winding states, which are dual and BPS. Following \cite{Polchinski:1995df}, we have that $m_{\rm KK, h}=R_{\rm h}^{-1}=D m_{\rm w, I'}$, so that
\begin{equation}\label{eq:BPS SO32}
	\frac{1}{R_{\rm h}}=\frac{D}{2\pi\alpha_{\rm I'}}2\int_0^{2\pi}\dd x^9\Omega^2
\end{equation}
in 10-dimensional Planck units, where we have used that $T_s=\frac{1}{2\pi\alpha'_{I'}}$, the area element is $\dd A=\Omega(x^9)^2\dd x^9\dd x^0$, and that we wrap the string from one orientifold to the other and back. 

Now, expressions \eqref{eq:grav}, \eqref{eq:gauge} and \eqref{eq:BPS SO32} mix both heterotic and Type I$'$ $\alpha'$ factors. Substituting $\alpha'_{\rm h}=g_{\rm h}^{-1}\alpha'_{\rm I'}$, we obtain that for the $SO(32)$ theories,
\begin{subequations}
	\begin{align}\label{D so32}
		D&=2^{1/4}(\alpha'_{\rm I'})^{-1/2}\left(\int_0^{2\pi}\dd x^9\hat\Omega^2\right)^{-1/4}
	\left(\int_0^{2\pi}\dd x^9\hat\Omega^8 e^{-2\Phi_{\rm I'}}\right)^{1/4}
	\left(\sum_{i=1}^{16}\left.\hat\Omega^5 e^{-\Phi_{\rm I'}}\right|_{x^9=x^9_i}\right)^{-1/2}\\
		R_{\rm h}&=\frac{\pi}{2^{1/4}}(\alpha'_{\rm I'})^{1/2}\left(\int_0^{2\pi}\dd x^9\hat\Omega^2\right)^{-3/4}
	\left(\int_0^{2\pi}\dd x^9\hat\Omega^8 e^{-2\Phi_{\rm I'}}\right)^{-1/4}
	\left(\sum_{i=1}^{16}\left.\hat\Omega^5 e^{-\Phi_{\rm I'}}\right|_{x^9=x^9_i}\right)^{1/2}\\
		g_{\rm h}&=\frac{\sqrt{2}}{\pi}\left(\int_0^{2\pi}\dd x^9\hat\Omega^2\right)^{3/2}
	\left(\int_0^{2\pi}\dd x^9\hat\Omega^8 e^{-2\Phi_{\rm I'}}\right)^{-2}
	\left(\sum_{i=1}^{16}\left.\hat\Omega^5 e^{-\Phi_{\rm I'}}\right|_{x^9=x^9_i}\right)^{-1/2},
	\end{align}
\end{subequations}
where we have introduced the dimensionless function $\hat{\Omega}=(\alpha'_{\rm I'})^{-1/2}\Omega$. The above expressions have the expected dimensions and units. While we could try to use duality relations to obtain the expression of the Type I$'$ radius and coupling, this will not be necessary. Using \eqref{eq: kappa 9 I} and \eqref{D so32}, we find
\begin{equation}
M_{\rm Pl; 9}=\frac{2^{25/4}}{\pi}(\alpha'_{\rm I'})^{-\frac{1}{2}}\biggl(\int_0^{2\pi}\dd x^9\hat\Omega^2\biggr)^{-\frac{1}{4}}
	\biggl(\int_0^{2\pi}\dd x^9\hat\Omega^8 e^{-2\Phi_{\rm I'}}\biggr)^{\frac{11}{28}}
	\biggl(\sum_{i=1}^{16}\left.\hat\Omega^5 e^{-\Phi_{\rm I'}}\right|_{x^9=x^9_i}\biggr)^{-\frac{1}{2}},
\end{equation}
On the other hand, in the $E_8\times E_8$ case, we have that the heterotic winding (where the winded heterotic strings at small $g_{\rm h}$ and large $R_{\rm h}$ are wrapped M2-branes from M-theory with large $R_9$ and small $R_{10}$) and Type I$'$ winding modes with the strings wrapped between one $O8$-plane and the D8-brane located further away inside the interval at $x_I=0,\,2\pi-B$ (where the winded strings are wrapping M2-branes from M-theory perspective). This way
\begin{equation}\label{eq: BPS E8E8}
m_{\rm w, h}=\frac{R_{\rm h}}{2\pi\alpha'_{\rm h}}=\frac{D}{2\pi\alpha'_{I'}}2\int_0^{2\pi-B}\dd x^9\Omega^2=m_{\rm w,I'}
\end{equation}
From the above equation and \eqref{eq:grav} and \eqref{eq:gauge} we obtain, using again that $\alpha'_{\rm h}=g_{\rm h}^{-1}\alpha'_{\rm I'}$,
\begin{subequations}
	\begin{align}
		D&=\sqrt{2}(\alpha_{\rm I'})^{-1/2}\left(\int_0^{2\pi-B}\dd x^9\hat\Omega^2\right)^{1/4}
	\left(\sum_{i=1}^{16}\left.\hat\Omega^5 e^{-\Phi_{\rm I'}}\right|_{x^9=x^9_i}\right)^{-1/4}\\
		R_{\rm h}&=\sqrt{2}(\alpha'_{\rm I'})^{1/2}\left(\int_0^{2\pi-B}\dd x^9\hat\Omega^2\right)^{3/4}
	\left(\int_0^{2\pi}\dd x^9\hat\Omega^8 e^{-2\Phi_{\rm I'}}\right)^{-1}
	\left(\sum_{i=1}^{16}\left.\hat\Omega^5 e^{-\Phi_{\rm I'}}\right|_{x^9=x^9_i}\right)^{5/4}\\
		g_{\rm h}&=\frac{2}{\pi}\left(\int_0^{2\pi-B}\dd x^9\hat\Omega^2\right)^{1/2}
	\left(\int_0^{2\pi}\dd x^9\hat\Omega^8 e^{-2\Phi_{\rm I'}}\right)
	\left(\sum_{i=1}^{16}\left.\hat\Omega^5 e^{-\Phi_{\rm I'}}\right|_{x^9=x^9_i}\right)^{-3/2},
	\end{align}
\end{subequations}
as well as
\begin{equation}
	M_{\rm Pl;9}=\frac{2^{13/2}}{\pi}(\alpha'_{\rm I'})^{-1/2}\Biggl(\int_0^{2\pi-B}\dd x^9\hat\Omega^2\Biggr)^{1/4}
	\Biggl(\int_0^{2\pi}\dd x^9\hat\Omega^8 e^{-2\Phi_{\rm I'}}\Biggr)^{1/7}
	\Biggl(\sum_{i=1}^{16}\left.\hat\Omega^5 e^{-\Phi_{\rm I'}}\right|_{x^9=x^9_i}\Biggr)^{-1/4}
\end{equation}

\subsection{Masses of BPS towers}\label{sec: BPS mass}
Using the expressions obtained in the above subsection one can finally compute the expression for the mass of the different towers:
\begin{subequations}\label{eqs: masses SO32}
\begin{align}
m_{\rm osc, I'}^{SO(32)}&\sim (\alpha'_{\rm I'})^{-1/2}\notag\\
&\sim \left(\int_0^{2\pi}\dd x^9\hat\Omega^2\right)^{1/4}
	\left(\int_0^{2\pi}\dd x^9\hat\Omega^8 e^{-2\Phi_{\rm I'}}\right)^{-11/28}
	\left(\sum_{i=1}^{16}\left.\hat\Omega^5 e^{-\Phi_{\rm I'}}\right|_{x^9=x^9_i}\right)^{1/2}M_{\rm Pl; 9}\\
m_{\rm KK, h}^{SO(32)}&\sim\frac{D}{\pi \alpha_{\rm I'}}\int_0^{2\pi}\dd x^9\Omega^2\notag\\
&\sim  \left(\int_0^{2\pi}\dd x^9\hat\Omega^2 \right) \left(\int_0^{2\pi}\dd x^9\hat\Omega^8e^{-2\hat\Phi_{\rm I'}}\right)^{-1/7}M_{\rm Pl; 9}\\
m_{\rm osc, h}^{SO(32)}&\sim(\alpha'_{\rm h})^{-1/2}=g_{\rm h}^{1/2}(\alpha'_{\rm I'})^{-1/2}\notag\\ &\sim\left(\int_0^{2\pi}\dd x^9\hat\Omega^8 e^{-2\Phi_{\rm I'}}\right)^{-5/14}
	\left(\sum_{i=1}^{16}\left.\hat\Omega^5 e^{-\Phi_{\rm I'}}\right|_{x^9=x^9_i}\right)^{-1/2}M_{\rm Pl; 9}\\
m_{\rm w, h}^{SO(32)}&\sim\frac{R_{\rm h}}{2\pi \alpha'_{\rm h}}=\frac{g_{\rm h}R_{\rm h}}{2\pi\alpha'_{\rm I'}}\notag\\&\sim \left(\int_0^{2\pi}\dd x^9\hat\Omega^2\right)^{-1}
	\left(\int_0^{2\pi}\dd x^9\hat\Omega^8 e^{-2\Phi_{\rm I'}}\right)^{-6/7}
	\left(\sum_{i=1}^{16}\left.\hat\Omega^5 e^{-\Phi_{\rm I'}}\right|_{x^9=x^9_i}\right)^{-1}M_{\rm Pl; 9}
,\end{align}
\end{subequations}

Similarly, the $E_8\times E_8$ towers are given by
\begin{subequations}\label{eqs: masses E8E8}
\begin{align}
	m_{\rm osc,I'}^{E_8\times E_8}&\sim(\alpha'_{\rm I'})^{-1/2}\notag\\
	&\sim \left(\int_0^{2\pi-B}\dd x^9\hat\Omega^2\right)^{-1/4}
	\left(\int_0^{2\pi}\dd x^9\hat\Omega^8 e^{-2\Phi_{\rm I'}}\right)^{-1/7}
	\left(\sum_{i=1}^{16}\left.\hat\Omega^5 e^{-\Phi_{\rm I'}}\right|_{x^9=x^9_i}\right)^{1/4} M_{\rm Pl; 9}\\
	m_{\rm KK,h}^{E_8\times E_8}&\sim\frac{1}{R_{\rm h}}\notag\\
	&\sim \left(\int_0^{2\pi-B}\dd x^9\hat\Omega^2\right)^{-1}
	\left(\int_0^{2\pi}\dd x^9\hat\Omega^8 e^{-2\Phi_{\rm I'}}\right)^{6/7}
	\left(\sum_{i=1}^{16}\left.\hat\Omega^5 e^{-\Phi_{\rm I'}}\right|_{x^9=x^9_i}\right)^{-1} M_{\rm Pl;9}\label{eq: mKK E8E8}\\
	m_{\rm osc,h}^{E_8\times E_8}&\sim(\alpha'_h)^{-1/2}=g_{\rm h}^{1/2}(\alpha'_{\rm I'})^{-1/2}\notag\\
	&\sim
	\left(\int_0^{2\pi}\dd x^9\hat\Omega^8 e^{-2\Phi_{\rm I'}}\right)^{5/14}
	\left(\sum_{i=1}^{16}\left.\hat\Omega^5 e^{-\Phi_{\rm I'}}\right|_{x^9=x^9_i}\right)^{-1/2}M_{\rm Pl; 9}\\
	m_{\rm w, h}^{E_8\times E_8}&\sim\frac{D}{\pi\alpha'_{\rm I'}}\int_0^{2\pi-B}\dd x^9\Omega^2\sim \left(\int_0^{2\pi-B}\dd x^9\hat\Omega^2\right)
	\left(\int_0^{2\pi}\dd x^9\hat\Omega^8 e^{-2\Phi_{\rm I'}}\right)^{-1/7}M_{\rm Pl;9}\label{eq: mw E8E8}
\end{align}
\end{subequations}
We must take into account that the $z(x^9)$ has different expressions for the $SO(32)$ and $E_8\times E_8$, respectively given by \eqref{eq zso32} and \eqref{eq zE8E8}.

\section{Kaluza-Klein modes for Type I$'$ in nine dimensions\label{app:KK}}
In this section, we compute the moduli-dependence of the scaling of the masses of the highly-excited KK modes for Type I$'$ string theory in 9d, and we demonstrate a universal formula governing the scaling. We consider both the $SO(32)$ case and the $E_8\times E_8$ cases. We first compute the masses of these modes from the dilaton, then the RR 1-form, and finally show that our formulas apply to all KK modes that come from massless 9d fields.

In this section, we express everything in terms of the I$'$ 9d Planck mass, which we set to 1, since this allows us see clearly the scaling of the masses in terms of the $B$ and $C$ fields. Only the scaling is important in our analysis, because that determines the $\vec{\zeta}$-vectors that are computed in subsection \ref{app.sliding}. We are free to do this here because all of our analysis is in terms of I$'$ string theory, unlike the situation in Appendix \ref{app:dual} where we do not set the type I$'$ 9d Planck to 1 as in that case we compare I$'$ string theory with heterotic string theory.
\subsection{Background fields}

As derived in Appendix \ref{app:eom}, the equations of motion for the 10-dimensional string frame metric and dilaton are given by
\begin{subequations}\label{e.background}
\begin{align}
        g_{\rm I'}{}_{MN}&=\Omega(x^9)^2\eta_{MN},&
        e^{\Phi(x^9)}&=z(x^9)^{-5/6},&
        \Omega(x^9)&=Cz(x^9)^{-1/6},
\end{align}
where indices $M,N$ run from 0 to 9, with
\begin{align}
	z_\text{SO$(32)$}(x^9)&=z_0C(B+8x^9),\\
	z_{E_8\times E_8}(x^9)&=\begin{cases}
		z_0 Cx^9&0\leq x^9\leq B\\
 	z_0CB&B\leq x^9\leq 2\pi-B\\
 	z_0C(2\pi-x^9)& 2\pi-B\leq x^9\leq 2\pi,
	\end{cases}
\end{align}
\end{subequations}
where $B$ and $C$ fields are dimensionless\footnote{In Appendix \ref{app:eom} we obtained that $C$ had dimensions of lenght, but we can rescale it $C\to C(\alpha'_{\rm I'})^{-1/2}$ so that it becomes adimensional, resulting in $z_0$ being a numerical factor too.} with $z_0$ a numerical constant which will not be important for the subsequent derivations. 
This solution is sufficient for computing the I$'$ KK modes.

To get the $(d=9)$-dimensional theory, we integrate over the $x^9$ direction in the 10d action \eqref{e.10daction} using the backgrounds in \eqref{e.background}. However, the resulting action is not in Einstein frame. To get into Einstein frame, we must Weyl-rescale to the metric $\mathsf{g}_{\mu \nu}$ (where $\mu$ and $\nu$ run from 0 to 8), defined as
\begin{align}
	\mathsf{g}_{\mu\nu}=\left(\int \dd x^9 e^{-2\Phi}\Omega^{D-2}\right)^{\frac 2{d-2}}\eta_{\mu\nu}=\left(\int \dd x^9 e^{-2\Phi}\Omega^{D-2}\right)^{\frac 2{d-2}}\Omega^{-2}g_{\rm I'}{}_{MN}.\label{e.9dEinsteinmetric}
\end{align}
As we will argue below, the highly excited KK mode masses from all 10d fields will universally scale with the moduli via
\begin{align}
	m_{\text{KK}}^{\mathrm I'}\sim \left(\int_0^{2\pi}\dd x^9\Omega^8e^{-2\Phi_{\rm I'}}\right)^{-1/7}
\end{align}

\subsection{I$'$ KK masses}

\subsubsection*{KK modes from the dilaton \label{KKmassappendix}}

We now compute the $\vec{\zeta}$-vectors for high-excitation KK modes from the dilaton in I$'$ string theory.

The strategy we employ is as follows. First, we expand the dilaton $\Phi$ as a mode expansion, $\Phi(x^M)=\hat \Phi(x^9)+\sum_n \phi_n(x^\mu)f_n(x^9)$, where $\hat \Phi(x^9)$ is a background field and the functions $f_n(x^9)$ are a basis of functions on $x^9$. For a wise choice of $f_n(x^9)$, we have that in Einstein frame in 9d $\mathrm I'$ string theory the modes have an action that takes the form
\begin{align}
	\frac{1}{2}\int \dd^9 x\sqrt{-\mathsf{g}}\left(R_\mathsf{g}-\sum_n\bigg[(\partial\phi_n)^2+m_n^2\phi_n^2\bigg]\right)+\dots.
\end{align}
Since this is in Einstein frame, the KK-mode masses are just $m_n$ (times the 9d I$'$ Planck mass, which in the above formula is set to 1). With this mass, and also a computation of the metric on moduli space, we can find the scalar charge-to-mass ratios of the dilaton's KK modes.

To find out the KK mode masses from the dilaton using the above prescription, let us decompose the dilaton into a background and some fluctations using the following expansion ansatz.
\begin{align}
	\Phi(x^M)=\hat \Phi(x^9)+\sum_n \phi_n(x^\mu)f_n(x^9),\label{e.dilatonexpansion}
\end{align}
where $\hat \Phi(x^9)$ is the background value of the dilaton from \eqref{e.background}, $\phi_n$ are the KK modes of the dilaton and are $x^9$ independent, and $f_n$ is $x^\mu$-independent and a basis for functions of $x^9$. When we plug the ansatz \eqref{e.dilatonexpansion} into the action \eqref{e.10daction}, we have that the dilaton's KK modes $\phi_n$ appear in the action in the following way,
\begin{align}
	S_{\phi_n}\sim \int \dd^{10} x\sqrt{-g_{\rm I'}}e^{-2\hat \Phi}\sum_{m,n}\nabla_M(\phi_mf_m)\nabla^M(\phi_nf_n).\label{e.10dkkaction}
\end{align}

To find out what the masses of the dilaton's highly-excited KK modes are in 9d, we would the basis $f_n$ to be such that, in 9d Einstein frame, the mode expansion takes the form
\begin{align}
	S_{\phi_n}\sim \int\dd^9 x\sqrt{-\mathsf{g}} \sum_n\left((\partial \phi_n)^2+m_{n,\text{KK}}^2 \phi_n^2\right).\label{e.9dkkaction}
\end{align}
That is, we would like to have the kinetic and mass parts of the modes $\phi_n$ to be diagonal. The diagonality in \eqref{e.9dkkaction} does not automatically follow from \eqref{e.10dkkaction}. To obtain it, we need the basis $f_n(x_9)$ to be carefully chosen so that both the kinetic and mass parts of \eqref{e.9dkkaction} are diagonal. Fortunately, the following approach for finding a basis $f_n$ accomplishes this job.

Let us write the metric $g_{\rm I'}{}_{MN}$ in the following way,
\begin{align}
	\dd s^2_{D}=e^{a\sigma}h_{\mu \nu}(x^\alpha)\dd x^\mu \dd x^\nu +e^{2\sigma}h_{99}(\dd x^9)^2,\label{e.hmetric}
\end{align}
where $h_{\mu \nu}$ is $x^9$-independent, $a$ is some yet-to-be-determined number, and $\sigma$ and $h_{99}$ satisfying
\begin{align}
	e^{\alpha \sigma}=e^{2\sigma}h_{99}=\Omega^2.\label{e.sigmahconditions}
\end{align}
Inserting the metric \eqref{e.hmetric} into the KK-mode action \eqref{e.10dkkaction}, and using the fact that $\phi_n$ are $x^9$-independent and $f_n$ is $x^\mu$-independent, we have that the dilaton's KK modes are governed by the action
\begin{align}
	S_{\phi_n}\sim \int \dd^D x\sqrt{-h_dh_{99}}e^{\frac 12(da+2)\sigma}\sum_{m,n}\left(e^{-a\sigma}f_mf_nh^{\mu \nu}\partial_\mu\phi_m \partial_\mu\phi_n+ e^{-2\sigma}\phi_m \phi_n h^{99} \partial_9 f_m \partial_9 f_n\right),\label{e.phinhansatz}
\end{align}
Let us have $a$ in the ansatz \eqref{e.hmetric} satisfy
\begin{align}
	0=\frac{1}{2}(da+2)-a\quad\Longrightarrow\quad a=-\frac 2{d-2},\label{e.avalue}
\end{align}
as this choice allows us to perform the following integration by parts,
\begin{align}
	S_{\phi_n}&\sim  \int\dd^D x\sqrt{-h_dh_{99}}\sum_{m,n}\left(f_mf_nh^{\mu \nu}\partial_\mu\phi_m \partial_\nu\phi_n+ e^{-2\frac{d-1}{d-2}\sigma}\phi_m \phi_n h^{99} \partial_9 f_m \partial_9 f_n\right)\nonumber\\
	&= \int\dd^D x\sqrt{-h_dh_{99}}\sum_{m,n}\left(f_mf_nh^{\mu \nu}\partial_\mu\phi_m \partial_\nu\phi_n- \phi_m \phi_n  f_m  h^{99} \nabla_9(e^{-2\frac{d-1}{d-2}\sigma}\nabla_9 f_n)\right).\label{e.phinphimaction}
\end{align}
Now one can check that the operator
\begin{align}
	h^{99}\nabla_9^{(h)}\left(e^{-\frac{d-1}{d-2}\sigma}\nabla_9^{(h)}\bullet\right)
\end{align}
is indeed self-adjoint with respect to both the integration measures $\dd^D x\sqrt{-h_dh_{99}}$ and $\dd x^9\sqrt{h_{99}}$, so that its eigenvectors $\{f_n\}_n$ are orthogonal. As a result, we define the basis $f_n$ in \eqref{e.dilatonexpansion} to satisfy the eigenvector equation
\begin{align}
	h^{99}\nabla_9^{(h)}\left(e^{-\frac{d-1}{d-2}\sigma}\nabla_9^{(h)}f_n\right)=-\lambda_n^2 f_n.\label{e.eigenvector}
\end{align}
This implies that the KK-mode action \eqref{e.phinphimaction} can be rewritten as
\begin{align}
	S_{\phi_n}&\sim  \int\dd^9 x\sqrt{-h_d}\sum_{m,n}\left(h^{\mu \nu}\partial_\mu\phi_m \partial_\nu\phi_n+\lambda_n^2 \phi_m \phi_n \right)\int \dd x^9\sqrt{h_{99}}f_m f_n\nonumber\\
	&=\int\dd^9 x\sqrt{-h_d}\sum_{n}\left(h^{\mu \nu}\partial_\mu\phi_n \partial_\nu\phi_n+\lambda_n^2 \phi_n ^2\right)\int \dd x^9\sqrt{h_{99}}f_n^2.\label{e.Sphinphimprogress}
\end{align}
In going from the first to second line, we used the fact that $\int \dd x^9 \sqrt{-h_{99}}f_mf_n\propto\delta_{mn}$, implied by the orthogonality of the $\{f_n\}_n$ basis with respect to the $\dd x^9\sqrt{-h_{99}}$ measure.

To proceed, we need to find out what $\lambda_n$ and $f_n$ are. Under the WKB approximation (where $\lambda_n$ is assumed to be very large), and using \eqref{e.sigmahconditions}, we have that the eigenvalue equation \eqref{e.eigenvector} for $f_n$ yields
\begin{align}
	-\lambda_n^2f_n\approx \partial_9^2f_n+\mathcal O(\lambda_n).
\end{align}
That is, under the WKB $\lambda_n\gg 1$ approximation, $f_n$ takes the form\footnote{Had we included the $\mathcal O(\lambda_n)$ term in the eigenfunction equation, an overall power of the warping factor would appear before the $\cos$ function, not affecting the $\lambda_n$ expression.}
\begin{align}
	f_n(x^9)=c_n \cos(\lambda_n x^9+k_n).
\end{align}
The constants $c_n$ and $k_n$ in the above equation are fixed by the boundary conditions, and are not important for our analysis, as we are not interested in the precise nature of these boundary conditions, just that $f_n$ has moduli-independent periodicity. The periodicity on $f_n$ results in $\lambda_n$ being an integer,
\begin{align}
	\lambda_n=n.
\end{align}
For very large $n$, we thus have
\begin{align}
	\int \dd x^9\sqrt{h_{99}}f_n^2\approx \frac 12c_n^2 \int \dd x^9\sqrt{h_{99}},
\end{align}
and thus for high excitation modes, \eqref{e.Sphinphimprogress} becomes
\begin{align}
	S_{\phi_n}\sim \int \dd^d x\sqrt{-h_d}\sum_n \left(h^{\mu \nu}\partial_\mu \phi_n\partial_\nu \phi_n+n^2\phi_n^2\right) \int \dd x^9\sqrt{h_{99}}.\label{e.Sphinh}
\end{align}

The metric in \eqref{e.Sphinh} is not in Einstein frame, so we cannot interpret the coefficient in front of $\phi_n^2$ as being the mass. When we express the metric $h_{\mu \nu}$ \eqref{e.hmetric} in terms of the Einstein frame metric $\mathsf{g}_{\mu\nu}$ \eqref{e.9dEinsteinmetric}, the action for the KK modes becomes
\begin{align}
	S_{\phi_n}= \int \dd^d x\sqrt{-\mathsf{g}}\sum_n \left((\partial\phi_n)^2+\left(\int \dd x^9 e^{-2\hat\Phi}\Omega^{D-2}\right)^{-\frac 2{d-2}}n^2\phi_n^2\right).\label{e.SKKdilaton}
\end{align}
Since \eqref{e.SKKdilaton} is in Einstein frame, we can read off the dilaton's high-excitation KK-mode mass,
\begin{align}
	m_{{\rm KK}}^{(\Phi)}=\left(\int \dd x^9 e^{-2\hat\Phi}\Omega^{D-2}\right)^{-\frac 1{d-2}} =\left(\int \dd x^9 e^{-2\hat\Phi}\Omega^{8}\right)^{-\frac 1{7}}
	,\label{e.phiKKmass}
\end{align}
in 9d I$'$ Planck units.

\subsubsection*{KK modes from the RR 1-form}
We now calculate the mass of the KK mode from the RR 1-form, and this approach is similar to the calculation for the KK modes from the dilaton. The RR 1-form is governed by the following action,
\begin{align}
	S_{F_2^2}=-\frac{1}{2}\int\dd^D x\sqrt{-g_{\rm I'}}F_2^2.
\end{align}
We decompose the 1-form into the following basis,
\begin{align}
	A_M(x^\mu,x^9)&=\sum_n A^{(n)}_M(x^\mu)g_n(x^9).
\end{align}
where $A^{(n)}_M$ are $x^9$ independent and $g_n(x^9)$ are $x^\mu$-independent. With this decomposition, we have
\begin{subequations}
\begin{align}
	F_{MN}F^{MN}&=F_{\mu \nu}F^{\mu \nu}+2F_{\mu 9}F^{\mu 9}F_{99}F^{99}\\
	F_{\mu \nu}F^{\mu \nu}&=\sum_{mn}F^{(n)}_{\mu \nu}F^{(m)\mu \nu}g_ng_m\\
	F_{\mu 9}F^{\mu 9}&=\sum_{mn}\left(\nabla_\mu A_9^{(n)}-A^{(n)}_\mu\partial_9 \log g_n\right)\left(\nabla^\mu A^{9(m)}-A^{(m)\mu} \partial^9 \log g_n\right)g_n g_m\\
	F_{99}F^{99}&=0.	
\end{align}
\end{subequations}

For highly excited modes,
\begin{align}
	F_{\mu 9}F^{\mu 9}&\approx \sum_{mn}A^{(n)}_\mu A^{(m)\mu} g^{99}\partial_9 g_n \partial_9 g_m.
\end{align}
Thus, under the WKB approximation,
\begin{align}
	S_{F^2}= -\frac{1}{2}\int\dd^D x\sqrt{-g_{\rm I'}}\sum_{mn}\left(F^{(n)}_{\mu \nu}F^{(m)\mu \nu}g_ng_m +A^{(n)}_\mu A^{(m)\mu} g^{99}\partial_9 g_n \partial_9 g_m \right)
\end{align}

To proceed, consider the following metric ansatz,
\begin{align}
	\dd s^2_D&=e^{a\varsigma}H_{\mu \nu}(x^\mu)\dd x^\mu \dd x^\nu+e^{2\varsigma}H_{99}(\dd x^9)^2,\label{e.Hansatz}
\end{align}
where $H_{\mu \nu}(x^\mu)$ is $x^9$-independent, and we can have this ansatz by having backgrounds $\varsigma$ and $H_{99}$ satisfy
\begin{align}
	e^{a\varsigma}=e^{2\varsigma}H_{99}=\Omega^2.
\end{align}

Using this metric, we have
\begin{align}
\begin{aligned}
	S_{F_2^2}&= -\frac{1}{2}\int\dd^D x\sqrt{-H_dH_{99}}e^{\frac{1}{2}(da+2)\varsigma}\\
	&\sum_{mn} H^{\mu \nu }\left(e^{-2a\varsigma}H^{\rho\sigma}F^{(n)}_{\mu\rho}F^{(m)}_{\nu\sigma}g_ng_m +e^{-(a+2)\varsigma}A^{(n)}_\mu A^{(m)}_\nu H^{99}\partial_9 g_n \partial_9 g_m \right).
\end{aligned}
\end{align}
If we have the ansatz \eqref{e.Hansatz} satisfy,
\begin{align}
	0=\frac 12(da+2)-2a\quad \Longrightarrow \quad a=-\frac{2}{d-4}.
\end{align}
then we can integrate by parts to get
\begin{align}
	S_{F^2}&= -\frac{1}{2}\int\dd^D x\sqrt{-h_dh_{99}}\times\notag\\
	&\quad\times\sum_{mn}H^{\mu \nu}\left(H^{\rho \sigma}F^{(n)}_{\mu \rho}F^{(m)}_{\nu \sigma}g_ng_m -A^{(n)}_\mu A^{(m)}_\nu g_n H^{99} \nabla_{9}^{(H)}\left(e^{-2\frac{d-3}{d-4}\varsigma}\nabla_{9}^{(H)} g_m \right)\right)_h.
\end{align}
Note that $H^{99}\nabla_9^{(H)}\left[e^{-2\frac{d-3}{d-4}\varsigma}\nabla_9^{(H)}\bullet\right]$ is self-adjoint with respect with the measures $\dd^D x\sqrt{-H_dH_{99}}$ and $\dd x^9\sqrt{H_{99}}$, so eigenvectors of this operator are orthogonal with respect to these measures. Thus, we choose our basis $g_m$ to satisfy
\begin{align}
	H^{99}\nabla_9^{(H)}\left[e^{-2\frac{d-3}{d-4}\varsigma}\nabla_9^{(H)}g_m\right]=-\lambda_m^2g_m.
\end{align}

For highly excited modes,
\begin{align}
	-\lambda^2_ng_{n}&\approx \partial_9^2g_n+\mathcal O(\lambda_n)
\end{align}

Imposing moduli-independent periodicity, this is satisfied by $\lambda_n= n$ and $g_n(x^9)= \sqrt 2 \sin(nx^9)$, and so for highly excited modes,

\begin{align}
	S_{F^2}= -\frac{1}{2}\int \dd^d x\sqrt{-H_d}\left(\int \dd x^9\sqrt{H_{99}}\right)\sum_{n}\left(F^{(n)}_{\mu \nu}F^{(m)\mu \nu} +n^2A^{(n)}_\mu A^{(n)\mu}\right)_h.
\end{align}

Let's now compare with Einstein frame. Switching from $H_{\mu \nu}$ \eqref{e.Hansatz} to $\mathsf{g}_{\mu \nu}$ \eqref{e.9dEinsteinmetric}, we get (after locally canonically normalizing the massive vector)
\begin{align}
	S_{F^2}= -\frac{1}{2}\int \dd^d x\sqrt{-\mathsf{g}}\sum_{n}\left(F^{(n)}_{\mu \nu}F^{(m)\mu \nu} -n^2\left(\int \dd x^9 e^{-2\hat\Phi}\Omega^{D-2}\right)^{-\frac 2{d-2}}A^{(n)}_\mu A^{(n)\mu}\right).\label{e.SF2Einstein}
\end{align}
Since the above action \eqref{e.SF2Einstein} is in Einstein frame, we can read off the mass as
\begin{align}
	m_{\text{KK}}^{(A_1)}&\sim \left(\int \dd x^9 e^{-2\hat\Phi}\Omega^{D-2}\right)^{-\frac 1{d-2}}=\left(\int \dd x^9 e^{-2\hat\Phi}\Omega^{8}\right)^{-\frac 1{7}},
\end{align}
again in the appropriate 9d Planck units. This is the same as the mass of the KK mode from the dilaton.

\subsubsection*{Generalization to KK modes from any massless I$'$ field}
In fact, all high-excitation KK modes from massless fields I$'$ string theory have masses that scale with the moduli in exactly the same way. In general, suppose we have, suppressing indices, some massless field $\Psi$ (e.g. the dilaton, a $p$-form, or the metric) with some number of Lorentz indices, and in 10d the action for this field is schematically (suppressing indices)
\begin{align}
	S[\Psi]\sim \frac 12\int\dd^{10} x\sqrt{-g_{\rm I'}}(R+ a(\hat\Phi)(\partial \Psi)^2),
\end{align}
where $a(\hat\Phi)$ is some function of the dilaton.

For high excitation modes, $\Psi$ behaves as
\begin{align}
	\Psi\sim \sum_n\psi_n(x^\mu)\sin(nx^9),
\end{align}
and upon integrating over $x^9$, we have schematically
\begin{align}
	S\sim \frac 12\int\dd^9 x\sqrt{-\eta} \left(\left(\int \dd x^9 e^{-2\hat\Phi}\Omega^{D-2}\right)R+b(\phi)\sum_n\left[(\partial \psi_n)^2+n^2\psi_n^2\right]\right),
\end{align}
where $b(\phi)$ some unspecified function of the moduli and $\eta$ is the Minkowski metric.

Now, note that this is not in Einstein frame. Moving to it, we obtain
\begin{align}
	S\sim \frac 12\int\dd^9 x\sqrt{-\mathsf{g}} \left(R+c(\phi)\sum_n\left[(\partial \psi_n)^2+\left(\int \dd x^9 e^{-2\hat\Phi}\Omega^{D-2}\right)^{-\frac 2{d-2}}n^2\psi_n^2\right]\right),
\end{align}
for some unimportant function $c(\phi)$. This way, switching to Einstein frame causes the kinetic term and the mass terms to always  differ by a frame-switching factor, namely $\left(\int \dd x^9 e^{-2\Phi}\Omega^{D-2}\right)^{-\frac 2{d-2}}$, due to both terms depending on different powers of the metric.

Thus, the highly excited I$'$ KK mode from any massless field in 10d has a mass that satisfies a universal dependence on the moduli, given in the 9-dimensional Planck units by
\begin{align}
	m_n\sim \left(\int \dd x^9 e^{-2\hat\Phi}\Omega^{D-2}\right)^{-\frac 1{d-2}}=\left(\int \dd x^9 e^{-2\hat\Phi}\Omega^{8}\right)^{-\frac 1{7}}.
\end{align}

\section{Moduli space metric, flat coordinates and sliding\label{app:metric}}
In order to compute the scalar charge-to-mass vectors associated to the different towers, we will need the moduli space metric $\mathsf{G}_{ij}$. Because of the warping of the internal dimension, the moduli space metric will not correspond to the usual hyperbolic metric $\mathsf{G}_{ij}=\frac{\mathsf{G}_{ij}^{(0)}}{\varphi^i\varphi^j}$ unless in the low warping limits. The easiest way to obtain it is by considering the expression of scalar charge-to-mass vectors of the masses,
\begin{equation}
\zeta^i_I=-\delta^{ij}e^{a}_j\partial_a\log m_I,
\end{equation}
and noting that is nothing but a linear transformation in $T_p\mathcal{M}$ to the flat frame described by normal coordinates (so that the moduli space metric is given by $\mathsf{G}_{ab}=\delta_{ab}$) $\partial_a\log m\to \zeta^i$ by the matrix $\delta^{ij}e^{a}_j$. Knowing how the elements of a basis of $T_p\mathcal{M}$ transform will give us the expression of $\delta^{ij}e^{a}_j$. Now, in \S \ref{sec:decompactification} it was argued that BPS states (such as the heterotic KK and winding modes) have fixed $\vec{\zeta}_I$. Denoting $\hat{e}^{ia}=\delta^{ij}e^a_j$, $(\zeta_{\rm BPS})^{i}_J= \zeta^i_J$ and $(M_{\rm BPS})_{aI}=\partial_a\log m_I$, then $\hat{e}=\zeta_{\rm BPS}M_{\rm BPS}^{-1}$, and from here $\mathsf{G}=(\hat{e}^\intercal\hat{e})^{-1}$. The BPS towers mass are given in Section \ref{sec: BPS mass}, and we can ask that in some normal coordinates of $T_p\mathcal{M}$ (all of them will be related by a $SO(\dim \mathcal{M})=SO(2)$ transformation)
$$
\vec{\zeta}_{\rm KK,h}=\left(1,-\frac{1}{\sqrt{7}}\right),\qquad 
		\vec{\zeta}_{\rm w,h}=\left(-1,-\frac{1}{\sqrt{7}}\right),		
		$$
		which corresponds with the expected result in the low-warping limit.

We first start with the $SO(32)$ moduli space metric. While the full bulk moduli space metric is slightly complicated,\footnote{The complete expression of the $BB$ component being
\begin{equation}\label{eq:full metBB}
\resizebox{1\hsize}{!}{$\mathsf{G}^{SO(32)}_{BB}=\frac{4}{63}\frac{\left(-800 \pi  B^{\frac 23}-50 B^{\frac 53}+3584 \pi ^2 B^{-\frac 13}\right) (B+16 \pi)^{-\frac 13}+\left(1792 \pi ^2 B^{-\frac{2}{3}}+25 B^{\frac{4}{3}}\right)+\left(25 B^2+800 \pi  B+8192 \pi ^2\right) (B+16 \pi )^{-\frac{2}{3}}}{\left((B+16 \pi )^{\frac{5}{3}}-B^{\frac 53}+16 \pi B^{1/3} (B+16 \pi)^{1/3} \right)^{2}}$}
\end{equation} }

we recover the following asymptotic expression:
\begin{subequations}\label{eq: GSO32}
\begin{align}
	\mathsf{G}^{SO(32)}_{BB}&=\left\{
	\begin{array}{ll}
	\frac{1}{2^{1/3}72\pi^{4/3}B^{2/3}}+\mathcal{O}(1)&\text{for }B\ll 1\\
	\frac{22}{63B^2}+\mathcal{O}(B^{-3})&\text{for }B\gg 1
	\end{array}
	\right.\\
	\mathsf{G}^{SO(32)}_{CC}&=\frac{100}{63 C^2}\\
	\mathsf{G}^{SO(32)}_{BC}=\mathsf{G}_{CB}^{SO(32)}&=\frac{100}{63 C}\frac{(B+16 \pi )^{2/3}-B^{2/3}}{(B+16 \pi )^{5/3}-B^{5/3}+16 \pi  {B}^{1/3} (B+16 \pi)^{1/3}}\notag\\
	&=\left\{
	\begin{array}{ll}
	\frac{25}{252\pi C}+\mathcal{O}(C^{-1}B^{1/3})&\text{for }B\ll 1\\
	\frac{25}{63CB}+\mathcal{O}(C^{-1}B^{-2})&\text{for }B\gg 1
	\end{array}
	\right.
\end{align}
\end{subequations}

We can use the above expression to obtain the geodesics of the moduli space. The $B\gg1$ case is pretty straightforward, with geodesic equations resulting in the usual
\begin{equation}
	(B,C)(\lambda)=(B_0\lambda^{b},C_0\lambda^c),\qquad \lambda\gg 1,\, b>0,
\end{equation}
which we could implicitly rewrite as $B\sim C^\alpha$ for some $\alpha>0$. Choosing $b=0$ results in a geodesic sending $C\to 0,\,\infty$ for fixed $B$. On the other hand, it is not difficult to show that 
\begin{equation}
	(B,C)(\lambda)=(B_0,C_0\lambda),
\end{equation}
also solve the geodesic equations in any point of the moduli space. This results in a $(B,C)\to(B_0,0),\,(B_0,\infty)$ limit, corresponding to trajectories with null tangent $B$ component.

On the other hand, for the $E_8\times E_8$ component, the moduli space metric is given by
\begin{subequations}\label{eq: GE8E8}
	\begin{align}
	\mathsf{G}^{E_8\times E_8}_{BB}&=\frac{4 \left(25 B^2-50 \pi  B+88 \pi ^2\right)}{63 B^2 (B-4 \pi )^2}=\left\{
	\begin{array}{ll}
	\frac{22}{63B^2}+\mathcal{O}(B^{-1})&\text{for }B\ll 1\\
	\frac{4}{9\pi^2}+\mathcal{O}(\pi-B)&\text{for }B\sim \pi
	\end{array}
	\right.\\
	\mathsf{G}^{E_8\times E_8}_{CC}&=\frac{100}{63 C^2}\\
	\mathsf{G}^{E_8\times E_8}_{BC}=\mathsf{G}_{CB}^{E_8\times E_8}&=\frac{100 (B-\pi )}{63 B (B-4 \pi ) C}=\left\{
	\begin{array}{ll}
	\frac{25}{63B C}+\mathcal{O}(C^{-1})&\text{for }B\ll 1\\
	\frac{100 (\pi-B) }{189 \pi ^2 C}+\mathcal{O}(C^{-1}(\pi-B)^{-2})&\text{for }B\sim \pi
	\end{array}
	\right.
\end{align}
 \end{subequations}
 As for the geodesics, they are analogous to the $SO(32)$ case, with $(B,C)(\lambda)=(B_0,C_0\lambda)$ being geodesic trajectories in every point of the moduli space, and $(B,C)(\lambda)=(B_0\lambda^b,C_0\lambda^c)$ for the $B,\lambda\ll 1$ and $b>0$, corresponding with the low warping limit.

Finally, in the two cases studied above it is not difficult to show, computing the only independent component of the Riemann tensor in a 2-dimensional manifold, that both moduli spaces are flat.

Another possible way of obtaining the moduli space metric is by dimensionally reducing the Einstein-dilaton terms in 10-dimensional Einstein frame of the action (with metric $\tilde{g}_{MN}=\Omega^2e^{-\hat{\Phi}_{\rm I'}/2}\eta_{MN}$), and inspecting the kinetic terms of the massless moduli, $\mathsf{G}_{ab}\partial_\mu \varphi^i\partial^\mu\varphi^j$ in the lower dimensional Einstein frame. 

This way, one obtains
\begin{align}
S_{\rm I'}&\supset\frac{1}{2\kappa_{10,\rm I'}^2}\int \dd^{10}x\sqrt{-\tilde{g}}\left\{R_{\tilde{g}}-\frac{1}{2}\left(\partial\hat{\Phi}_{\rm I '}\right)^2\right\}=\frac{1}{2\kappa_{\rm 9,I'}^2}\int\dd^9x\sqrt{-\mathsf{g}}\left\{R_{\mathsf{g}}-\mathsf{G}_{ab}\partial_\mu\varphi^a\partial^\mu\varphi^b\right\},
\end{align}
where
\begin{align}\label{eq: kin term}
	\mathsf{G}_{ab}\partial_\mu\varphi^a\partial^\mu\varphi^b&=\frac{1}{r}\int_0^{2\pi}\dd x^9\Omega^8e^{-2\hat{\Phi}_{\rm I'}}\left\{\frac{7}{8}\left[\partial\log\left(\frac{\Omega^8e^{-2\hat{\Phi}_{\rm I'}}r_0^{1/7}}{r^{8/7}}\right)\right]^2+\frac{1}{2}(\partial\hat{\Phi}_{\rm I'})^2\right\}+\delta_{\rm kin}^{(2)},
\end{align}

where $\delta_{\rm kin}^{(2)}$ is an extra, second order term, coming from the Ricci scalar reduction, given by
\begin{equation}
	\frac{1}{2\kappa_{9,\rm I'}^2}\int\dd^9x\sqrt{-\mathsf{g}}\delta_{\rm kin}^{(2)}=\frac{1}{2\kappa_{9,\rm I'}^2}\int\dd^9x\sqrt{-\mathsf{g}}\left\{\frac{2}{r}\int_0^{2\pi}\dd x^9\Omega^8e^{-2\hat{\Phi}_{\rm I'}}\Delta_{\mathsf{g}}\log\left(\frac{\Omega^8e^{-2\hat{\Phi}_{\rm I'}}r_0^{1/7}}{r^{8/7}}\right)\right\}
\end{equation}
In the low warping limit, the $x^9$ integral in the above expression factorizes and the above term vanishes, as then $\delta_{\rm kin}^{(2)}$ corresponds to a total derivative. It can be then checked that from \eqref{eq: kin term} the low warping limits of \eqref{eq: GSO32} and \eqref{eq: GE8E8} are recovered. However, this is not the case for points of the moduli space for which there is warping, as the above term does not vanish and the moduli space metric does not correspond with the expression obtained from \eqref{eq: kin term}. We then need to integrate by parts and substract the $B\to\infty$ or $B\to 0$ expressions, depending on whether we are considering the $SO(32)$ or $E_8\times E_8$ (which correspond to a total derivative, so that we recover the appropriate $B\to \infty$ or $B\to 0$ behavior), so that we can rewrite
\begin{align}
\delta_{\rm kin}^{(2)}=\hat{\delta}-\frac{1}{\sqrt{-\mathsf{g}}}\lim_{B\to \infty,0}\left[\sqrt{-\mathsf{g}}\hat{\delta}\right],
\end{align}
with 
\begin{equation}
\hat{\delta}=-\frac{2}{r}\int_0^{2\pi}\dd x^9\Omega^8e^{-2\hat\Phi_{\rm I'}}\left\{\left[\partial\log\left(\frac{\Omega^8e^{-2\hat{\Phi}_{\rm I'}}r_0^{1/7}}{r^{8/7}}\right)\right]^2+\frac{1}{7}\partial_\mu\log\left(\frac{r}{r_0}\right)\partial^\mu\log\left(\frac{\Omega^8e^{-2\hat{\Phi}_{\rm I'}}r_0^{1/7}}{r^{8/7}}\right)\right\}
\end{equation}
It can be checked that the only term in the metrics receiving contributions from $\delta_{\rm kin}^{(2)}$ is $\mathsf{G}_{BB}$. This way, in the $SO(32)$ case, from this two terms we obtain that for $B\ll 1$, $\mathsf{G}_{BB}^{SO(32)}\sim\frac{1}{2^{1/3}36\pi^{4/3}B^{2/3}}-\frac{1}{2^{1/3}72\pi^{4/3}B^{2/3}}=\frac{1}{2^{1/3}72\pi^{4/3}B^{2/3}}$, as found in \eqref{eq: GSO32}, while the contributions from $\delta_{\rm kin}^{(2)}$ are subleading with respect to $\frac{22}{63B^2}$ in the $B\gg 1$ limit, with $\delta_{\rm kin}^{(2)}$ vanishing. As for the $E_8\times E_8$ case one obtains that for any value of $B\in(0,\pi)$, now $\mathsf{G}^{E_8\times E_8}_{BB}=\frac{16 (B-22 \pi ) (B-\pi )}{63 B^2 (B-4 \pi )^2}-\frac{4 (B-\pi )}{3 B (B-4 \pi )^2}=\frac{4 \left(25 B^2-50 \pi  B+88 \pi ^2\right)}{63 B^2 (B-4 \pi )^2}$, as in \eqref{eq: GE8E8}. 

In any case, as it have been shown above, the moduli space metric is more straightforwardly computed by fixing the BPS towers.
\subsection{Flat coordinates\label{app:flat}}
Once the explicit expression of $\mathsf{G}^{SO(32)}$ and $\mathsf{G}^{E_8\times E_8}$ is known in terms of $\{B,C\}$, we can obtain a set of flat coordinates $\{\phi_B,\phi_C\}$ for which $\mathsf{G}_{\phi_i\Phi_j}=\delta_{ij}$.

We will start with the $SO(32)$. For this we take \eqref{eq: GSO32} and, after completing squares, impose
\begin{align}
\dd s^2_{\mathcal{M}_{SO(32)}}&=\frac{100}{63}\left[\frac{\dd C}{C}+\frac{(B+16\pi)^{1/3}-B^{1/3}}{(B+16\pi)^{4/3}-B^{4/3}}\dd B\right]^2\notag\\
&\qquad +\left[\frac{32\pi}{3B^{1/3}(B+16\pi)^{1/3}[(B+16\pi)^{4/3}-B^{4/3}]}\dd B\right]^2\notag\\
&=\dd\phi_C^2+\dd\phi_B^2,
\end{align}
which result in the following system:
\begin{subequations}
	\begin{align}
		\dd \phi_C&=\frac{10}{3\sqrt{7}}\left[\frac{\dd C}{C}+\frac{(B+16\pi)^{1/3}-B^{1/3}}{(B+16\pi)^{4/3}-B^{4/3}}\dd B\right]\\
		\dd\phi_B&=\frac{32\pi}{3B^{1/3}(B+16\pi)^{1/3}[(B+16\pi)^{4/3}-B^{4/3}]}\dd B\;.
	\end{align}
\end{subequations}
Note that each of the above equations are unique up to a $\pm$ sign, which we have the freedom to choose (the relation between different flat coordinates is only a $O(2)$ transformation that includes reflections along some axis). The above equations can be integrated (up to integration constants we choose to be zero) to
\begin{subequations}\label{eq: flat SO32}
	\begin{align}
		\phi_C&=\frac{10}{3\sqrt{7}}\log C+\frac{5}{2\sqrt{7}}\log \left[(B+16\pi)^{4/3}-B^{4/3}\right]\\
		\phi_B&=\frac{1}{2}\log\frac{(B+16\pi)^{2/3}+B^{2/3}}{(B+16\pi)^{2/3}-B^{2/3}}\,,
	\end{align}
\end{subequations}
with $\phi_c\in\mathbb{R}$ and $\phi_B\in(0,+\infty)$, corresponding with the $B\to 0$ and $B\to\infty$ limits.

\begin{figure}
\begin{center}
\begin{subfigure}{0.475\textwidth}
\center
\includegraphics[width=65mm]{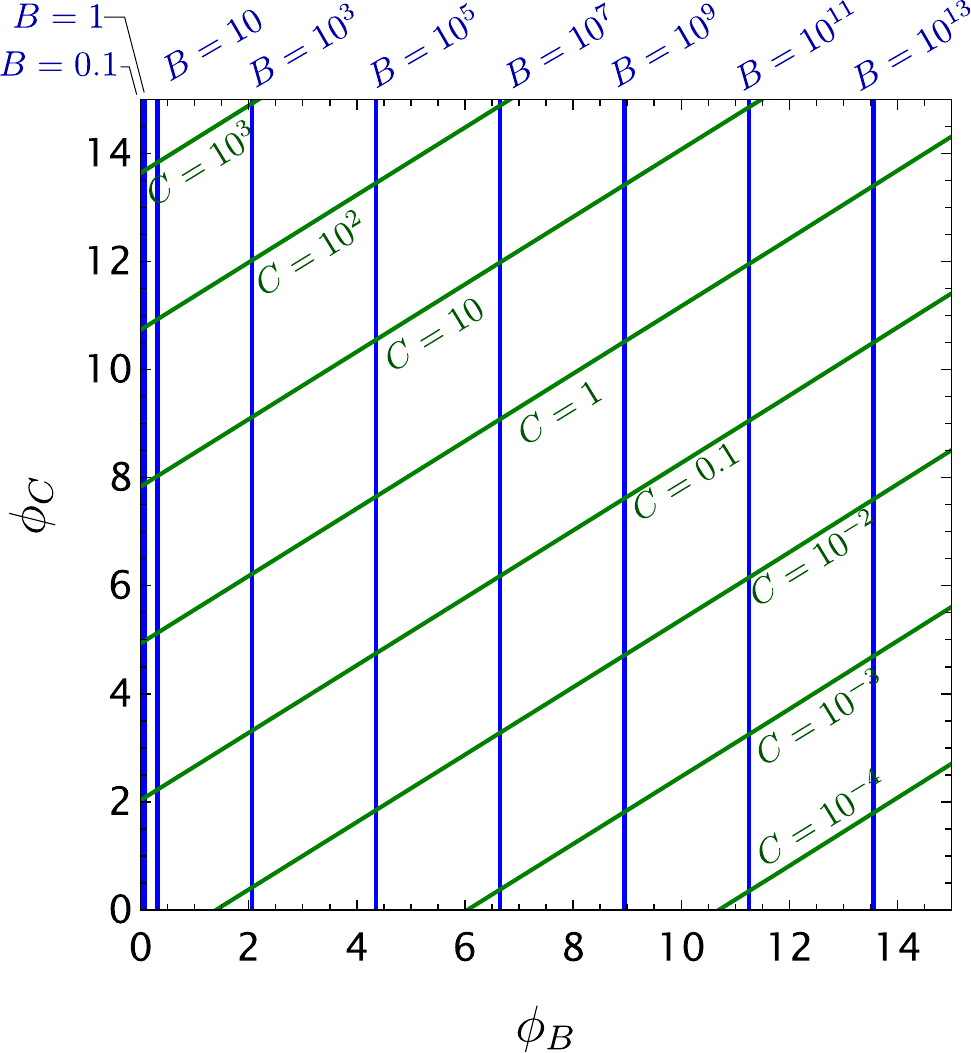}
\caption{$SO(32)$} \label{flatso32}
\end{subfigure}
\begin{subfigure}{0.475\textwidth}
\center
\includegraphics[width=72mm]{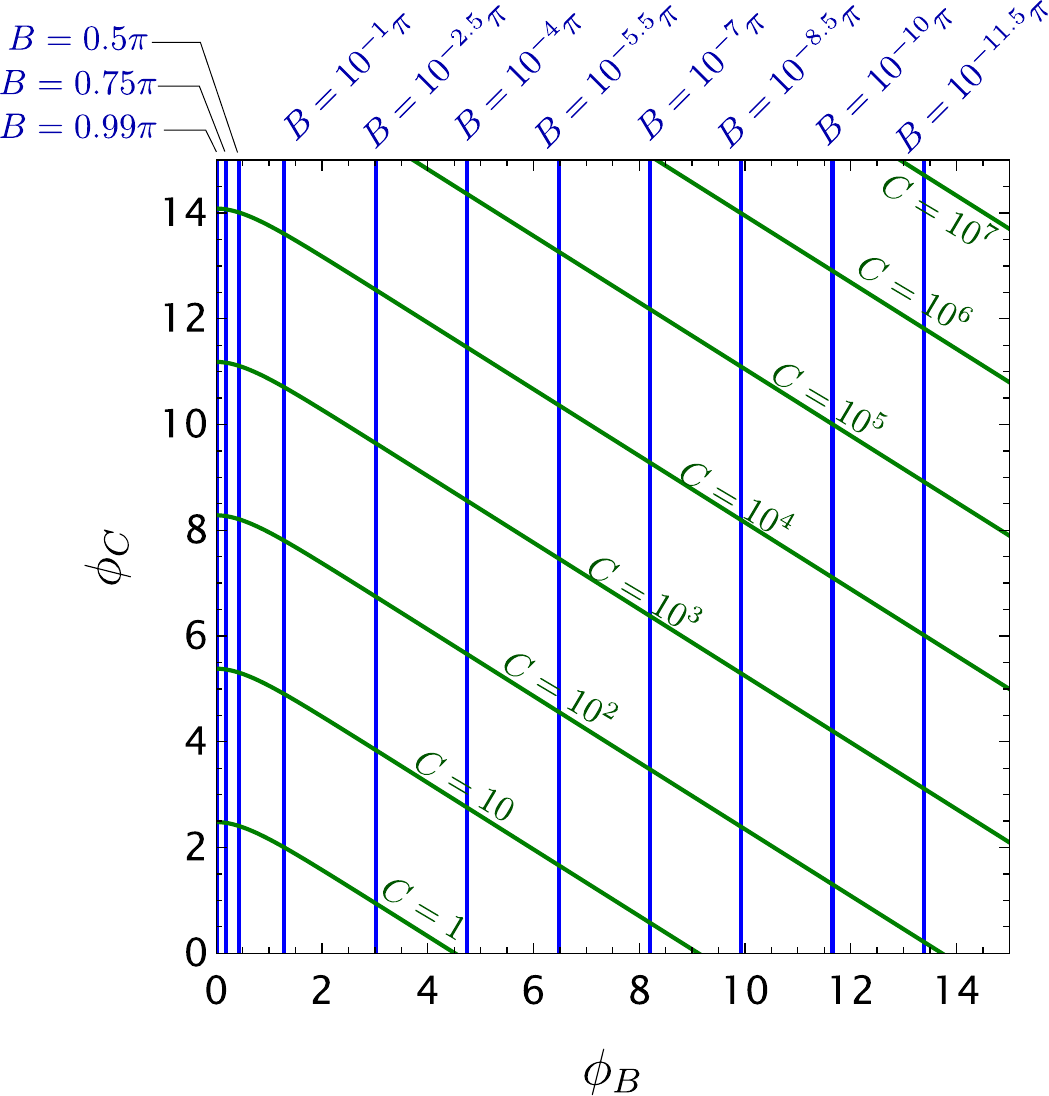}
\caption{$E_8\times E_8$} \label{flate8e8}
\end{subfigure}
\caption{Plot of the coordinate curves for constant $B$ (in blue) and $C$ (in green) in the $(\phi_B,\phi_C)$ flat chart for both the $SO(32)$ and $E_8\times E_8$ configurations. Note that $\phi_C\in\mathbb{R}$, so that it would continue for $\phi_c<0$, even though only the positive values are plotted.\label{fig:flat curves}}
\end{center}
\end{figure}

On the other hand, for $E_8\times E_8$ we take \eqref{eq: GE8E8} and impose
\begin{align}
	\dd s^2_{\mathcal{M}_{E_8\times E_8}}&=\frac{100}{63}\left[\frac{\dd C}{C}+\frac{\pi-B}{B(4\pi-B)}\dd B\right]^2+\left[\frac{2\pi}{B(4\pi-B)}\dd B\right]^2=\dd\phi_C^2+\dd\phi_B^2,
\end{align}
resulting in the following differential equations:
\begin{subequations}
	\begin{align}
		\dd \phi_C&=\frac{10}{3\sqrt{7}}\left[\frac{\dd C}{C}+\frac{\pi-B}{B(4\pi-B)}\dd B\right]\\
		\dd\phi_B&=-\frac{2\pi}{B(4\pi-B)}\dd B\,,\label{eq:dphiB B E8E8}
	\end{align}
\end{subequations}
where here we have chosen the $-$ sign for $\dd \phi_B$ equation, for reasons that will become clear soon. Upon integration (and setting constants to zero) they yield
\begin{subequations}
	\begin{align}\label{eq: flat E8E8}
		\phi_C&=\frac{10}{3\sqrt{7}}\log C+\frac{5}{6\sqrt{7}}\log\left[B(4\pi-B)^3\right]\\
		\phi_B&=-\frac{1}{2}\log\frac{3B}{4\pi-B}
	\end{align}
\end{subequations}
here $\phi_C\in\mathbb{R}$ and $\phi_B\in(0,+\infty)$, corresponding with the $B\to \pi$ and $B\to 0$ limits. In Figure \ref{fig:flat curves}, the coordinate curves for $B$ and $C$ are depicted in teh $(\phi_B,\phi_C)$ frame for $SO(32)$ and $E_8\times E_8$.

\subsection{KK mode sliding \label{app.sliding}}
Once we have the expression for flat coordinates $\{\phi_B,\phi_C\}$ in terms of $B$ and $C$ we can invert the relation and rewrite the heterotic KK and winding and Type I$'$ KK masses in terms of these expressions and see whether they remain constant as we move along moduli space. 
\begin{figure}
\begin{center}
\begin{subfigure}{0.475\textwidth}
\center
\includegraphics[width=70mm]{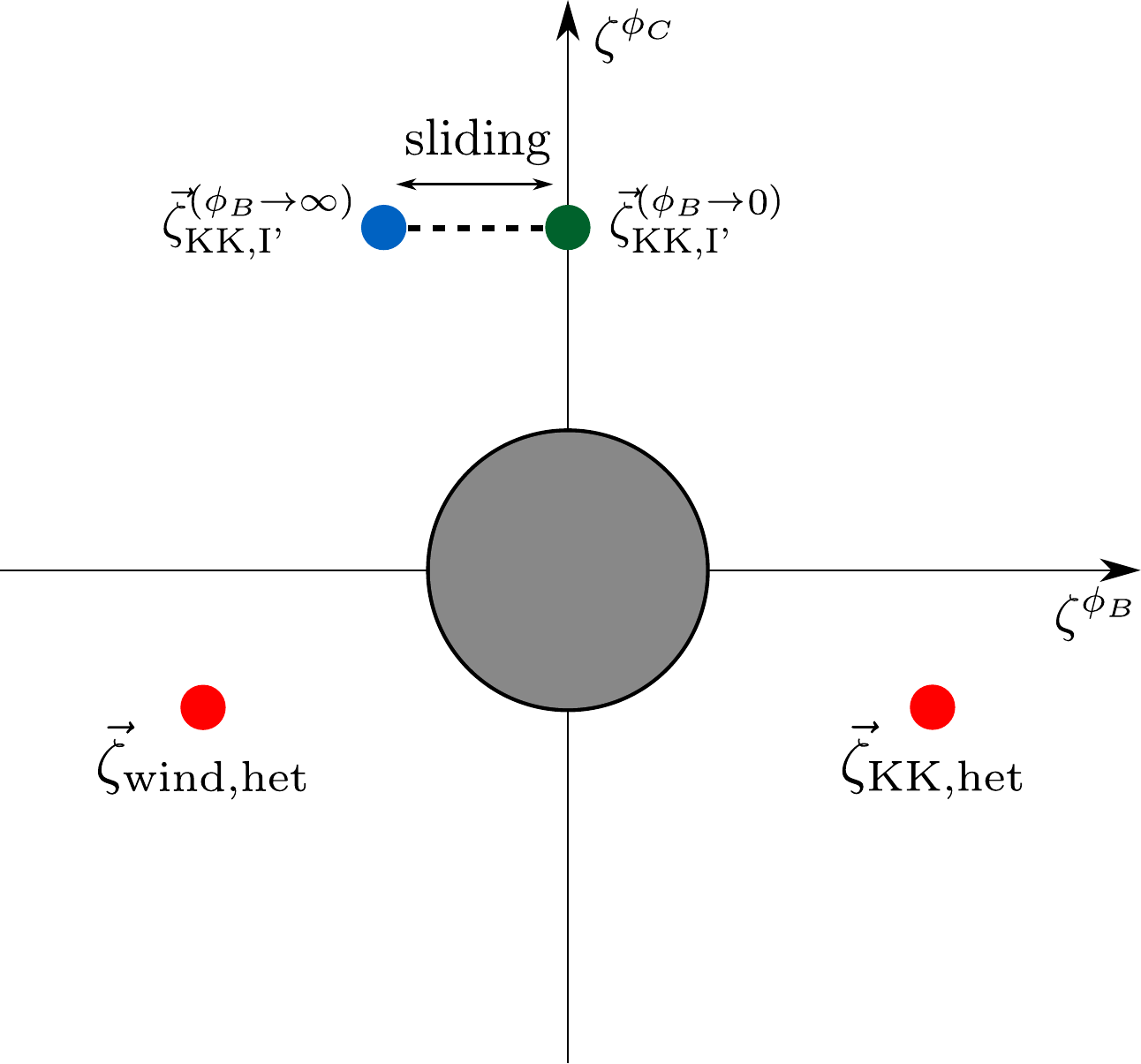}
\caption{$SO(32)$} \label{t.KKslidingSO32}
\end{subfigure}
\begin{subfigure}{0.475\textwidth}
\center
\includegraphics[width=70mm]{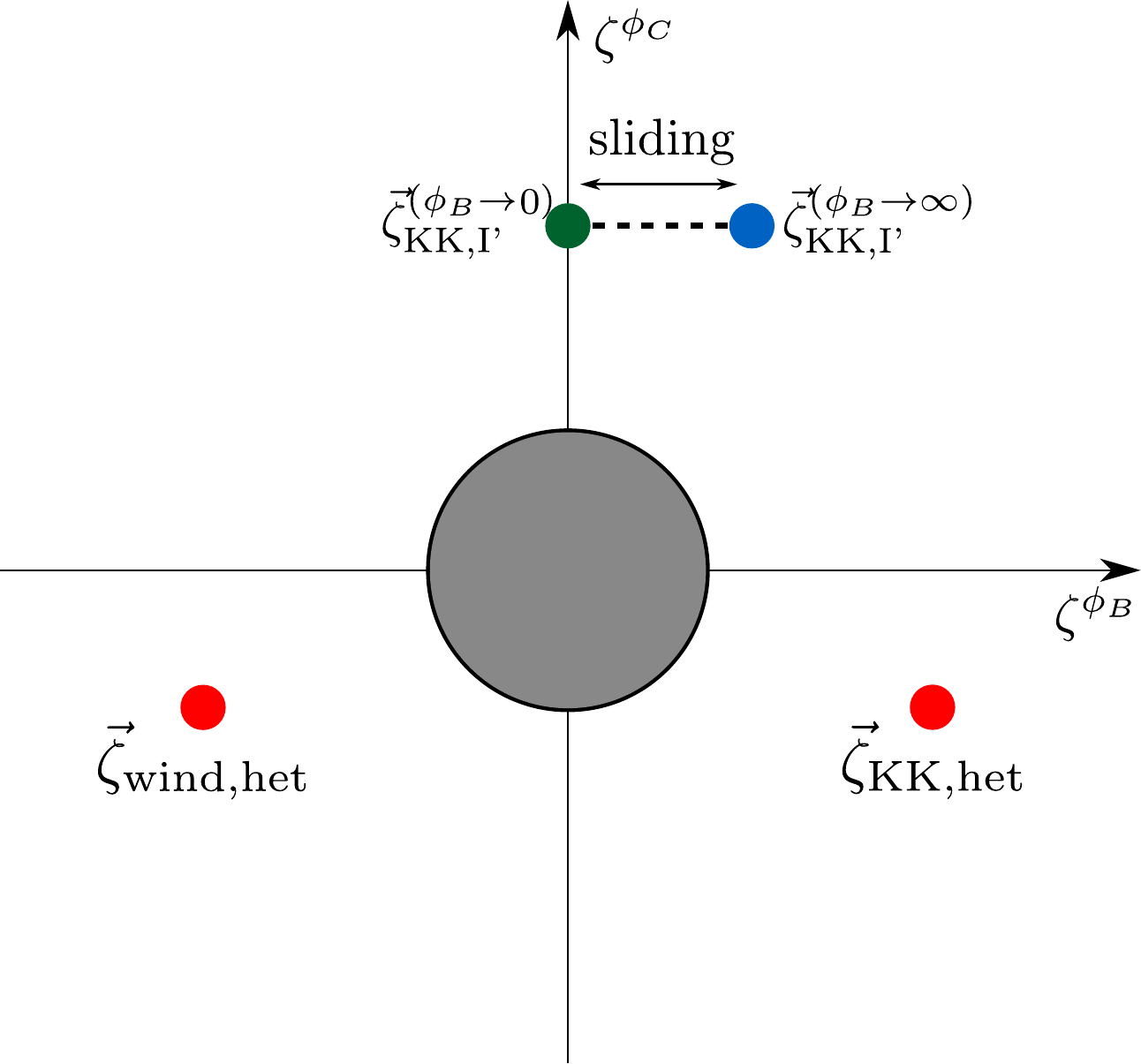}
\caption{$E_8\times E_8$} \label{t.KKslidingE8E8}
\end{subfigure}
\caption{The sliding of the I$'$ KK modes in the $SO(32)$ and $E_8\times E_8$ cases, in terms of the normalized $\phi_B$ and $\phi_C$ moduli. The sliding of I$'$ string theory's high excitation KK mode occurs across the dashed region. The region in gray is the ball of radius $\frac{1}{\sqrt{7}}$. Note that in this basis, the $\phi_C$-axis is the self-dual line. Recall that from $SO(32)$ the limits $\phi_B\to 0,+\infty$ correspond with $B\to 0,+\infty$, while for $E_8\times E_8$, $\phi_B\to 0,+\infty$ are given by $B\to \pi,\,0$.\label{fig: KKsliding}}
\end{center}
\end{figure}
First of all, for $SO(32)$, we find (after some algebraic effort) from \eqref{eqs: masses SO32} and \eqref{e.phiKKmass}
\begin{subequations}
	\begin{align}
	m_{\rm w,h}^{SO(32)}&\sim e^{\frac{1}{\sqrt{7}}\Phi_C+\phi_B}\\
	m_{\rm KK,h}^{SO(32)}&\sim e^{\frac{1}{\sqrt{7}}\Phi_C-\phi_B}\\
	m_{\rm KK,I'}^{SO(32)}&\sim\frac{(e^{2\phi_B}+1)^{3/2}+(e^{2\phi_B}-1)^{3/2}}{3e^{4\phi_B}+1}e^{\frac{3}{2}\phi_B-\frac{5}{2\sqrt{7}}\phi_C}
	\end{align}
\end{subequations}
resulting in the following scalar charge-to-mass vectors in the $\{\phi_B,\phi_C\}$ basis:
\begin{align}
\vec{\zeta}_{\rm w,h}=\left(-1,-\frac{1}{\sqrt{7}}\right),\qquad \vec{\zeta}_{\rm KK,h}=\left(-1,\frac{1}{\sqrt{7}}\right),\notag\\
\vec{\zeta}_{\rm KK,I'}=\left(-\frac{3}{2}\left[\frac{2}{\sqrt{1-e^{-4\phi_B}}}+1\right]^{-1},\frac{5}{2\sqrt{7}}\right)\;.
\end{align}
We see that all of the above components are constant but $\zeta^{\phi_B}_{\rm KK,I'}$, which is a monotonic function of $\phi_B$, with a sliding occurring from $\zeta^{\phi_B}_{\rm KK,I'}=-\frac{1}{2}$ for $\phi_B=\infty$ to $\zeta^{\phi_B}_{\rm KK,I'}=0$ at $\phi_B=0$.

On the other hand, for $E_8\times E_8$ we take \eqref{eqs: masses E8E8} and \eqref{e.phiKKmass} and invert \eqref{eq: flat E8E8} to find
\begin{subequations}
	\begin{align}
		m_{\rm w,h}&\sim e^{\frac{1}{\sqrt{7}}\phi_C+\phi_B}\\
		m_{\rm KK,h}&\sim e^{\frac{1}{\sqrt{7}}\phi_C-\phi_B}\\
		m_{\rm KK,I'}&\sim e^{-\frac{5}{2\sqrt{7}}\phi_C+\frac{3}{2}\phi_B}\left(1+3e^{2\phi_B}\right)^{-1}
	\end{align}
\end{subequations}
Note that now it is evident the implications of choosing the $-$ sign in \eqref{eq:dphiB B E8E8}, as this way the $m_{\rm w,h}$ and $m_{\rm KK,h}$ coincide for both $SO(32)$ and $E_8\times E_8$. Now the scalar charge-to-mass vectors have the following expressions:
\begin{align}
\vec{\zeta}_{\rm w,h}=\left(-1,-\frac{1}{\sqrt{7}}\right),\qquad \vec{\zeta}_{\rm KK,h}=\left(-1,+\frac{1}{\sqrt{7}}\right),\notag\\
\vec{\zeta}_{\rm KK,I'}=\left(\frac{1}{2}-\frac{2}{1+3e^{2\phi_B}},\frac{5}{2\sqrt{7}}\right)\;.
\end{align}
Here again all the components but $\zeta^{\phi_B}_{\rm KK,I'}$ are constant, with it being monotonic as a function of $\phi_B$ and sliding between $\zeta^{\phi_B}_{\rm KK,I'}=\frac{1}{2}$ for $\phi_B=\infty$ and $\zeta^{\phi_B}_{\rm KK,I'}=0$ for $\phi_B= 0$. This is depicted, for both $SO(32)$ and $E_8\times E_8$, in Figure \eqref{fig: KKsliding}.

\bibliographystyle{JHEP}
\bibliography{ref}
\end{document}